\documentclass[aoas]{imsart}

\RequirePackage{amsthm,amsmath,amsfonts,amssymb}
\RequirePackage[authoryear]{natbib}
\RequirePackage[colorlinks,citecolor=blue,urlcolor=blue]{hyperref}
\RequirePackage{graphicx}
\usepackage{enumitem}
\usepackage{slashbox}
\usepackage{booktabs}

\startlocaldefs
\numberwithin{equation}{section}
\theoremstyle{plain}
\newtheorem{theorem}{Theorem}

\theoremstyle{remark}

\newcommand{\Rmnum}[1]{\expandafter\@slowromancap\romannumeral #1@}

\def\xx{\mbox{xx}}
\def\xy{\mbox{xy}}
\def\fxx{\footnotesize\xx}
\def\fxy{\footnotesize\xy}
\def\real{\mbox{real}}

\def\sr{\scriptsize\real}
\def\oracle{\mbox{oracle}}
\def\so{\scriptsize\oracle}
\def\vt{\text{var}}
\def\wt{\widetilde}

\def\wh{\widehat}

\def\mD{\mathcal D}
\def\mQ{\mathcal Q}
\def\mR{\mathbb{R}}

\def\mL{\mathcal L}

\def\mK{\mathcal K}
\def\maE{\mathcal E}
\def\mC{\mathcal C}

\def\1{\mbox{\boldmath $1$}}

\def\mT{\mathcal T}

\def\cov{\mbox{cov}}

\def\argmin{\mbox{argmin}}
\def\argmax{\mbox{argmax}}

\def\BIC{\mbox{BIC}}

\usepackage{underscore}

\endlocaldefs

\begin{document}

\begin{frontmatter}
\title{Mixture Conditional Regression with Ultrahigh \\ Dimensional Text Data for Estimating \\ Extralegal Factor Effects}
\runtitle{Mixture Conditional Regression for Estimating Extralegal Factors}

\begin{aug}
\author[A]{\fnms{Jiaxin}~\snm{Shi}\ead[label=e1]{jxshi@stu.pku.edu.cn}},
\author[B]{\fnms{Fang}~\snm{Wang}\ead[label=e2]{wangfang226@sdu.edu.cn}},
\author[A]{\fnms{Yuan}~\snm{Gao}\ead[label=e3]{ygao_stat@outlook.com}},
\author[A]{\fnms{Xiaojun}~\snm{Song}\ead[label=e4]{sxj@gsm.pku.edu.cn}}
\and
\author[A]{\fnms{Hansheng}~\snm{Wang}\ead[label=e5]{hansheng@gsm.pku.edu.cn}}
\address[A]{Guanghua School of Management, Peking University, Beijing, China\printead[presep={,\ }]{e1,e3,e4,e5}}

\address[B]{Data Science Institute, Shandong University, Jinan, China\printead[presep={,\ }]{e2}}
\end{aug}

\begin{abstract}
Testing judicial impartiality is a problem of fundamental importance in empirical legal studies, for which standard regression methods have been popularly used to estimate the extralegal factor effects. However, those methods cannot handle control variables with ultrahigh dimensionality, such as found in judgment documents recorded in text format. To solve this problem, we develop a novel mixture conditional regression (MCR) approach, assuming that the whole sample can be classified into a number of latent classes. Within each latent class, a standard linear regression model can be used to model the relationship between the response and a key feature vector, which is assumed to be of a fixed dimension. Meanwhile, ultrahigh dimensional control variables are then used to determine the latent class membership, where a Na\"ive Bayes type model is used to describe the relationship. Hence, the dimension of control variables is allowed to be arbitrarily high. A novel expectation-maximization algorithm is developed for model estimation. Therefore, we are able to estimate the interested key parameters as efficiently as if the true class membership were known in advance. Simulation studies are presented to demonstrate the proposed MCR method. A real dataset of Chinese burglary offenses is analyzed for illustration purpose.
\end{abstract}


\end{frontmatter}

\section{Introduction}

Our research is empirically motivated by studies of equality, impartiality, and justice in jurisprudence \citep{2001myths,2006Jurisprudence}. Fairness and justice have defined features of the judicial role, and include aspects such as substantive decision-making by judges \citep{weiler1968two}, procedural justice \citep{2000procedure}, judicial independence \citep{meron2005}, and the proper assessment of scientific evidence \citep{2002evidence}. We emphasize criminal substantive justice in this paper. This means that judicial decisions must follow the principle of legality and should not be affected by prejudice regarding races, incomes, and other extralegal factors \citep{bright2008failure,2011Mapping}. 
Substantive justice represents the ultimate good of judicial impartiality. It is the core standard of good conduct and is essentially crucial for public confidence in the courts. In this regard, countries around the world have been promoting sustained reforms to unravel miscarriages of justice and safeguard judicial impartiality \citep{1993unravelling,stith1998fear,2010impact}. By doing so, people wish the extralegal factor effects on judicial impartiality can be controlled and minimized.

Despite the fact that judicial impartiality has been universally promoted over the world for a long history, bias and prejudice due to extralegal factors do exist in practice. In fact, this is one of the most important research areas in empirical legal studies \citep{1995miscarriages,2003Unsafe,2012exonerations}. Researchers have made enormous efforts to document cases and analyze the reasons behind them. For example, \cite{1993supreme} studied about 4,000 cases from the U.S. Supreme Court database in the period of 1956--1989 and found that the Court has been highly responsive to public opinion. Its decisions have not only reflected the American public's overall policy preferences, but reinforced and legitimized emerging majoritarian concerns. \cite{1998Steffensmeier} investigated 139,000 criminal conviction cases from Pennsylvania in 1989--1992 and found blacks and males to be more likely to be incarcerated and to receive longer sentences, even after controlling for the type and severity of the offense and the offender's prior record.
\cite{2001judging} analyzed 14,633 sentenced offenders from the state of Maryland and reported that African Americans had 20\% longer sentences on average than whites, holding constant age, gender, and recommended sentence length from the guide. \cite{2014judicial} studied 2,078 death penalty decisions issued by the U.S. state courts under four judicial selection systems between 1980 and 2006 and found that judges were significantly more responsive to majority opinions on capital sentences in nonpartisan election systems than partisan systems. \cite{2015identifying} examined 2,674 unique votes cast by 244 appeal judges from the U.S. Courts of Appeals on gender-related cases. They found that judges with daughters consistently tended to vote in a more feminist fashion on gender issues than judges with only sons. \cite{bielen2021prosecution} focused on 766 violent criminal cases that occured during the 12-week interval around the day of Theo van Gogh's assassination (November 2, 2004). They found that immediately afterward, the prospects of prosecution for unrelated violent crimes with male suspects born in Muslim-majority countries increased by about 19$\%$.

To summarize, there have been ample amount of empirical studies of a possible dependent relationship between judicial decisions and the interested extralegal factors, after controlling for a number of legal factors. 
Such problems can be nicely formulated as a regression problem with both the main covariates of interest and a set of control variables.
Specifically, the dependent variable ($Y$) is a judicial decision, the key variables ($X_1$) are extralegal factors (e.g., race, gender, public opinion), and the control variables ($X_2$) are legal factors (e.g., severity of the current offense, the offense type, the previous criminal record). Then, the interested problem becomes one of testing the statistical significance of the conditional regression relationship between $ Y $ and $ X_1$, after controlling for the effect of $ X_2 $. Assuming that the law is perfectly just and self-contained and that no partiality or prejudice exists, we should expect $ Y $ and $ X_1 $ to be conditionally uncorrelated with each other, after controlling for the effect of $ X_2 $.

To fix the idea, consider for example the study of Pennsylvania criminal conviction cases in 1989--1992 \citep{1998Steffensmeier}, where the response variable ($Y$) is the decision whether to incarcerate an offender; the extralegal factors of interest (the main covariates $X_1$) include race, gender, age, and their interaction; and the legal characteristic variables (the control variables $X_2$) include the type and severity of the offense and the offender's criminal record. To test the conditional regression relationship between $ Y $ and $ X_1 $ after controlling for $ X_2 $, a standard logistic regression model was employed. In a recent study of criminal cases happened before and after Theo van Gogh's assassination \citep{bielen2021prosecution}, the dependent variable ($Y$) is the decision whether to prosecute a charge; the interested extralegal factors (the main covariates $X_1$) are the unrelated extraneous events (e.g., the Theo van Gogh's assassination); and the legal factors (the control variables $X_2$) include the criminal history and type of charge. In this case, the difference-in-differences (DID) regression approach was used to study the conditional regression relationship between $ Y $ and $ X_1 $, after controlling for the effect of $X_2$. The literature suggests that testing conditional regression relationships should be of great importance for examining the effects of extralegal factors on judicial impartiality.

As one can see, to test the conditional regression relationship between the judicial decision ($Y$) and the extralegal factors ($X_1$), after controlling for the effects of legal factors ($X_2$), various standard regression models have been adopted for $ Y $ and ($X_1, X_2$). Those regression methods are easy to implement and have a nice interpretation. However, they also suffer from one serious limitation. That is they can only handle control variables ($X_2$) with a relatively low dimension. For our empirical study of Chinese judicial decisions, the legal factors ($X_2$) contain features extracted from legal documents, which are lengthy text documents. Each element of the $X_2$ vector is then of a binary form, representing the existence or not of one particular keyword in the judgments. Since the original judgments are written in Chinese and contain a large number of legal issue-related keywords, the dimension of $X_2$ is very high, which precludes the immediate use of standard regression methods. Hence, how to test the conditional regression relationship between $ Y $ and $ X_1 $ with an ultrahigh dimensional $ X_2 $ becomes a problem of great importance.

Before we formally solve this problem, we consider splitting the original ultrahigh dimensional vector $X_2$ into two parts. The first part contains a subvector of $X_2$, which has a fixed dimension and is strongly correlated with $Y$. In our case, this corresponds to those keywords in judgments that are not only high in frequency but highly correlated with the response variable $Y$.
This is merged with $X_1$ to form a new feature vector $X$, which has a fixed dimension and a strong correlation with $Y$. Therefore, a standard regression model can be used to describe the regression relationship for $Y$ and $X$. The remaining part of $X_2$ is formed as another new vector, $Z$, which has a very high dimension and is weakly correlated with the response variable $Y$. Since it is related to the response (even weakly), it does carry useful information for predicting it. However, since it is of ultrahigh dimension and is weakly related to the response, it can hardly be directly incorporated into a usual regression model structure. Then, how to effectively model the regression relationship between $Y$ and ($X,Z$) with a fixed dimensional $ X $ and an ultrahigh dimensional $Z$ becomes a key problem. 

To solve this problem, we develop here a novel mixture conditional regression approach. Our method contains two important components. The first component is a mixture model. We assume that all the samples (i.e., legal cases) can be grouped into different classes. Within each group, a standard linear regression model can be assumed for $ Y $ and $ X $. We allow the intercepts of those regression models to be class-specific, so that inter-class heterogeneity can be modeled. We force the regression coefficients of the main covariates to be the same across classes so that the overall main covariate effect can be quantified. As one can see, the main regression model assumed between $ Y $ and $ X $ for every class is low-dimensional. This makes the subsequent parameter estimation and statistical inference very easy. However, the main challenge here is that the class membership for every sample is a latent variable that is not directly observed. For our cases, the class membership is mainly determined by the judgments, which are represented by an ultrahigh dimensional binary vector. Therefore, it is theoretically appealing to assume for $Z$ a Na\"ive Bayes type mixture model, so that the rich information contained in $ Z $ can be fully utilized to identify the latent membership for each sample.

To summarize, our model allow a feature vector to affect the response by two different mechanisms as follows. The first mechanism is simply including a feature as an usual explanatory variable. The strength of this mechanism is that it can provide the best explanatory power from $X$ to $Y$ in a very direct way. The weakness of this approach is that the dimension of $X$ cannot be too large. Otherwise, we should suffer from the curse of dimensionality. The second mechanism is to relate a feature $Z$ with the response but indirectly through the latent class membership. The weakness of this approach is that the feature cannot affect the response directly. Therefore, the  explanatory power is sacrificed to some extend. However, the strength of this approach is that it can easily accommodate as many features as possible. This leads to an interesting phenomenon, that is the blessing of dimensionality. Simply speaking, each mechanism has its own strength and weakness theoretically. Therefore, they need to be treated differently in practice. 

For convenience, we refer to our model as a mixture conditional regression (MCR) model. To estimate it, we develop here a novel estimation method. It contains four steps. We show theoretically that the resulting estimators can be statistically as efficient as the oracle estimators, which are obtained by assuming that the latent class membership is known in advance. Extensive simulation studies are presented to demonstrate the finite sample performance of this method. A real data example of 6,118 judgments is analyzed for illustration purpose. The rest of this article is organized as follows. Section 2 develops the MCR model, including the four-step estimators and their asymptotic statistical properties. Simulation studies are presented in Section 3 and a real data example of judgments is analyzed in Section 4. Finally, Section 5 concludes with a brief discussion. All technical details are relegated to the Appendix.

\section{Methodology}
\subsection{A statistical model} 
Let $(Y_i, X_i)$ with $ 1\leq i\leq n$ be the observation collected from the $ i $-th subject. Here $ Y_i\in \mathbb{R}^1$ is the response of interest. In our case, it is the log-transformed sentence length and it is assumed to follow a continuous distribution. In the meanwhile, $ X_i=(X_{i1},\ldots,X_{iq})^{\top}\in\mathbb{R}^q$ is the associated main covariates. To model their regression relationship, we assume that 
\begin{equation}\label{2.1}
	Y_i=\sum_{k=1}^{K}I(\mathcal{K}_i=k)\gamma_k+X_i^{\top}\theta+\varepsilon_i, 
\end{equation}
where $ I(\cdot)$ is an indicator function, $\mathcal{K}_i\in\{1,2,\ldots,K\}$ is a latent categorical variable identifying the latent class membership of the $i$-th legal document, $\gamma_k$ is an associated unknown coefficient, $\theta=(\theta_1,\ldots,\theta_q)^{\top}\in\mR^q$ is the regression coefficient associated with confounding factors, and $\varepsilon_i$ is random noise, which we assume to follow a normal distribution with mean $ 0 $ and variance $\sigma^2$. It is noteworthy that we allow different intercepts $ \gamma_{k} $ according to class, so that the inter-class heterogeneity can be modeled. In the meanwhile, we assume the same coefficient $ \theta $ for different classes, so that the overall main covariate effect can be quantified.

We next consider how to model the dependence relationship between the latent class membership and the ultrahigh dimensional binary feature vector $Z_i=(Z_{i1},\ldots,Z_{ip})^{\top}\in\mathbb{R}^p$ with $Z_{ij}\in\{0,1\}$. Specifically, $Z_{ij}$ is defined to be 1 if a prespecified keyword appears in the $i$-th legal document and otherwise is 0. To model the dependence relationship between $\mathcal{K}_i$ and $Z_i$, a classical Na\"ive Bayes model \citep{1984Spiegelhalter} is assumed. Specifically, we assume that $ Z_i $ and $ (Y_i,X_i) $ are conditionally independent with $\mathcal{K}_i$ given. We also assume that $Z_{ij}$s for $1\leq j\leq p$ are mutually conditionally independent given $\mathcal{K}_i$. Mathematically, this jointly means that
\begin{equation*}
	P(Z_i|\mathcal{K}_i=k,Y_i, X_i)= P(Z_i|\mathcal{K}_i=k)=\prod_{j=1}^{p}P\Big(Z_{ij}|\mathcal{K}_i=k\Big)=\prod_{j=1}^{p}p_{kj}^{Z_{ij}}\Big(1-p_{kj}\Big)^{1-Z_{ij}},
\end{equation*}
where $p_{kj}=P(Z_{ij}=1|\mathcal{K}_i=k)$. Finally, we assume the class prior probability $P(\mathcal{K}_i=k)=\pi_{k}$. Recall that $\varepsilon_i$ follows a normal distribution with mean $ 0 $ and variance $\sigma^2$. Write $ \phi\big(\gamma_{k}+X_i^{\top}\theta,\sigma^2\big)=\big(2\pi\sigma^2\big)^{-1}\exp\Big\{-\big(Y_i-\gamma_k-X_i^{\top}\theta\big)^2/\big(2\sigma^2\big)\Big\}$. Then, a log-likelihood function can be written as
\begin{equation}\label{2.2}
	\mathcal{L}(\Theta)= \sum_{i=1}^{n}\log\bigg\{\sum_{k=1}^K\pi_k\phi\Big(\gamma_{k}+X_i^{\top}\theta,\sigma^2\Big)\prod_{j=1}^pp_{kj}^{Z_{ij}}\Big(1-p_{kj}\Big)^{1-Z_{ij}}\bigg\},
\end{equation}
where $\Theta=(\pi^{\top},\gamma^{\top},\theta^{\top},\sigma^2,\text{vec}(P)^{\top})^{\top}\in\mR^{2K+q+1+pK}$, $\pi=(\pi_1,\ldots,\pi_K)^{\top}\in\mR^K $, $\gamma=(\gamma_1,\ldots,\gamma_K)^{\top}\in\mR^K $, and $P=(p_{kj})\in\mR^{K\times p} $. 
For an arbitrary matrix $A$ with dimension $M\times N$ , vec($A$) stands for an $ MN\times 1$ column vector defined by stacking the columns of the matrix $A$ on top of one another. Then the maximum likelihood estimator (MLE) can be obtained as $\wh{\Theta}_{\text{mle}}=(\wh{\pi}_{\text{mle}}^{\top},\wh{\gamma}_{\text{mle}}^{\top},\wh{\theta}_{\text{mle}}^{\top},\wh{\sigma}^2_{\text{mle}}, \text{vec}(\wh{P}_{\text{mle}})^{\top})^{\top}=\argmax_{\Theta}\mL(\Theta)$. 

\subsection{Identifiability, interestingness and relevance}

To estimate the the model (\ref{2.2}), it is crucial to ensure its identifiability. As one can see, if the response $Y$ is integrated out, we are then left with binary feature $Z$ and the covariate $X$ only. In this case, the identifiability becomes a serious issue. Non-identifiable examples can be easily constructed.
Therefore, for model identification purpose, we cannot integrate $Y$ out. Instead, we need to make full use of $Y$ for model identification purpose. 
To fix this idea, consider for example a highly simplified case with $Y$ observed only (even without the binary feature vector $Z$), then the model (\ref{2.2}) reduces to a standard mixture regression model. As pointed out by \cite{shalabh2008finite}, this model can be nicely identified under appropriate regularity conditions; see section 3 in \cite{shalabh2008finite}. This discussion suggests that even with the information $Y$ only, we are able to identify the latent class membership in a probabilistic way. By supplying $Y$ with additional information from $Z$, the identifiability can be further improved. Once the latent class membership is identified, the parameters $p_{kj}s$ associated $Z$ can be estmated. This explains why when we develop our initial estimator for $\Omega$ in the next subsection, the response $Y$ must be always involved.

As one can see, the statistical model (\ref{2.2}) developed in Section 2.1 is a natural extension of the classical mixture linear regression model \citep{de1989mixtures,wedel2000mixture}. We modify this model slightly so that a large number of binary features can be included for a more accurate identification of the latent class membership. This extension is theoretically interesting due to the following reasons. First, this is an extension of the classical model from fixed-dimensional data to high-dimensional ones. Second, this allows us to extend the application of the classical mixture linear regression from structured data to unstructured text data, which are represented by a ultrahigh dimensional binary feature vector. Lastly, while the mainstream of the statistical literature complains about the curse of dimensionality, our model setup makes the high dimensionality a blessing. That is higher feature dimension leads to more accurate identification of the latent class membership.

We then apply our methodology to the study of judicial imparitiality. This is a problem of fundamental importance for empirical legal studies. In this regard, a lot of statistical methods have been developed \citep{1998Steffensmeier,2001judging,2015identifying,2022ethnic}. The key feature of all those methods is to quantify the effect of the primary covariate of interest (e.g., ethnic, gender, age), after controlling for the confounding effects of legal factors. It is remarkable that most traditional statistical methods cannot handle ultrahigh dimensional data. Therefore, only a fixed number of legal factors can be included for controlling their confounding effects. That leaves ample amount of information contained in the legal documents in text format completely ignored. On the other side, this part of information is extremely useful for controlling the confounding effects of legal factors. Then how to solve this problem becomes practically important or even emergent. That inspires our methodology.

\subsection{An initial estimator}

We next consider how to practically estimate the model (\ref{2.1}), which has a rather sophisticated structure and a large number of unknown parameters. It can hardly be optimized in a straightforward way by for example a standard Newton-Raphson algorithm. Thus, directly optimizing the joint log-likelihood function (\ref{2.2}) might be practically extremely challenge or even infeasible. To solve this problem, we develop here an interesting estimation method, which starts with an initial estimator for the linear regression part and then progresses to a more sophisticated and also accurate final estimator. We are to show that this is a computationally more feasible solution with guaranteed statistical property.
Specifically, if we focus on observations $ \big\{(Y_i, X_i)\big\}_{i=1}^n$ only, we should have a log-likelihood function given by 
\begin{equation}\label{2.3}
	\mathcal{L}(\Omega)= \sum_{i=1}^{n}\log\bigg\{\sum_{k=1}^K\pi_k\phi\Big(\gamma_{k}+X_i^{\top}\theta,\sigma^2\Big)\bigg\},
\end{equation} 
where $\Omega=(\pi^{\top},\gamma^{\top},\theta^{\top},\sigma^2)^{\top}\in\mR^{2K+q+1}$ is a finite dimensional parameter and $\mL(\cdot)$ is different from (\ref{2.2}) with a slight abuse of notation. Therefore, an estimator for $\Omega$ can be defined as $ \wh{\Omega}=\big(\wh{\pi}^{\top},\wh{\gamma}^{\top},\wh{\theta}^{\top},\wh{\sigma}^2\big)^{\top}=\argmax_{\Omega}\mathcal{L}(\Omega)$. To compute this estimator, we wish to obtain its first-order conditions. Note that $ \pi_k$ must to be optimized under the constraint $ \sum_{k=1}^K \pi_k=1 $. We conduct the classical method of Lagrange multipliers \citep{1980lagrange,1984wald}. 
Then the first-order conditions for $\Omega$ are given by
\begin{align}\label{2.4}
	\begin{split}
		\pi_k&=\Big(\sum_{i=1}^n\omega_{ik}\Big)\big/n\\
		\gamma_k&=\bigg\{\sum_{i=1}^n\Big(Y_i-X_i^{\top}\theta\Big)\omega_{ik}\bigg\}\Big/\bigg(\sum_{i=1}^n\omega_{ik}\bigg)\\
		\theta&= \left(X^{\top}X\right)^{-1}X^{\top}\left(Y-V\right)\\
		\sigma^2&=\sum_{i=1}^n\sum_{k=1}^K\Big(Y_i-\gamma_k-X_i^{\top}\theta\Big)^2\omega_{ik}\big/ n,
	\end{split}
\end{align}
where $X=(X_1^{\top},\ldots,X_n^{\top})^{\top}\in\mR^{n\times q}$, $ Y=(Y_1,\dots,Y_n)^{\top}\in\mR^{n}$, $ V=(V_1,\dots,V_n)^{\top}\in\mR^{n} $, $V_i = \sum_{k=1}^K\omega_{ik}\gamma_k $, and $\omega_{ik}=P(\mathcal{K}_i=k|Y_i,X_i)=\pi_k\exp\Big\{-\big(Y_i-\gamma_k-X_i^{\top}\theta\big)^2/\big(2\sigma^2\big)\Big\}$  $\big/\Big[\sum_{k=1}^K\pi_k\exp\Big\{-\big(Y_i-\gamma_k-X_i^{\top}\theta\big)^2/\big(2\sigma^2\big)\Big\}\Big]$ is the posterior probability of the $i$-th  observation being the $k$-th class with limited information and we have $ \sum_{k=1}^K\omega_{ik}=1$. 

Next, we turn this set of first-order conditions (\ref{2.4}) into a classical expectation-maximization (EM) type of algorithm \citep{Dempster1977,1998gentle,2009gaussian}. 
Specifically, let $ \wh{\Omega}^{(0)}=(\wh{\pi}^{(0)\top},\wh{\gamma}^{(0)\top},\wh{\theta}^{(0)\top},\wh{\sigma}^{2(0)})^{\top} $ be an arbitrarily specified initial estimator. For example, we can set $\wh{\pi}_k^{(0)}=1/K $, $ \wh{\gamma}^{(0)}=0 $, $ \wh{\theta}^{(0)}=0 $, and $ \wh{\sigma}^{2(0)}=1$. Write $ \wh{\Omega}^{(t)}=(\wh{\pi}^{(t)\top},\wh{\gamma}^{(t)\top},\wh{\theta}^{(t)\top},\wh{\sigma}^{2(t)})^{\top}$ as the estimator obtained in the $t$-th step. Following the idea of the classical EM algorithm, the next step, to update $\wh{\Omega}^{(t+1)}$ is given by
\begin{align}\label{2.5}
	\begin{split}
		\wh{\pi}_k^{(t+1)}&=\Big(\sum_{i=1}^n\wh{\omega}_{ik}^{(t)}\Big)\big/n\\
		\wh{\gamma}_k^{(t+1)}&=\bigg\{\sum_{i=1}^n\Big(Y_i-X_i^{\top}\wh{\theta}^{(t)}\Big)\wh{\omega}_{ik}^{(t)}\bigg\}\Big/\bigg(\sum_{i=1}^n\wh{\omega}_{ik}^{(t)}\bigg)\\
		\wh{\theta}^{(t+1)} &= \left(X^{\top}X\right)^{-1}X^{\top}\left(Y-V^{(t)}\right)\\
		\wh{\sigma}^{2(t+1)} &=\sum_{i=1}^n\sum_{k=1}^K\Big(Y_i-\wh{\gamma}_k^{(t)}-X_i^{\top}\wh{\theta}^{(t)}\Big)^2\wh{\omega}_{ik}^{(t)}\big/n,
	\end{split}
\end{align}
where $ V^{(t)}=(V_1^{(t)},\dots,V_n^{(t)})^{\top} $, $V_i^{(t)} = \sum_{k=1}^K\wh{\omega}_{ik}^{(t)}\wh{\gamma}_k^{(t)} $, and $ \wh{\omega}_{ik}^{(t+1)}=\wh{\pi}_k^{(t)}\exp\Big\{-\big(Y_i-\wh{\gamma}_k^{(t)}-X_i^{\top}\wh{\theta}^{(t)}\big)^2/\big(2\wh{\sigma}^{2(t)}\big)\Big\}\big/\Big[\sum_{k=1}^K\wh{\pi}_k^{(t)}\exp\Big\{-\big(Y_i-\wh{\gamma}_k^{(t)}-X_i^{\top}\wh{\theta}^{(t)}\big)^2/\big(2\wh{\sigma}^{2(t)}\big)\Big\}\Big]$. 

\subsection{Estimating the response probability}

We next consider how to estimate the response probability $p_{kj}$ for every $ k $ and $ j $. To this end, we consider the joint log-likelihood function for $ (Y_i,X_i) $ and $ Z_{ij}$ as
\begin{equation}\label{2.6}
	\mathcal{L}^{(j)}(\Theta)= \sum_{i=1}^{n}\log\bigg\{\sum_{k=1}^K\pi_k\phi\Big(\gamma_{k}+X_i^{\top}\theta,\sigma^2\Big)p_{kj}^{Z_{ij}}\Big(1-p_{kj}\Big)^{1-Z_{ij}}\bigg\},
\end{equation}
where the superscript ``$j$'' means that the log-likelihood is related to $Z_{ij}$.
Ideally, we should optimize $ \mathcal{L}^{(j)}(\Theta) $ with respect to all unknown parameters (i.e., $ \pi_{k}$, $\gamma_k$, $\theta$, $\sigma^2$, and $ p_{kj}$). It can be directly optimized by for example a Newton-Raphson type iterative algorithm. The associate computational cost should be practically very acceptable for a fixed $j$. However, if the feature dimension $p$ is large, the total computational cost becomes much heavier. We are then inspired to search for computationally more efficient alternative. In fact, with the help of the initial estimator $\wh{\Omega}=\big(\wh{\pi}^{\top},\wh{\gamma}^{\top},\wh{\theta}^{\top},\wh{\sigma}^2\big)^{\top}$ given in the previous subsection, we can simply replace $\Omega=(\pi^{\top},\gamma^{\top},\theta^{\top},\sigma^2)^{\top}$ by its initial estimator $\wh{\Omega}$. This leads to a simplified log-likelihood function
\begin{equation}\label{2.7}
	\wh{\mL^{(j)}}(p_j)= \sum_{i=1}^{n}\log\bigg\{\sum_{k=1}^K\wh{\pi}_k\phi\Big(\wh{\gamma}_{k}+X_i^{\top}\wh{\theta},\wh{\sigma}^2\Big)p_{kj}^{Z_{ij}}\Big(1-p_{kj}\Big)^{1-Z_{ij}}\bigg\},
\end{equation}
where $ p_j=(p_{1j},\ldots,p_{Kj})^{\top}\in\mR^K$, and $ \wh{\Omega} $ is fixed. Therefore, an estimator for $p_{j}$ can be defined as $ \wh{p}_j=\big(\wh{p}_{1j},\ldots,\wh{p}_{Kj}\big)^{\top}=\argmax_{p} \wh{\mL^{(j)}}(p)$. To compute this estimator, we can obtain the first-order condition for $ p_{kj}$ as follows,
\begin{equation}\label{2.8}
	p_{kj}=\Big\{\sum_{i=1}^n\wh{\pi}_{ik}^{(j)}Z_{ij}\Big\}\big/\Big\{\sum_{i=1}^n\wh{\pi}_{ik}^{(j)}\Big\},
\end{equation}
where $ \wh{\pi}_{ik}^{(j)}=\wh{\pi}_k\exp\Big\{-\big(Y_i-\wh{\gamma}_k-X_i^{\top}\wh{\theta}\big)^2/\big(2\wh{\sigma}^2\big)\Big\}\Big(\wh{p}_{kj}\Big)^{Z_{ij}}\Big(1-\wh{p}_{kj}\Big)^{1-Z_{ij}}\big/\Big[\sum_{k=1}^K\wh{\pi}_k$  $\exp\Big\{-\big(Y_i-\wh{\gamma}_k-X_i^{\top}\wh{\theta}\big)^2/\big(2\wh{\sigma}^2\big)\Big\}\Big(\wh{p}_{kj}\Big)^{Z_{ij}}\Big(1-\wh{p}_{kj}\Big)^{1-Z_{ij}}\Big]$. Then, we can immediately obtain the EM algorithm for $ p_{kj}$ as follows. By doing so, we can make full use of this analytical formula (\ref{2.8}) so that the algorithm is no longer iterative beween $\Omega$ and $ p_j$. Consequently, the computational cost can be significantly reduced. Similar to the previous subsection, let $\wh{p}_{kj}^{(0)}$ be an arbitrarily specified initial estimator, such as $\wh{p}_{kj}^{(0)}=1/2$ for every $k$ and $j$. Write $\wh{p}_{kj}^{(t)}$ be the estimator obtained in the $t$-th step. Then, the next step to update $\wh{p}_{kj}^{(t+1)}$ is
\begin{equation}\label{2.9}
	\wh{p}_{kj}^{(t+1)}=\Big\{\sum_{i=1}^n\wh{\pi}_{ik}^{(j,t)}Z_{ij}\Big\}\big/\Big\{\sum_{i=1}^n\wh{\pi}_{ik}^{(j,t)}\Big\},
\end{equation}
where $ \wh{\pi}_{ik}^{(j,t+1)}=\wh{\pi}_k\exp\Big\{-\big(Y_i-\wh{\gamma}_k-X_i^{\top}\wh{\theta}\big)^2/\big(2\wh{\sigma}^2\big)\Big\}\Big(\wh{p}_{kj}^{(t)}\Big)^{Z_{ij}}\Big(1-\wh{p}_{kj}^{(t)}\Big)^{1-Z_{ij}}\big/\Big[\sum_{k=1}^K\wh{\pi}_k$  $\exp\Big\{-\big(Y_i-\wh{\gamma}_k-X_i^{\top}\wh{\theta}\big)^2/\big(2\wh{\sigma}^2\big)\Big\}\Big(\wh{p}_{kj}^{(t)}\Big)^{Z_{ij}}\Big(1-\wh{p}_{kj}^{(t)}\Big)^{1-Z_{ij}}\Big]$. 

Note that the initial estimator $\wh{\Omega}$ is the standard M-estimator, which we know is $\sqrt{n}$-consistent under appropriate regularity conditions \citep{vaart1998,2003MS}. 
Then, we can further prove that $ \wh{p}_j=\big(\wh{p}_{1j},\ldots,\wh{p}_{Kj}\big)^{\top}\in\mR^K$ is also $\sqrt{n}$-consistent for every $j$. To study the theoretical properties of $\wh{p}_j$ over every $j$, we shall focus on the log-likelihood function $\mL^{(j)}(p_j)$. The first- and second- order partial derivatives of $\mL^{(j)}(p_j)$ with respect to $p_j$ are given by $\dot{\mL}^{(j)}(p_j)=\partial\mL^{(j)}(p_j)\big/\partial p_j\in\mR^K$ and $\ddot{\mL}^{(j)}(p_j)=\partial^2\mL^{(j)}(p_j)\big/\big(\partial p_j\partial p_j^{\top}\big)\in\mR^{K\times K}$, respectively. Write $ \dot{\mL}^{(j)}(p_j)=\big(\dot{\ell_1}^{(j)},\ldots,\dot{\ell_K}^{(j)}\big)^{\top}\in\mR^K$ and  $\ddot{\mL}^{(j)}(p_j)=\big(\ddot{\ell}_{k_1k_2}^{(j)}\big)\in\mR^{K\times K}$. Let $\alpha_{ik}^{(j)}=c_{ik}p_{kj}^{Z_{ij}}\big(1-p_{kj}\big)^{1-Z_{ij}}\big/\sum_{k=1}^Kc_{ik}p_{kj}^{Z_{ij}}\big(1-p_{kj}\big)^{1-Z_{ij}}$, and $c_{ik}=\pi_k\big(\sqrt{2\pi}\sigma\big)^{-1}$  $\exp\Big\{-\Big(Y_i-\gamma_k-X_i^{\top}\theta\Big)^2\big/\big(2\sigma^2\big)\Big\}$. Define $ s(Z_{ij},p_{kj})=Z_{ij}/p_{kj}-\big(1-Z_{ij}\big)/\big(1-p_{kj}\big)$. It can be verified that $ \dot{\ell_k}^{(j)}=\sum_{i=1}^n\alpha_{ik}^{(j)}s(Z_{ij},p_{kj})$ and $\ddot{\ell}_{k_1k_2}^{(j)}=-\sum_{i=1}^n\alpha_{ik_1}^{(j)}s(Z_{ij},p_{k_1j})\alpha_{ik_2}^{(j)}$ $s(Z_{ij},p_{k_2j})$. Define $I(p_j)=-E\big\{n^{-1}\ddot{\mL}^{(j)}(p_j)\big\}=-E\big\{n^{-1}\big(\ddot{\ell}_{k_1k_2}^{(j)}\big)\big\}=\big(\maE_{k_1k_2}^{(j)}\big)\in\mR^{K\times K}$, where $\maE_{k_1k_2}^{(j)}=n^{-1}\sum_{i=1}^nE\big(c_{ik_1}c_{ik_2}/M_{ij}\big)$, and $M_{ij}=\sum_{k=1}^Kc_{ik}p_{kj}$  $\sum_{k=1}^Kc_{ik}\big(1-p_{kj}\big)$.

Let $A$ be an arbitrary matrix with dimension $M\times N$. Define its norm as $\Vert A\Vert=\lambda_{\max}^{1/2}(A^{\top}A)=\lambda_{\max}^{1/2}(AA^{\top})$, where $\lambda_{\max}(B)$ stands for the maximal eigenvalue of an arbitrary symmetric matrix $B$. Similarly, write $\lambda_{\min}(B)$ as the minimal eigenvalue of $B$.  Write $\pi_{\min}=\min_k\pi_k$ and $\pi_{\max}=\max_k\pi_k$ with $0<\pi_{\min}\leq\pi_{\max}<1$. Write $\gamma_{\max}^2=\max_k\gamma_k^2>0$. Then, the uniform consistency for $ \wh{p}_j $ over every $j$ can be established by Theorem \ref{thm1}, which is proved in Appendix B.1. By Theorem \ref{thm1}, we known that $ \wh{p}_{j}$ is uniformly consistent for $p_{j} $ over $ 1\leq j\leq p $. The uniform convergence rate is slightly slower than the standard rate of $ \sqrt{n} $ by a slowly diverging factor $ C_n $.
\begin{theorem}\label{thm1}
	Assume the technical conditions (C1), (C2) and (C3) as given in Appendix A hold. Furthermore, assume that $ C_n>0$ is an arbitrary positive sequence such that: (\Rmnum{1}) $ C_n/\sqrt{n}\to 0 $ and (\Rmnum{2}) $C_n^2/\log(p)\to \infty$ as $ n\to\infty$. We then have $ \max_{j}\left\Vert\wh{p_j}-p_j\right\Vert=O_p\big(C_n/\sqrt{n}\big)$.
\end{theorem}
\subsection{Class membership identification}

With the help of the response probability estimators, we are able to estimate the latent class membership with extra-ordinarily high accuracy. This is mainly because the feature dimension $p$ is extremely high. That leads to an ample amount of information for class membership identification. Specifically, we are still interested in estimating the posterior probability for $\mK_i=k$ but with $(Y_i,X_i,Z_i)$ information given. Write $ a_{ik}=I(\mK_i=k)$ and $\pi_{ik}=P\Big(\mathcal{K}_i=k\Big|Y_i,X_i,Z_i\Big)$. Then direct computation suggests that
\begin{equation}\label{2.10}
	\pi_{ik}=\dfrac{\pi_k\exp\bigg\{-\dfrac{1}{2\sigma^2}\Big(Y_i-\gamma_k-X_i^{\top}\theta\Big)^2\bigg\}\prod_{j=1}^pp_{kj}^{Z_{ij}}\Big(1-p_{kj}\Big)^{1-Z_{ij}}}{\sum_{k=1}^K\pi_k\exp\bigg\{-\dfrac{1}{2\sigma^2}\Big(Y_i-\gamma_k-X_i^{\top}\theta\Big)^2\bigg\}\prod_{j=1}^pp_{kj}^{Z_{ij}}\Big(1-p_{kj}\Big)^{1-Z_{ij}}}.
\end{equation}
Next, we can replace the unknown parameters in (\ref{2.10}) by the estimators given in Sections 2.3 and 2.4. This leads to an another estimator for the posterior probability for each observation $i$ as given by
\begin{equation}\label{2.11}
	\wh{\pi}_{ik}=\dfrac{\wh{\pi}_k\exp\bigg\{-\dfrac{1}{2\wh{\sigma}^2}\Big(Y_i-\wh{\gamma}_k-X_i^{\top}\wh{\theta}\Big)^2\bigg\}\prod_{j=1}^p\wh{p}_{kj}^{Z_{ij}}\Big(1-\wh{p}_{kj}\Big)^{1-Z_{ij}}}{\sum_{k=1}^K\wh{\pi}_k\exp\bigg\{-\dfrac{1}{2\wh{\sigma}^2}\Big(Y_i-\wh{\gamma}_k-X_i^{\top}\wh{\theta}\Big)^2\bigg\}\prod_{j=1}^p\wh{p}_{kj}^{Z_{ij}}\Big(1-\wh{p}_{kj}\Big)^{1-Z_{ij}}}.
\end{equation}
Note that the same posterior probability was also evaluated in Section 2.3 but with $(Y_i,X_i)$ information only, which is denoted by $\omega_{ik}$. Note that, the feature involved in $\omega_{ik}$ has a fixed dimension and thus there is very limited information. Therefore, the posterior estimator $\omega_{ik}$ provided in (\ref{2.4}) cannot be consistent for $a_{ik}$. In other words, the difference between $\omega_{ik}$ and $a_{ik}$ will not converge to $0$ even if $n\to\infty$. However, the story dramatically changes for the new posterior estimator $\wh{\pi}_{ik}$, for which we assume that $ p\to\infty$ as $n\to\infty$. Consequently, a sufficient amount of information can be accumulated for the latent class membership $\mK_i$. As a direct conseuqence, it is more likely to be consistent for $a_{ik}$. 
In fact, this conjecture is formally verified by Theorem \ref{thm2}, whose proof is given in Appendix B.2. From this, we know that the posterior probability estimator $\wh{\pi}_{ik}$ is extremely close to the true membership indicator function $a_{ik}=I(\mK_i=k)$, with a tiny error of order $ O\big(\exp(-\nu p)\big) $. This makes the subsequent estimator and inference for the main regression model of $ Y $ and $ X $ very easy.
\begin{theorem}\label{thm2}
	Assume the technical conditions (C1)$-$(C6) as given in Appendix A hold. Then, for an arbitrary constant $0<\nu<\Delta_{\min}/2$, where $\Delta_{\min}$ is defined in Condition (C5),
	there exists a positive constant $M>0$ such that $ P\Big\{\max_{i,k}\big|\wh{\pi}_{ik}-I(\mK_i=k)\big|>\exp\big(-\nu p\big)\Big\}=o(1)$ as long as $p>M$.
\end{theorem}
\subsection{The final main estimator}

By Theorem \ref{thm2}, we know that the true class membership can be consistently estimated by $\wh{\pi}_{ik}$ with super excellent accuracy. As a consequence, we should be able to re-estimate the interested regression parameter $\Omega=\big(\pi^{\top},\gamma^{\top},\theta^{\top},\sigma^2\big)^{\top}\in\mR^{2K+q+1}$ with much-improved estimation accuracy. Define $ X_i^a=(a_i^{\top},X_i^{\top})^{\top}\in\mR^{K+q}$, where $ a_i=(a_{i1},\ldots,a_{iK})^{\top}\in\mR^K$. Recall that $ \sum_{k=1}^Ka_{ik}=1$. Specifically, if the latent class membership $\mK_i $ is known in advance, we should estimate the linear regression model parameters by minimizing the following classical least squares (OLS) objective function as $ \mL\big(\Phi\big)=\sum_{i=1}^n\sum_{k=1}^Ka_{ik}\Big(Y_i-\gamma_{k}-X_i^{\top}\theta\Big)^2=\sum_{i=1}^n\Big(Y_i-X_i^{a\top}\Phi\Big)^2 $,
where $\Phi=(\gamma^{\top},\theta^{\top})^{\top}\in\mR^{K+q}$. By optimizing $\mL\big(\Phi\big)$ with respect to $ \Phi $, we obtain $ \wh{\Phi}_{\so}=\argmin_{\Phi}\mL\big(\Phi\big)=\Big(\wh{\Sigma}^{\fxx}_{\so}\Big)^{-1}\wh{\Sigma}^{\fxy}_{\so}$, where $ \wh{\Sigma}^{\fxx}_{\so}=n^{-1}\sum_{i=1}^nX_i^aX_i^{a\top}$ and $ \wh{\Sigma}^{\fxy}_{\so}=n^{-1}\sum_{i=1}^nX_i^aY_i $. 

As one can see, $\wh{\Phi}_{\so}$ is an oracle estimator, which cannot be practically computed, since the latent class membership $\mK_i $ is not directly observed. However, by Theorem \ref{thm2}, we know that this binary indicator random variable $a_{ik}=I(\mK_i=k)$ can be estimated consistently and accurately by $\wh{\pi}_{ik}$. We are then motivated to approximate $ X_i^a $ by $ X_i^{\pi} $, where $ X_i^{\pi}=(\wh{\pi}_i^{\top},X_i^{\top})^{\top}\in\mR^{K+q}$, and $ \wh{\pi}_i=(\wh{\pi}_{i1},\ldots,\wh{\pi}_{iK})^{\top}\in\mR^K$. Then, a practically feasible estimator can be constructed as $ \wh{\Phi}_{\sr}= \Big(\wh{\Sigma}^{\fxx}_{\sr}\Big)^{-1}\wh{\Sigma}^{\fxy}_{\sr} $, where $ \wh{\Sigma}^{\fxx}_{\sr}=n^{-1}\sum_{i=1}^nX_i^{\pi}X_i^{\pi\top}$ and $ \wh{\Sigma}^{\fxy}_{\sr}=n^{-1}\sum_{i=1}^nX_i^{\pi}Y_i $. Thereafter, an oracle estimator for $\sigma^2$ can be defined as $\wh{\sigma}^2_{\so}= \sum_{i=1}^n\Big(Y_i-X_i^{a\top}\wh{\Phi}_{\so}\Big)^2\big/ n$. Once $ \wh{\Phi}_{\sr}$ is obtained, $\wh{\sigma}^2_{\so}$ can be estimated as $\wh{\sigma}^2_{\sr} =\sum_{i=1}^n\Big(Y_i-X_i^{\pi\top}\wh{\Phi}_{\sr}\Big)^2\big/ n$.
Lastly, we define the final estimator for $\pi_k$ as $\wh{\pi}_{\sr}=\big(\wh{\pi}_{\sr,1},\ldots,\wh{\pi}_{\sr,K}\big)$, where  $\wh{\pi}_{\sr,k}= \Big(\sum_{i=1}^n\wh{\pi}_{ik}\Big)\big/n $, and its ideal counterpart as $\wh{\pi}_{\so}=\big(\wh{\pi}_{\so,1},\ldots,\wh{\pi}_{\so,K}\big)$, where $\wh{\pi}_{\so,k} = \Big(\sum_{i=1}^n a_{ik}\Big)\big/n$. 
Recall that $\Omega=(\pi^{\top},\gamma^{\top},\theta^{\top},\sigma^2)^{\top}\in\mR^{2K+q+1}$. Then, we can immediately have the real estimator $\wh{\Omega}_{\sr}=(\wh{\pi}^{\top}_{\sr},\wh{\gamma}^{\top}_{\sr},\wh{\theta}^{\top}_{\sr},\wh{\sigma}_{\sr}^2)^{\top}$ and the oracle estimator $\wh{\Omega}_{\so}=(\wh{\pi}^{\top}_{\so},\wh{\gamma}^{\top}_{\so},$  $\wh{\theta}^{\top}_{\so},\wh{\sigma}_{\so}^2)^{\top}$.
Theorem \ref{thm3} characterizes the difference between the real and oracle estimator, from which we find that the resulting estimator enjoys the same convergence rate and asymptotic distribution as its ideal counterpart, which is defined by assuming that $\mK_i$ is known in advance. The proof of Theorem \ref{thm3} is given in Appendix B.3.
\begin{theorem}\label{thm3}
	Assume that the technical conditions (C1)$-$(C6) as given in Appendix A hold. Then, we have 
	$ \big\Vert\wh{\Omega}_{\sr}-\wh{\Omega}_{\so}\big\Vert=o_p\big(1/\sqrt{n}\big) $.
\end{theorem}
\section{Simulation studies}
\subsection{The simulation setup}

To demonstrate the finite sample performance of the proposed MCR method, we performed a number of simulation studies. Specifically, we would like to study the finite sample performance of (a) the initial estimator $\wh{\Omega}$; (b) the response probability estimators $\wh{p}_js$; (c) the class membership identification estimators $ \wh{\pi}_{ik}s$; and (d) the final main estimator $\wh{\Omega}_{\sr}$. For the entire simulation study, we considered various sample sizes with $n=$1,000, 2,000, or 5,000. For each $n$, the dimension of the binary feature vector was set to be $p=n/10$, $n/5$, $n/2$, $n$, or $2n$. Once $ n $ and $ p $ are given, following \cite{1996tibshirani}, we generated $X_i\in\mR^q$ with $q=8$ from a multivariate normal distribution with mean 0 and $ \cov(X_{ij_1},X_{ij_2})=\rho^{|j_1-j_2|} $ with $ \rho=0.5$ for $1\leq j_1,j_2\leq 8$. The number of classes was fixed at $K=5$ \citep{2002latent}. The true value of $\Theta=\big(\pi^{\top},\gamma^{\top},\theta^{\top},\sigma^2,\text{vec}(P)^{\top}\big)^{\top}\in\mR^{2K+q+1+Kp}$ was set at $\pi=(0.15,0.2,0.3,0.25,0.1)^{\top}$, $ \gamma=\big(-4,-1,2,5,8\big)^{\top}$, $\theta=(3,1.5,0,0,2,0,0,0)^{\top}$, $\sigma^2=1$, and $ P=(p_{kj})\in\mR^{K\times p}$. Here the matrix $ P $ was divided into $K^2$ block matrices, where the block diagonal elements were generated from a uniform distribution between 0.8 and 0.95, and other elements were generated from a uniform distribution between 0.01 and 0.3. Then, $ Z_i=(Z_{ij})\in\mR^p$ could be generated. The residual term $\varepsilon_i$ was independently generated from a standard normal distribution. This leads to the final response variable $Y_i$ according to the model (\ref{2.1}).
\subsection{The initial estimator $\wh{\Omega}$}

\begin{figure}
	\centering
	\includegraphics[width=5in]{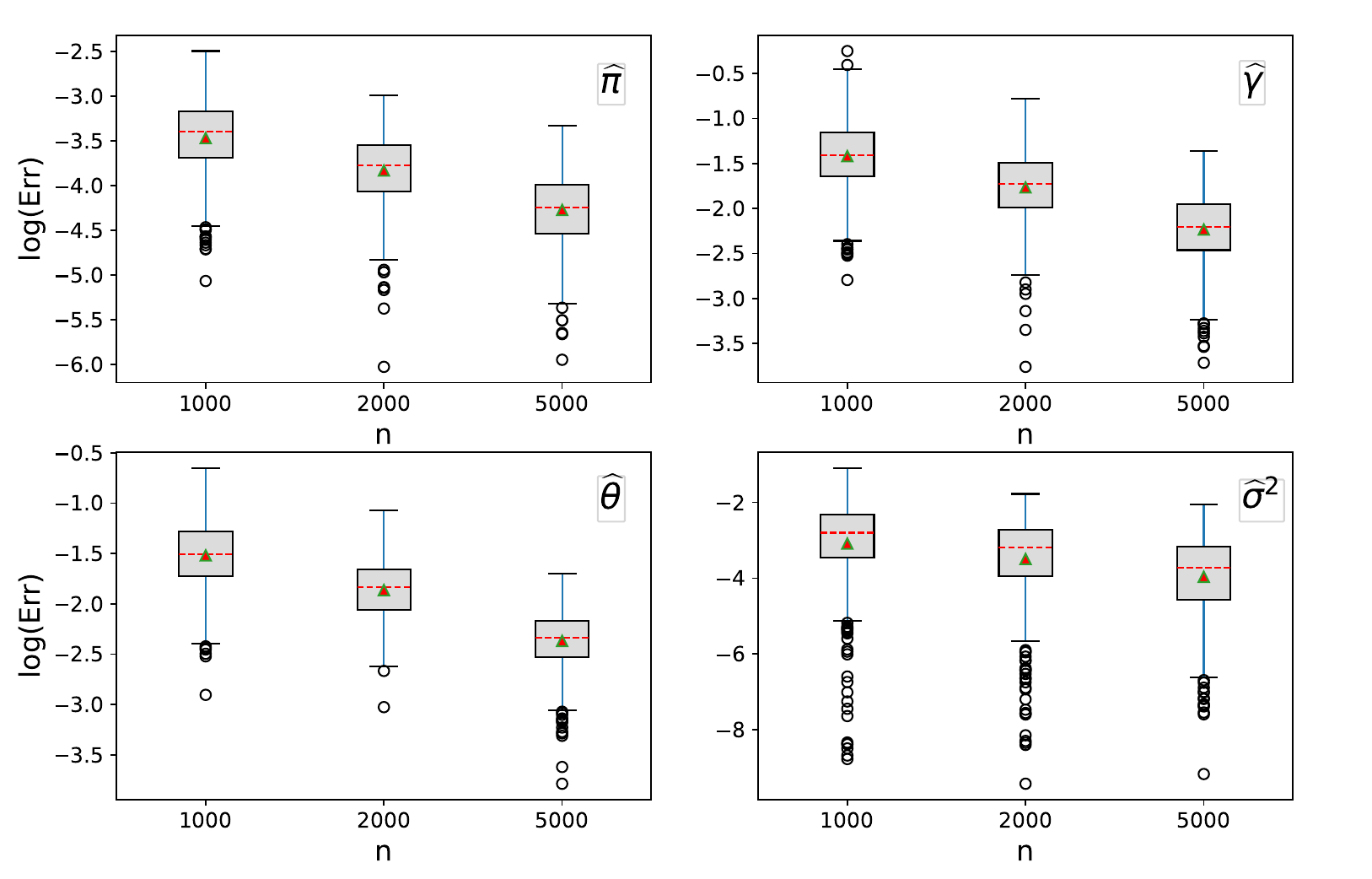}
	\caption{Log(Err) values for various initial estimators: $\wh{\pi}$ (the upper left panel), $\wh{\gamma}$ (the upper right panel), $\wh{\theta}$ (the lower left panel), and $\wh{\sigma}^2$ (the lower right panel). Three different sample sizes are considered. They are $n=1000$, $2000$ and $5000$ respectively.}
	\label{f1}
\end{figure}

We start with the initial estimator $\wh{\Omega}=\big(\wh{\pi}^{\top},\wh{\gamma}^{\top},\wh{\theta}^{\top},\wh{\sigma}^2\big)^{\top}\in\mR^{2K+q+1}$. For a given sample size $n$ and binary feature dimension $ p $, the experiment was randomly replicated for a total of $R=500$ times. We use $\wh{\tau}^{(r)}$ to represent one particular estimator (e.g., $\wh{\theta}^{(r)}$) obtained in the $r$-th replication ($1\leq r\leq R$). The true parameter is denoted by $\tau$. Then, the estimator error (Err) can be evaluated as Err$=\big\Vert\wh{\tau}^{(r)}-\tau\big\Vert$ for every $1\leq r\leq R$. This leads to a total of $R$ Err values, which are then log-transformed and box-plotted in Figure \ref{f1}. By Figure \ref{f1}, we find that, for essentially every estimator of interest (i.e., $ \wh{\pi}$, $\wh{\gamma}$, $\wh{\theta}$, and $\wh{\sigma}^2$), large sample sizes always lead to smaller estimation errors. This numerical finding confirms that the initial estimator $\wh{\Omega}$ is indeed consistent. 
\subsection{The response probability estimator $\wh{p}_j$}

\begin{figure}
	\centering
	\includegraphics[width=5in]{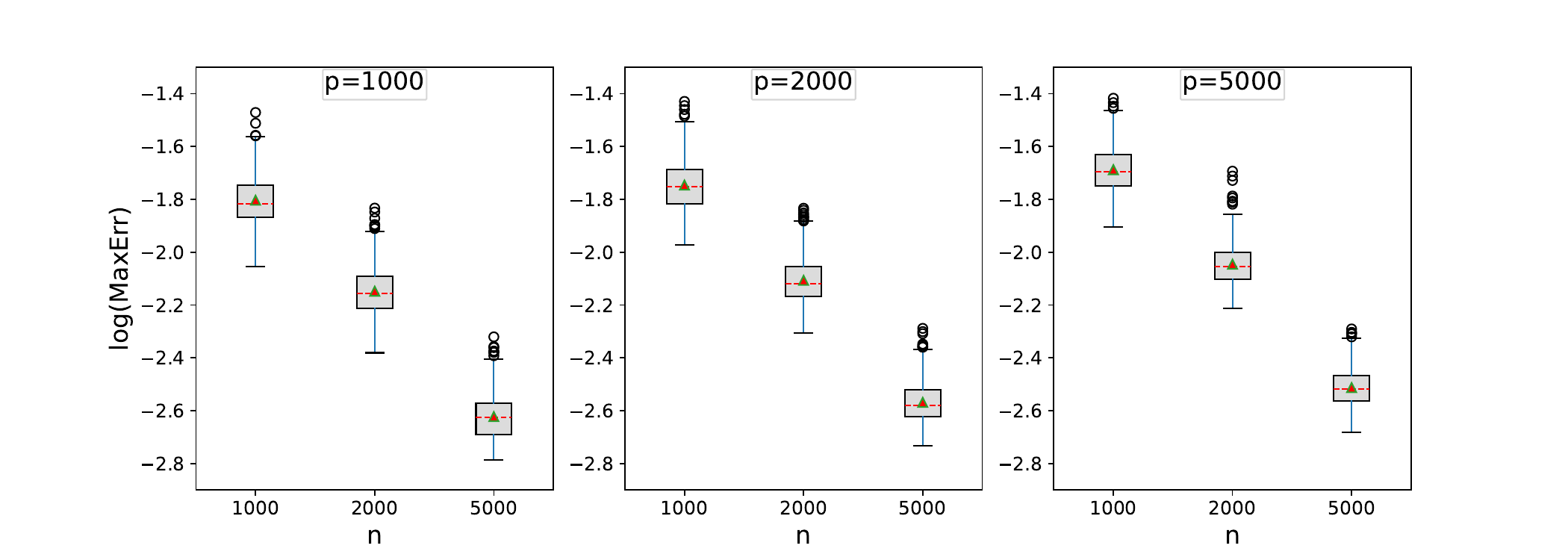}
	\caption{Log(MaxErr) values for the responce probability estimator $\wh{p}_j$. Different panels correspond to different feature dimensions: $p=1000$ (the left panel), $2000$ (the middle panel), and $5000$ (the right panel). For a given panel, different boxplots correspond to different sample sizes with $n=1000$, $2000$ and $5000$, respectively.}
	\label{f2}
\end{figure}

Next, we study $\wh{p}_j\in\mR^K$. Similarly, we can compute for each $\wh{p}_j$ an Err value decoded by Err$_j$ ($1\leq j\leq p$) and its maximum error (MaxErr) over $j$ is given by MaxErr$=\max_j$Err$_j$. Recall that we have $ R=500$ random replications. This leads to a total of $R$ MaxErr values, which are then log-transformed and box-plotted in Figure \ref{f2}. By Figure \ref{f2}, we obtain the following two interesting findings. First, for a fixed $p$, we find that the larger the sample size $n$, the smaller the maximum error (MaxErr). This confirms that $ \wh{p}_{j}$ is uniformly consistent for $p_{j} $ over $1\leq j\leq p$. Second, with a fixed sample size $n$, the maximum error (MaxErr) seems to be slightly larger as $p$ increases. This interesting numerical finding suggests that the uniform convergence rate of $\wh{p}_j$ is slightly slower than the standard rate of $ \sqrt{n} $ if $p\to\infty$ as $ n\to\infty$. All these results are in line with our theoretical findings in Theorem \ref{thm1}.
\subsection{The posterior probability estimator $\wh{\pi}_{ik}$}

\begin{figure}
	\centering
	\includegraphics[width=5in]{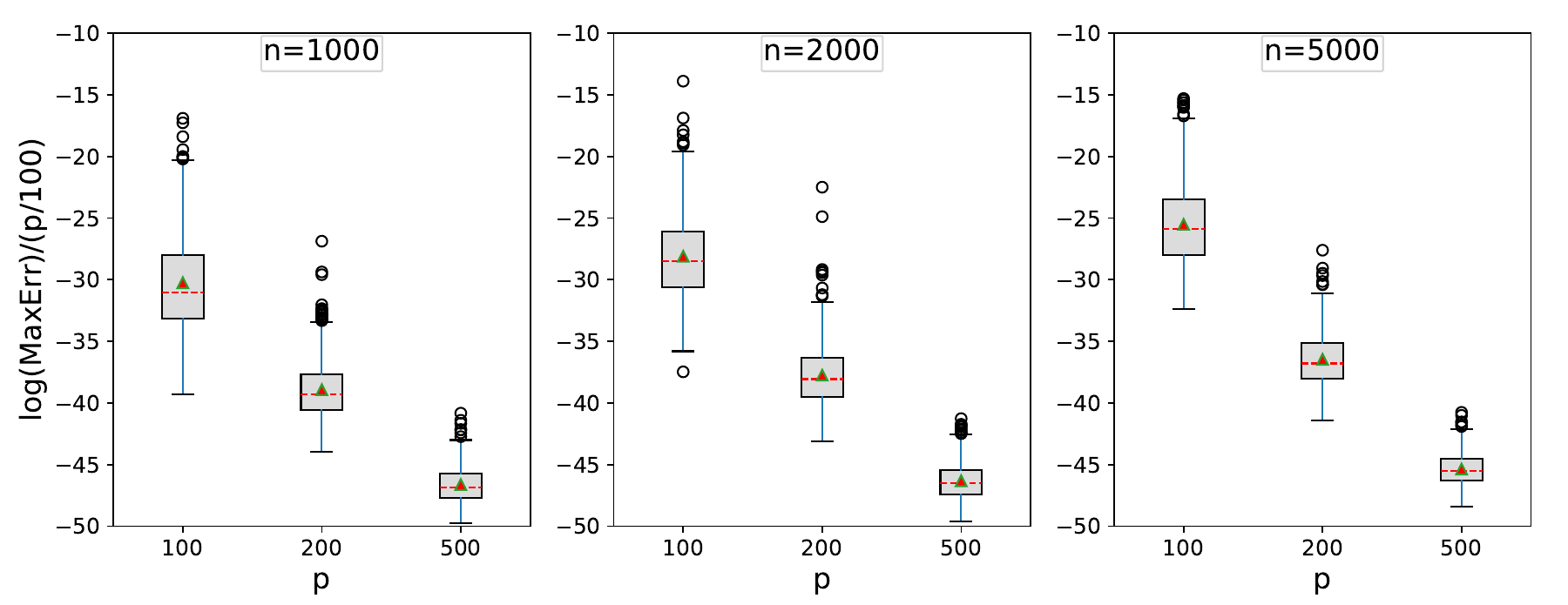}
	\caption{Log(MaxErr)$/(p/100)$ values for the posterior probability estimator $\wh{\pi}_{ik}$. Different panels correspond to different sample sizes: $n=1000$ (the left panel), $2000$ (the middle panel), and $5000$ (the right panel). For a given panel, different boxplots correspond to different feature dimensions  with $p=100$, $200$ and $500$, respectively.}
	\label{f3}
\end{figure}

We then study $\wh{\pi}_{ik}$. Recall that $\wh{\pi}_{ik}$ estimates the latent class membership. We are extremely interested in evaluating the difference between $\wh{\pi}_{ik}$ and the true membership indicator function $a_{ik}=I(\mK_i=k)$. Thus, we can compute Err$_{i,k}=\big|\wh{\pi}_{ik}-a_{ik}\big|$ for every $ i $ ($1\leq i\leq n$) and $ k $ ($1\leq k\leq K$), whose maximum error (MaxErr) over $i$ and $k$ is given by MaxErr$=\max_{i,k}$Err$_{i,k}$. Similarly, this leads to a total of $R$ MaxErr values, which are then log-transformed and box-plotted in Figure \ref{f3}. By Figure \ref{f3}, we find that with a fixed sample size $n$, a larger $p$ leads to smaller MaxErr values. The larger the $ p $ value is, the more feature information can be provided and thus the more accurate the posterior probability could be. This results verify that the feature information helps us to estimate the latent class membership with extra-ordinarily high accuracy, which is in line with our theoretical findings in Theorem \ref{thm2}.
\subsection{The final main estimator $\wh{\Omega}_{\sr}$}

\begin{figure}
	\centering
	\includegraphics[width=4in]{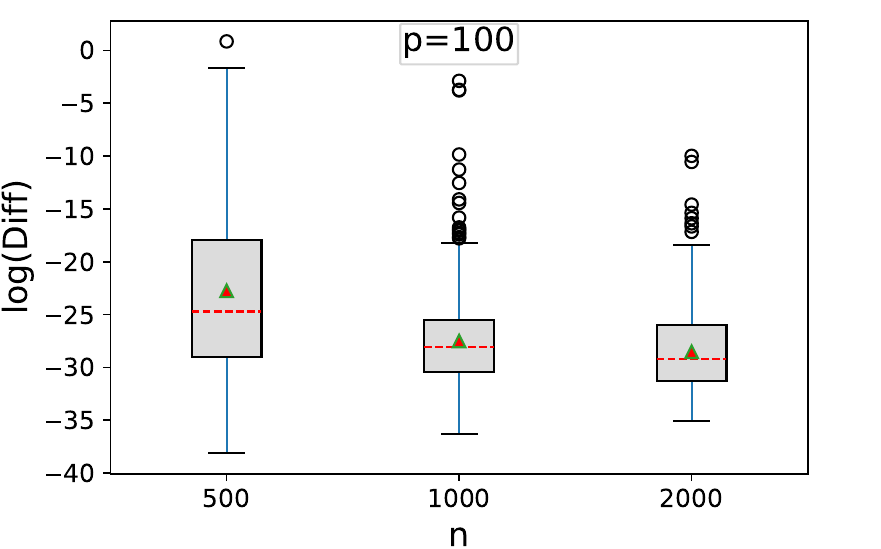}
	\caption{Log(Diff) values between $\wh{\Omega}_{\sr}$ and $\wh{\Omega}_{\so}$ with the feature dimension $ p=100 $. Different boxplots correspond to different sample sizes.}
	\label{f4}
\end{figure}

Finally, recall that $\wh{\Omega}_{\sr}$ is a practically feasible estimator to approximate the oracle estimator $\wh{\Omega}_{\so}$. Thus, we would like to study the difference between $\wh{\Omega}_{\sr}$ and $\wh{\Omega}_{\so}$, as evaluated by Diff=$\big\Vert\wh{\Omega}_{\sr}-\wh{\Omega}_{\so}\big\Vert$.
Similarly, this leads to a total of $R$ Diff values in log-scale, which are then box-plotted in Figure \ref{f4}. By Figure \ref{f4}, we find that for a fixed small feature dimension $p$ (e.g., $p=100$), the larger the sample size $n$, the smaller the mean error. Furthermore, the difference between $\wh{\Omega}_{\sr}$ and $\wh{\Omega}_{\so}$ rapidly shrinks to an extremely tiny value as $n$ increases. In fact, for a slightly large $p$ (e.g., $p=200$) and a reasonably large sample size (e.g., $n=500$), the Diff values are too tiny to be distinguished from 0 due to the limited precision of a computer system. This indicates that $\wh{\Omega}_{\sr}$ is almost identical to $\wh{\Omega}_{\so}$. All these results are in line with our theoretical findings in Theorem \ref{thm3}.
\subsection{A BIC method for $K$}

The simulation results presented in the previous subsections are based on the assumption that the true number of latent classes (i.e., $K$) is known in advance. Unfortunately, this is an unknown parameter that need be estimated. To this end, we follow the idea of \cite{schwarz1978} and develop here a BIC method. Specifically, let $K_{\max}$ be the maximum number of latent classes to be considered. For example, various $K_{\max}$ values (e.g., 10 and 20) have been considered. The resulting numerical performance is nearly identical. Therefore, we fix $K_{\max}=10$ for the rest of the simulation study. Next, for any $ 1\leq K\leq K_{\max} $, the interested model parameters can be estimated and denoted as $\wh{\Theta}^{(K)}=\Big(\wh{\pi}^{(K)\top},\wh{\gamma}^{(K)\top},\wh{\theta}^{(K)\top},\wh{\sigma}^{2(K)\top},\text{vec}\big(\wh{P}^{(K)}\big)\Big)^{\top}\in\mR^{d^*}$ with $d^*=K+K+q+1+pK=2K+q+1+pK$. Then, a BIC selection criterion can be developed as
\begin{align}
	\begin{split}
	\BIC(K) = -2&\sum_{j=1}^p\Bigg[\sum_{i=1}^n\log\bigg\{\sum_{k=1}^K\wh{\pi}_k^{(K)}\phi\Big(\wh{\gamma}_k^{(K)}+X_i^{\top}\wh{\theta}^{(K)}, \wh{\sigma}^{2(K)}\Big)\\
	\label{3.1}&\Big(\wh{p}_{kj}^{(K)}\Big)^{Z_{ij}}\Big(1-\wh{p}_{kj}^{(K)}\Big)^{1-Z_{ij}}\bigg\}\Bigg]+\text{df}\times\log(n)
\end{split}
\end{align}
where $\text{df} = d^*-1 = 2K+q+pK$ is the degree of freedom due to the whole parameter $ \Theta\in\mR^{d^*}$ and $\sum_{k=1}^K\pi_k =1$, and the penalization factor ($\text{df}\times\log(n)$) is due to the seminal work of \cite{schwarz1978}. Therefore, the optimum $K$ can be estimated as $\wh{K} = \argmin_{1\leq K\leq K_{\max}}\BIC(K)$. Following the simulation setting in the previous subsections, this experiment was randomly replicated $R=500$ times. The percentage of the experiments with $\wh{K}=K=5$ is shown in Table \ref{tt}. As one can see, for any fixed feature dimension $p$, the percentage of experiments with $\wh{K}=K=5$ converges to 100\% rapidly as the sample size $n$ increases. This suggests that $\wh{K}$ should be a consistent estimator of $ K $. 
\begin{table}
	\caption{Percentage (\%) of experiments with different $n$ and $p$.}
	\centering
	\begin{tabular}{ccccccc}
		\toprule
		\hline
		\backslashbox{$n$}{$p$} & 10 & 50 &100 & 200 & 500 & 1000\\
		\hline
		\specialrule{0em}{2pt}{0pt}   200 &36.2\%  & 28.2\% & 28.0\% & 25.0\%& 26.0\%  & 25.4\%  \\
		\specialrule{0em}{0pt}{2pt} 500 & 89.2\%& 85.0\% & 84.0\% & 83.8\%&	83.4\% & 84.4\% \\
		\specialrule{0em}{0pt}{2pt} 1000 &98.0\% &97.6\% & 98.0\% &97.0\%	&97.2\%	& 97.2\% \\
		\specialrule{0em}{0pt}{2pt} 2000 &100\%& 99.6\%& 99.8\% & 99.8\%&	99.8\%& 99.8\%\\
		\bottomrule
	\end{tabular}
	\label{tt}
\end{table}

\section{Real data analysis}
\subsection{The China Judgments Online data}

We present here a real case study. The dataset is obtained from \textit{China Judgments Online} (CJO). The full dataset contains a total of 1,361,354 cases that happened in China from 2017 to 2018. For illustration purpose, we study here the criminal cases only. This is mainly because the CJO dataset is a highly unbalanced dataset, with sample sizes varying considerably by crime. Obviously, we cannot work on crimes with extremely tiny sample sizes. In the meanwhile, past literature suggests that theft is one of the most common crimes worldwide \citep{1981crime,2010crime}. 
It happens that this is also the case for our CJO dataset, where theft accounts for about 24.44\% of all cases \citep{2002crime,2022crime}. For illustration purpose, we take burglaries in theft-related cases as an example. It accounts for about 13.59\% of all theft cases. Moreover, to render our analysis in a more straightforward way, only those first trials and fixed-term imprisonment cases without any missing information are kept. This leads to a final sample size of $n=6,118$ cases.

\begin{figure}
	\centering
	\includegraphics[width=5in]{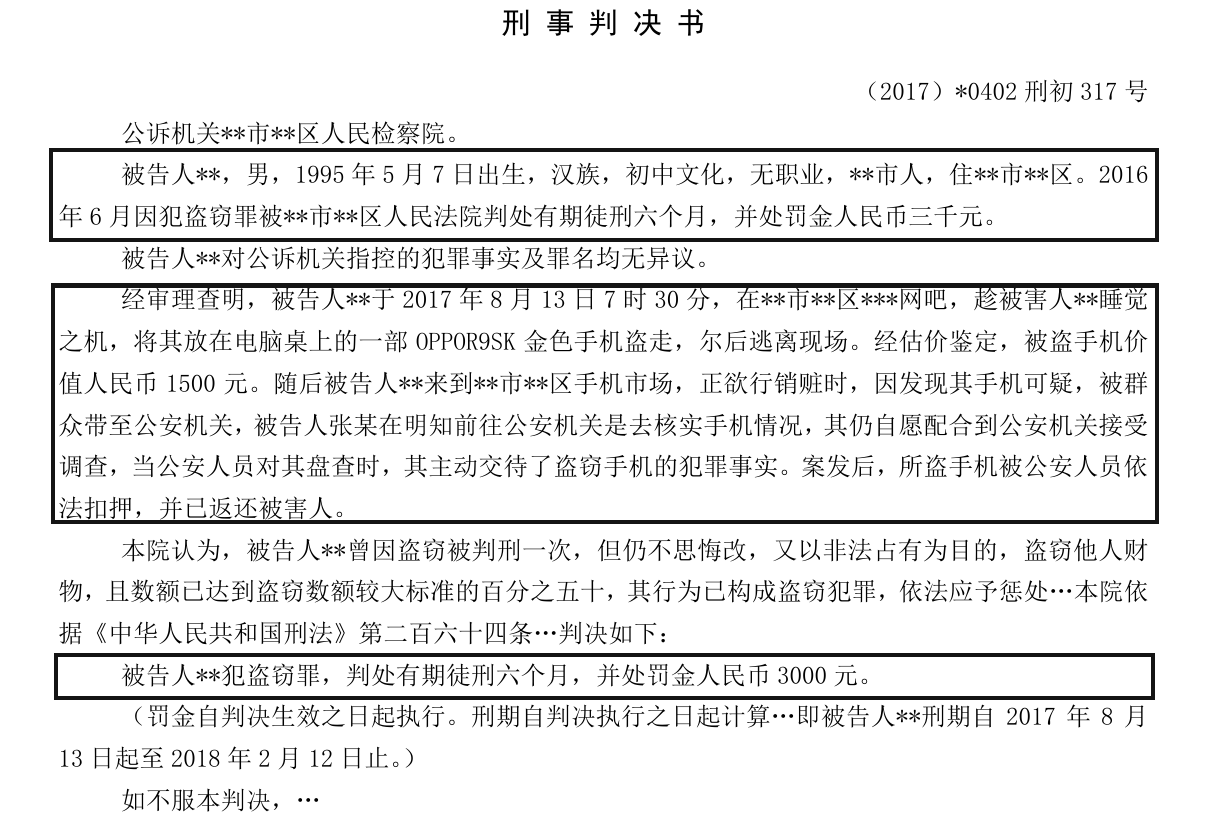}
	\caption{An arbitrarily selected judgment example. The top box includes the defendant's demographic characteristics. The middle box presents the court's findings of major facts. The bottom box shows the court's sentencing decisions.}
	\label{f5}
\end{figure}

For each case, the CJO dataset collects a judgment document written in Chinese, including the defendant's demographic characteristics, the court's findings of major facts, and the court's sentencing decisions \citep{2002crime}. The defendant's demographic characteristics are typically included in the first paragraph of the judgement documents. It typically contains important information, such as age, gender, and ethnicity. See for example the top box in Figure \ref{f5}. The court's findings of major facts locate in the judgment between `This trial is now ended' and `Our court holds that'. See for example the middle box in Figure \ref{f5}. Lastly, the court's sentencing decisions are shown in the paragraph beginning with `The judgement is as follows'. See for example the bottom box in Figure \ref{f5}.
\subsection{Variable description}

The primary variable of interest in our study is the length of the prison sentence, reported in months. We take the log-transformed length of the prison sentence as our response variable $Y_i$ for every $ 1\leq i\leq n $. We next consider a set of five extralegal factors (i.e., $X_{1i}$) available in our CJO dataset. These are mainly the demographic variables of the defendants, including age, gender, ethnicity, employment status, and education level. All variables are coded as dummies except for age. Age of the defendant ($X_{1i1}$) is measured in years, with values ranging from 16 to 78 years. The mean age is about 33.7 years, with a standard deviation of 9.8 years. Gender ($X_{1i2}$) is coded as $X_{1i2}=1$ for males and $X_{1i2}=0$ for females. More than ninety percent (97.7\%) of the defendants in our sample are males. For ethnicity ($X_{1i3}$), we represent Han Chinese by $ X_{1i3}=1$, and other minorities by $ X_{1i3}=0$. More than three-quarters (85.4\%) of the defendants are Han Chinese. Employment status ($X_{1i4}$) is coded as $ X_{1i4}=1$ if the defendant is employed and $ X_{1i4}=0$ otherwise. About 38.4\% of the defendants are employed. Lastly, the education level is coded as $ X_{1i5}=1$ if the defendant is in elementary school (42.7\%), $ X_{1i6}=1$ for junior middle school (40.6\%), and $ X_{1i7}=1$ for high school or above (7.7\%).  Illiterate defendants (9.0\%) are coded as $ X_{1i5}=X_{1i6}=X_{1i7}=0 $. This leads to the extralegal factor vector $ X_{1i}\in\mR^7$ for every $ 1\leq i\leq n $.

Next, we extract keywords from criminal facts as legal factors, i.e., control variable $X_{2i}=(X_{2ij})$ with $X_{2ij}\in\{0,1\}$. To this end, we first cut the Chinese judgment documents into keywords and then compute their frequencies. For illustration purpose, only those keywords with a frequency of more than 10 times are kept. These account for about 4.8\% of the total number of keywords but 95.99\% of the total frequency. Among those keywords, there are many keywords, which are very high in frequency but have little actual meaning. Those keywords are then excluded from our subsequent analysis. Those excluded keywords include for example `the defendant', `the victim', `the plaintiff', and `the Public Prosecution Service'. This leads to a final set of $6,578$ keywords. We then code for each keyword as a dummy variable $X_{2ij}$, whose value is 1 if the $j$-th keyword actually appears in the $i$-th document and is 0 otherwise. Follow the idea of the proposed MCR method, we then split $X_{2i}$ into two parts. The first part contains a set of keywords that are not only high in frequency but also high in correlation with $Y$. In the meanwhile, the size of the first part is determined by the BIC score in the standard linear regression, with a final size of 64. This subvector of $ X_{2i}$ is then merged with $ X_{1i}$, so that the main covariates $X_i\in\mR^q$ with $ q=71 $ can be formed. Then, the rest of $ X_{2i} $ is formed as $ Z_i\in\mR^p $, with $ p=6,514 $ for every $ 1\leq i\leq n $.

\subsection{The estimation results}

To apply the proposed MCR method, we first estimate the number of the latent classes (i.e., $K$), using the BIC selection criterion (\ref{3.1}) proposed in Section 3.6. Specifically, we fix $ K_{\max}=20$. This leads to the final estimate $\wh{K}=\argmin_{1\leq K\leq K_{\max}}\text{BIC}(K)=7$. Subsequently, we fix $K=\wh{K}=7$ so that the main parameters of interest can be estimated. As we are most interested in estimating the effects of extralegal factors on judicial impartiality, our interpretation should focus on $X_{j}$ ($1\leq j\leq 7$) only, since they are related to these factors. The detailed estimation results are summarized in the left panel of Table \ref{t2}. Note that the reported standard errors (SE) in Table \ref{t2} are computed by simply treating $ \wh{\pi}_{ik}s$ as fixed. By Theorem \ref{thm2}, we know that the estimated posterior probability $\wh{\pi}_{ik}$ should converge to the true membership indicator function $a_{ik}=I\big(\mathcal{K}_i=k\big)$ with super fast convergence rate. This fact has been numerically verified in Section 3.5, see Figure \ref{f4}. Therefore, the difference between $ \wh{\pi}_{ik}$ and $a_{ik}$ becomes asymptotically ignorable. Consequently, the ``simple'' standard errors estimator as reported in Table \ref{t2} is indeed a statistically valid estimator for the asymptotic variance. By Table \ref{t2} and focusing on the 5\% level of significance,  we find that $X_1$ (Age) and $ X_3$ (Ethnicity) seem to be statistically significant. Consider, for example, the age effect. The corresponding coefficient of $ X_1$ is 0.0012, indicating that those at an older age tend to receive longer sentences, even after controlling for the effects of legal factors, as reflected in $Z_i$. Specifically, holding all other factors fixed, if the age of the defendant is increased by 10 years, the average sentence is expected to be about 1.21\% longer. 

For the sake of comparison, the results of ordinary linear regression (OLR) model is also presented. For the OLR model, a linear regression model is directly fitted for $Y_i$ and $ X_i$, when the information contained in $Z_i$ is completely ignored. The detailed results are summarized in the right panel of Table \ref{t2}. The OLR results seem to suggest that the gender of the defendant (i.e., $X_2$) and the education level (i.e., $ X_6$, junior middle school) are also statistically significant. Consider, for example, the gender effect. By OLR results, we find that males tend to receive longer average sentences than females. However, after conditioning on the legal factors, as reflected in $Z_i$ by MCR, this effect becomes no longer statistically significant. Therefore, it seems to us that the seemingly significant gender effect as detected by the OLR method is very questionable. It might due to the fact that male defendants are often involved in more severe criminal acts. Once those legal factor effects are well controlled by $Z_i$, this seemingly significant gender effect disappears.
\begin{table}
	\caption{Estimation results of MCR model and OLR model.}
	\vspace{0.5em}
	\renewcommand\arraystretch{1.2}
	\centering
	\begin{tabular}{c|ccc|ccc}
		\toprule
		\hline
		& \multicolumn{3}{c|}{MCR} & \multicolumn{3}{c}{OLR} \\ \hline
		$\ \ $Variable & Estimate & SE & $ P $-value & Estimate & SE & $ P $-value\\
		\hline
		$X_1$ (Age)   & $\ $0.0012 & 0.001 & 0.042  & $\ $0.0016 & 0.001 & 0.032\\
		$X_2$ (Male) & $\ $0.0662 & 0.039 & 0.088	 & $\ $0.1081 & 0.048 & 0.025\\
		$X_3$ (Han) & -0.0633 & 0.017 & 0.000	 & -0.0761 & 0.021 & 0.000\\
		$X_4$ (Employed) & -0.0118 & 0.012 & 0.324	 & -0.0002 & 0.015 & 0.988\\
		$X_5$ (Elementary) & $\ $0.0361 & 0.021 & 0.092	 & $\ $0.0574 & 0.027 & 0.031\\
		$X_6$ (Middle)& $\ $0.0214 & 0.022 & 0.326	 & $\ $0.0432 & 0.027 & 0.110 \\
		$X_7$ (High) & -0.0381 & 0.029 & 0.184 & $\ $0.0040 & 0.036 & 0.911\\
		\bottomrule
	\end{tabular}
	\label{t2}
\end{table}

\begin{figure}
	\centering
	\includegraphics[width=4in]{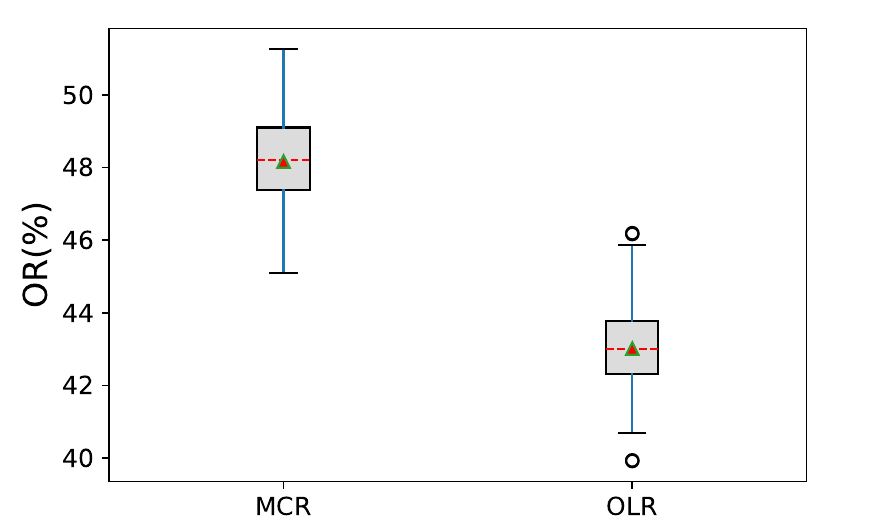}
	\caption{Out-of-sample $R$-squared (OR) values for the two competing models. The left and right boxplots represent the MCR and OLR methods, respectively.}
	\label{f6}
\end{figure}

To further support the MCR method, we next demonstrate that the MCR method also leads to more accurate prediction results than the typically used OLR method. To this end, we randomly split the whole CJO dataset into a training dataset (50\%) and a testing dataset (50\%). Here we use $\mT=\big\{\big(X_i^*, Y_i^*, Z_i^*\big): 1\leq i\leq n^*\big\}$ to represent the testing dataset. Consider an arbitrary testing sample $\big(X_i^*, Y_i^*, Z_i^*\big)\in\mT $. By model (\ref{2.1}), we should have $E\big(Y_i^*\big|X_i^*, Z_i^*\big)=\sum_{k=1}^{K}I(\mathcal{K}_i=k)\gamma_k+X_i^{*\top}\theta$. Since the value of $Y_i^*$ should not be known in advance, we need to define a new posterior probability estimator as $\wh{\pi}_{ik}^*=\wh{\pi}_k\prod_{j=1}^p\wh{p}_{kj}^{Z_{ij}}\Big(1-\wh{p}_{kj}\Big)^{1-Z_{ij}}\big/\sum_{k=1}^K\wh{\pi}_k$  $\prod_{j=1}^p\wh{p}_{kj}^{Z_{ij}}\Big(1-\wh{p}_{kj}\Big)^{1-Z_{ij}}$.
Thereafter, a predictor for $Y_i^*$ can be constructed as $ \wh{Y_i^*}=\sum_{k=1}^{K}\wh{\pi}_{ik}^*\wh{\gamma}_k+X_i^{*\top}\wh{\theta} $, where the unknown parameters are estimated on the training dataset by the final estimator $\wh{\Phi} = (\wh{\gamma}^{\top}, \wh{\theta}^{\top})^{\top}\in\mR^{K+q}$. Accordingly, the out-of-sample $R$-squared (OR) can be evaluated as
\begin{equation*}
	\text{OR} = \bigg\{1-\sum_{i=1}^{n^*}\Big(Y_i^*-\wh{Y_i^*}\Big)^2\Big/\sum_{i=1}^{n^*}\Big(Y_i^*-\overline{Y_i^*}\Big)^2\bigg\}\times 100\%,
\end{equation*}
where $\overline{Y_i^*}=\sum_{i=1}^{n^*}Y_i^*/n^*$. For a reliable evaluation, this experiment was randomly replicated for $M=100$ times. This leads to a total of $M$ OR values, which are then box-plotted in Figure \ref{f6}; see the left boxplot in Figure \ref{f6}. Repeating the experiment for the OLR method with the interested response $Y_i^*$ predicted by $\wh{Y_i}^{*(\text{OLR})}=\wh{\gamma}^{(\text{OLR})}+X_i^{*\top}\wh{\theta}^{(\text{OLR})}$, where $ \wh{\gamma}^{(\text{OLR})} $ and $\wh{\theta}^{(\text{OLR})}$ are the ordinary least squared estimators obtained on the training dataset. That leads to the right boxplot in Figure \ref{f6}. We find that the MCR method outperforms the OLR method clearly. The median of OR in OLR method is about 43.00\%, while that of MCR is about 48.19\%, which is almost 5.2\% better than OLR. Therefore, it supports that, by utilizing information provided by $ Z_i$ appropriately, the MCR method should be a very useful regression tool for testing judicial impartiality.
\section{Concluding remarks}

To summarize, we aim to provide here two important contributions to the existing literature. First, we provide the statistics literature a new regression tool for testing the interested conditional independence, when there exists an ultrahigh dimensional and binary control variable. Second, we provide the legal study literature, a new perspective for testing judicial impartiality and demonstrate its usefulness on a large-scale Chinese burglary judgment dataset. To conclude this article, we wish to discuss a few interesting new topics for future study. First, the MCR method assumes that the rich information contained in $ Z_i $ is represented by a binary feature vector. By doing so, the existence of a bag of keywords can be well represented. Nevertheless, the associated frequency information is completely ignored. Then, how to take the frequency information into consideration should be a good topic for further study \citep{kononenko1991,Sang2006}. Second, we treat an EM algorithm as if it sufficiently converge if the difference between two consecutive estimates is sufficiently small. Our numerical experiments suggest that this simple method works fairly well. However, whether the final estimator obtained by our EM algorithm indeed converges numerically to the global optimizer is not theoretically investigated and thus not guaranteed in this work. A further research along this line seems quite involved and should be a good topic for future studies \citep{xu1996convergence,1998gentle,balakrishnan2017statistical}. Third, the number of latent classes $ K $ is estimated by a BIC selection criterion here, which our preliminary numerical experiments suggest that this BIC method works fairly well. However, its theoretical properties remain unknown. Then how to fill this important theoretical gap is another interesting direction for future exploration \citep{biernacki2000,zhao2015}. Finally, the current MCR method can be viewed as a natural extension of the ordinary linear regression models. How to develop similar methods for many other popularly used generalized regression models (e.g., logistic regression) is also worth pursuing \citep{jansen1993,sedghi2016}.

\bibliographystyle{imsart-nameyear} 
\bibliography{reference}       

\begin{appendix}
	\section{Technical conditions}
	
	In this Appendix, we give some useful technical conditions for the subsequent theorems. To establish the sophisticated asymptotic theory for the proposed method, the following technical conditions are necessarily needed. Specifically, condition \ref{con1} is a standard regularity condition, which assumes that the fisher information matrix $I(p_j)$ is positive definite over $1\leq j\leq p$. Condition \ref{con2} assumes that the true parameters $p_{kj}s$ are uniformly bounded. A similar condition can be found in \cite{kononenko1991} and \cite{Sang2006}. By conditions \ref{con3} and \ref{con4}, we allow the feature dimension $p$ to diverge as $n\to\infty$ in a moderate rate. Condition \ref{con5} requires that the distributions of binary features belonging to two different classes exhibit certain differences. Condition \ref{con6} is a standard distribution assumption in high-dimensional data analysis \citep{wainwright2019high}.
	\begin{enumerate}[label=\textbf{(C\arabic*)}]
		\item\label{con1} Assume that there exists some fixed constant $ \tau_{\min}>0$ such that $ \min\limits_{j}\lambda_{\min}\big\{I(p_j)\big\}\geq\tau_{\min}$.
		\item\label{con2} Assume that there exist some positive constants $0<p_{\min}\leq p_{\max}<1$ such that $ p_{\min}\leq\min\limits_{k,j}p_{k,j}\leq\max\limits_{k,j}p_{k,j}\leq p_{\max}$.
		\item\label{con3} Assume that $p\to\infty$ and $ \log(p)/n\to 0$ as $ n\to\infty$.
		\item\label{con4} Assume that $ \log(n)/p\to 0$ as $ n\to\infty$.
		\item\label{con5} Assume that $\Delta_{\min}>0$, where $ \Delta_{\min}=\min_{k_1\neq k_2}\Big[-p^{-1}\sum_{j=1}^p\Big\{p_{k_1j}\log\big(p_{k_2j}/p_{k_1j}\big)+(1-p_{k_1j})$  $\log\big((1-p_{k_2j})/(1-p_{k_1j})\big)\Big\}\Big]$.
		\item\label{con6} Assume that $X_{ij}$ follows a sub-Gaussian distribution for each $ 1\leq j\leq q$, where $X_i=(X_{i1},\ldots,X_{iq})\in\mR^q$.
	\end{enumerate}
	
	\section{Proof of the main theoretical results}
	\subsection{Proof of Theorem 1}
	
	Consider the following simplified log-likelihood function for every $p_j$ as
	\begin{equation*}
		\wh{\mL^{(j)}}(p_j)=\sum_{i=1}^{n}\log\bigg\{\sum_{k=1}^K\wh{c_{ik}}p_{kj}^{Z_{ij}}\Big(1-p_{kj}\Big)^{1-Z_{ij}}\bigg\},
	\end{equation*}	
	where $\wh{c_{ik}}=\wh{\pi}_k\big(\sqrt{2\pi}\wh{\sigma}\big)^{-1}\exp\Big\{-\Big(Y_i-\wh{\gamma}_k-X_i^{\top}\wh{\theta}\Big)^2\big/\big(2\wh{\sigma}^2\big)\Big\}$. Recall that $ C_n>0$ is a positive constant such that $ C_n/\sqrt{n}\to 0 $ as $ n\to\infty$. Then following the idea of \cite{Fan2001}, we know that there must exists a local maximizer $ \wh{p_j}$ in $(p_j-C_n/\sqrt{n},p_j+C_n/\sqrt{n})$, if we can prove that
	\begin{equation*}
		\mathop{\text{lim inf}}\limits_{n\to\infty} P\bigg[C_n^{-2}\sup\limits_{\lVert u\lVert=1} \Big\{\wh{\mL^{(j)}}\Big(p_j+uC_n/\sqrt{n}\Big)-\wh{\mL^{(j)}}\big(p_j\big)\Big\}<0\bigg]\geq 1-\varepsilon.
	\end{equation*} 
	for any given $\varepsilon>0$. It can be verified that $-\wh{\mL^{(j)}}(p_j)$ is a strictly convex function for $ p_j $. The verification details are given in \sc{Part 1}\normalfont of Appendix C.1. Then, we know that the local maximizer $\wh{p_j}$ is also the uniquely defined global maximizer as $\wh{p_j}=\argmax_{p}\wh{\mL^{(j)}}(p)$. Since $ C_n/\sqrt{n}\to 0$ as $ n\to\infty$, we know that $ \left\Vert\wh{p_j}-p_j\right\Vert=O_p(C_n/\sqrt{n})$ and therefore $ \wh{p_j} $ is a consistent estimator for $ p_j $.
	
	Unfortunately, the conclusion needs to be proved here is not the statistical consistency of $ \wh{p_j} $ for any given $1\leq j\leq p$. The desired theorem conclusion is the uniform consistency for $ \wh{p_j} $ over every $ j $. In other words, we wish to prove $ \max_{j}\left\Vert\wh{p_j}-p_j\right\Vert=O_p(C_n/\sqrt{n})$. The theorem conclusion follows, if we can show that
	\begin{equation}\label{aa1}
		\mathop{\text{lim inf}}\limits_{n\to\infty} P\bigg[C_n^{-2}\max\limits_{j}\sup\limits_{\lVert u\lVert=1} \Big\{\wh{\mL^{(j)}}\Big(p_j+uC_n/\sqrt{n}\Big)-\wh{\mL^{(j)}}\big(p_j\big)\Big\}<0\bigg]\geq 1-\varepsilon,
	\end{equation} 
	for any given $\varepsilon>0$. To this end, we apply Taylor's expansion and decompose $\wh{\Delta^{(j)}}=\wh{\mL^{(j)}}(p_j+uC_n/\sqrt{n})-\wh{\mL^{(j)}}(p_j)=\wh{W_a^{(j)}}+\wh{W_b^{(j)}}+\wh{W_c^{(j)}}$, where $ \wh{W_a^{(j)}}=\big\{n^{-1/2}\wh{\dot{\mL}^{(j)}}(p_j)\big\}^{\top}\big(uC_n\big)$, $ \wh{W_b^{(j)}}=2^{-1}\big(uC_n\big)^{\top}\big\{n^{-1}$  $\wh{\ddot{\mL}^{(j)}}(p_j)\big\}\big(uC_n\big)$, $\wh{W_c^{(j)}}=2^{-1}\big(uC_n\big)^{\top}n^{-1}\big\{\wh{\ddot{\mL}^{(j)}}(\wt{p}_j)-\wh{\ddot{\mL}^{(j)}}(p_j)\big\}$  $\big(uC_n\big)$, and $ \wt{p}_j=\alpha_jp_j+(1-\alpha_j)(p_j+uC_n/\sqrt{n}) $ for some $ \alpha_j\in (0,1)$. We then have $ C_n^{-2}\max\limits_{j}\sup\limits_{\lVert u\lVert=1}\wh{\Delta^{(j)}}\leq C_n^{-2}\max\limits_{j}\sup\limits_{\lVert u\lVert=1}\wh{W_a^{(j)}}+C_n^{-2}\max\limits_{j}\sup\limits_{\lVert u\lVert=1}\wh{W_b^{(j)}}+C_n^{-2}\max\limits_{j}\sup\limits_{\lVert u\lVert=1}\wh{W_c^{(j)}}$. Define $\nu_{\min}=\tau_{\min}/2>0$. Recall that $\log\big(p\big)/C_n^2\to 0$ as $ p\to\infty$, $\log\big(p\big)/n\to 0$ as $ p\to\infty$ and $n\to\infty$. Note that $\wh{\Omega}$ is $\sqrt{n}$-consistent. Then, it suffices to show that 
	\begin{gather}
		P\bigg\{C_n^{-2}\max\limits_{j}\sup\limits_{\lVert u\lVert=1}\Big|\wh{W_a^{(j)}}\Big|>\varepsilon\bigg\}\nonumber\\
		\label{aa2}\leq 2K\exp\bigg\{-C_n^2\Big(\varepsilon^2C_1-\log(p)/C_n^2\Big)\bigg\}+
		KP\bigg\{\Big\Vert\wh{\Omega}-\Omega\Big\Vert>\varepsilon C_2C_n/\sqrt{n}\bigg\}+\frac{C_3}{n} \\
		P\bigg\{C_n^{-2}\max\limits_{j}\sup\limits_{\lVert u\lVert=1}\Big(\wh{W_b^{(j)}}+C_n^2\nu_{\min}\Big)>\varepsilon\bigg\}\nonumber\\
		\label{aa3} \leq 2K^2\exp\bigg\{-n\Big(U_1\varepsilon^2-\log(p)/n\Big)\bigg\}+ K^2P\bigg\{\Big\Vert\wh{\Omega}-\Omega\Big\Vert>U_2\varepsilon\bigg\}+\frac{U_3}{n} \\
		\label{aa4}P\bigg\{C_n^{-2}\max\limits_{j}\sup\limits_{\lVert u\lVert=1}\Big|\wh{W_c^{(j)}}\Big|>\varepsilon\bigg\}
		\leq 2K^2P\bigg\{\Big\Vert\wh{\Omega}-\Omega\Big\Vert>2U_2\varepsilon/3\bigg\}+\frac{2U_3}{n},
	\end{gather}
	for any given $\varepsilon>0$. Here $C_1$, $C_2$, $C_3$, $U_1$, $U_2$ and $U_3$ are some fixed and positive constants. Those three conclusions are to be proved in the following three steps.
	
	\sc{Step 1.}\normalfont$\ $We start with (\ref{aa2}). Define $ W_a^{(j)}=\big\{n^{-1/2}\dot{\mL}^{(j)}(p_j)\big\}^{\top}\big(uC_n\big)$ and $\Delta_a^{(j)}=\wh{W_a^{(j)}}-W_a^{(j)}$. Note that $ \wh{W_a^{(j)}}=W_a^{(j)}+\Delta_a^{(j)}$. Then, we have $P\Big\{C_n^{-2}\max\limits_{j}\sup\limits_{\lVert u\lVert=1}\big|\wh{W_a^{(j)}}\big|>\varepsilon \Big\}\leq\mD_{a1}^{(j)}+\mD_{a2}^{(j)}$, where $\mD_{a1}^{(j)}=P\Big\{C_n^{-2}\max\limits_{j}\sup\limits_{\lVert u\lVert=1}\big|W_a^{(j)}\big|>\varepsilon /2\Big\}$ and $\mD_{a2}^{(j)}=P\Big\{C_n^{-2}$  $\max\limits_{j}\sup\limits_{\lVert u\lVert=1}$  $\big|\Delta_a^{(j)}\big|>\varepsilon /2\Big\}$. To prove (\ref{aa2}), it suffices to upper bound $\mD_{a1}^{(j)}$ and $\mD_{a2}^{(j)}$ separately. The details are given in the following two sub-steps.
	
	\sc{Step 1.1.}\normalfont$\ $We start with $\mD_{a1}^{(j)}$.
	Note that $ \sup\limits_{\lVert u\lVert=1}\big|W_a^{(j)}\big|\leq C_n\sqrt{n}\big\Vert n^{-1}\dot{\mL}^{(j)}(p_j)\big\Vert$. Then, we obtain the following inequality as
	\begin{equation}\label{aa5}
		\mD_{a1}^{(j)}\leq \sum_{j=1}^pP\bigg\{\Big\Vert n^{-1}\dot{\mL}^{(j)}(p_j)\Big\Vert>\dfrac{\varepsilon C_n}{2\sqrt{n}}\bigg\}.
	\end{equation}
	To upper bound the right hand side of (\ref{aa5}), we shall focus on $n^{-1}\dot{\mL}^{(j)}(p_j)\in\mR^K$. Recall that $ \dot{\mL}^{(j)}(p_j)=\big(\dot{\ell_1}^{(j)},\ldots,\dot{\ell_K}^{(j)}\big)^{\top}\in\mR^K$, $\dot{\ell_k}^{(j)}=\sum_{i=1}^n\dot{\ell_k}^{(j,i)}$, $\dot{\ell_k}^{(j,i)}=\alpha_{ik}^{(j)}s(Z_{ij},p_{kj})$, $\alpha_{ik}^{(j)}=c_{ik}p_{kj}^{Z_{ij}}\big(1-p_{kj}\big)^{1-Z_{ij}}\big/\sum_{k=1}^Kc_{ik}p_{kj}^{Z_{ij}}\big(1-p_{kj}\big)^{1-Z_{ij}}$, and $ s(Z_{ij},p_{kj})$  $=Z_{ij}/p_{kj}-\big(1-Z_{ij}\big)/\big(1-p_{kj}\big)$. Then, we have $\big\Vert n^{-1}\dot{\mL}^{(j)}(p_j)\big\Vert=\Big\{\sum_{k=1}^K\big(n^{-1}\dot{\ell_k}^{(j)}\big)^2\Big\}^{1/2}\leq\sqrt{K}\max_k$  $\left|n^{-1}\dot{\ell_k}^{(j)}\right|$. Thus, $P\Big\{\big\Vert n^{-1}\dot{\mL}^{(j)}(p_j)\big\Vert>\varepsilon C_n/\big(2\sqrt{n}\big)\Big\}\leq\sum_{k=1}^KP\Big\{\left|n^{-1}\sum_{i=1}^n\dot{\ell_k}^{(j,i)}\right|>\varepsilon C_n\big/\Big(2\sqrt{nK}\Big)\Big\}$. One can verify that $\big|\dot{\ell_k}^{(j,i)}\big|\leq p_m$, $ E\big(\dot{\ell_k}^{(j,i)}\big)=0$ and $\vt\big(\dot{\ell_k}^{(j,i)}\big)\leq p_m^2$, where $p_m=p_{\min}^{-1}+(1-p_{\max})^{-1}$ is a fixed constant. The derivation details are given in \sc{Part 2 }\normalfont of Appendix C.1. Thus, we can apply the Bernstein's Inequality \citep{bernstein1926} as
	\begin{equation*}
		P\left\{\bigg|\frac{1}{n}\sum_{i=1}^n\dot{\ell}_k^{(j,i)}\bigg|>\dfrac{\varepsilon C_n}{2\sqrt{nK}}\right\}\leq 2\exp\left[-\dfrac{n\Big\{\varepsilon C_n\big/\Big(2\sqrt{nK}\Big)\Big\}^2/2}{p_m^2+p_m\varepsilon C_n\big/\Big(6\sqrt{nK}\Big)}\right].
	\end{equation*}
	Recall that $ C_n/\sqrt{n}\to 0$ as $n\to\infty$. This means that for any given $ \eta>0$, there exists a sufficient large but fixed constant $N_a>0$, such that $ C_n/\sqrt{n}<\eta$ if $n>N_a$. Then by (\ref{aa5}) and for sufficient large $n>N_a$, we should have
	\begin{equation}\label{aa6}
		\mD_{a1}^{(j)}\leq  2K\exp\bigg\{-C_n^2\Big(\varepsilon^2C_1-\log(p)/C_n^2\Big)\bigg\},
	\end{equation}
	where $C_1=1\big/\Big(8p_m^2K+4p_m\varepsilon\eta\sqrt{K}/3\Big)>0$ is a fixed constant.
	
	\sc{Step 1.2.}\normalfont$\ $Next, we study $\mD_{a2}^{(j)}$. Recall that $\mD_{a2}^{(j)}=P\Big\{C_n^{-2}\max\limits_{j}\sup\limits_{\lVert u\lVert=1}\big|\Delta_a^{(j)}\big|>\varepsilon /2\Big\}$. Define $\dot{\mL}^{(j)}_{\Delta}=n^{-1}\wh{\dot{\mL}^{(j)}}(p_j)-n^{-1}\dot{\mL}^{(j)}(p_j)=(\delta_1^{(j)},\ldots,\delta_K^{(j)})^{\top}\in\mR^K$, where $\delta_k^{(j)}=n^{-1}\sum_{i=1}^n\delta_k^{(j,i)}$ and $\delta_k^{(j,i)}=\wh{\alpha_{ik}^{(j)}}s(Z_{ij},p_{kj})-\alpha_{ik}^{(j)}s(Z_{ij},p_{kj})$. Note that $ \sup\limits_{\lVert u\lVert=1}\big|\Delta_a^{(j)}\big|=\sup\limits_{\lVert u\lVert=1}\big|\wh{W_a^{(j)}}-W_a^{(j)}\big|\leq C_n\sqrt{n}\big\Vert\dot{\mL}^{(j)}_{\Delta}\big\Vert$.  Then, we have 
	\begin{equation}\label{aa7}
		\mD_{a2}^{(j)}\leq P\bigg\{\max\limits_{j}\Big\Vert \dot{\mL}^{(j)}_{\Delta}\Big\Vert>\dfrac{\varepsilon C_n}{2\sqrt{n}}\bigg\}.
	\end{equation}
	To upper bound the right hand side of (\ref{aa7}), we shall focus on $\dot{\mL}^{(j)}_{\Delta}\in\mR^K$. Similar to \sc{Step 1.1,}\normalfont$\ $, we have $\big\Vert \dot{\mL}^{(j)}_{\Delta}\big\Vert=\Big\{\sum_{k=1}^K\big(\delta_k^{(j)}\big)^2\Big\}^{1/2}\leq\sqrt{K}\max_k$  $\left|\delta_k^{(j)}\right|$. Thus, $P\Big\{\max\limits_{j}\big\Vert \dot{\mL}^{(j)}_{\Delta}\big\Vert>\varepsilon C_n\big/\Big(2\sqrt{n}\Big)\Big\}\leq\sum_{k=1}^KP\Big\{\max\limits_{j}\left|\delta_k^{(j)} \right|>\varepsilon C_n\big/\Big(2\sqrt{nK}\Big)\Big\}$. By the Taylor's expansion about $ \delta_k^{(j)}$ for $\wh{\Omega}$ at $\Omega=(\pi^{\top},\gamma^{\top},\theta^{\top},\sigma)^{\top}\in\mR^{2K+q+1}$, we obtain that
	\begin{equation*}
		\frac{1}{n}\sum_{i=1}^n\Big(\wh{\alpha_{ik}^{(j)}}-\alpha_{ik}^{(j)}\Big)s\Big(Z_{ij},p_{kj}\Big)=\bigg\{\frac{1}{n}\sum_{i=1}^n\dot{\alpha}_{ik}^{(j)}(\wt{\Omega})s\Big(Z_{ij},p_{kj}\Big)\bigg\}^{\top}\Big(\wh{\Omega}-\Omega\Big),
	\end{equation*}	
	where $\dot{\alpha}_{ik}^{(j)}(\Omega)=\partial\alpha_{ik}^{(j)}/\partial\Omega\in\mR^{2K+q+1}$ and $\wt{\Omega}=\alpha\Omega+\big(1-\alpha\big)\wh{\Omega}$ for some $ \alpha\in(0,1)$.
	One can verify that $\big\Vert \dot{\alpha}_{ik}^{(j)}\big(\wt{\Omega}\big)s\big(Z_{ij},p_{kj}\big)\big\Vert\leq C_{K,q}M_a(\varepsilon_i)$ with $C_{K,q}=\sqrt{2K+q+1}>0$ and
	$M_a(\varepsilon_i)=M_1^a\varepsilon_i^2+M_2^a|\varepsilon_i|+M_3^a$. Here $M_1^a>0$, $M_2^a>0$ and $M_3^a>0$ are some fixed constants.
	The verification details are given in \sc{Part 3 }\normalfont of Appendix C.1.
	Then, we have $\Big\Vert\Big\{\sum_{i=1}^n\dot{\alpha}_{ik}^{(j)}\big(\wt{\Omega}\big)s\big(Z_{ij},p_{kj}\big)/n\Big\}^{\top}\big(\wh{\Omega}-\Omega\big)\Big\Vert\leq \Big\{\sum_{i=1}^nC_{K,q}M_a(\varepsilon_i)/n\Big\}$  $\big\Vert\wh{\Omega}-\Omega\big\Vert$, which is independent of $j$. Consequently, we can obtain that 
	\begin{equation}\label{aa8}
		P\left\{\max\limits_{j}\left|\delta_k^{(j)} \right|>\dfrac{\varepsilon C_n}{2\sqrt{nK}}\right\}\leq P\left[\bigg\{\frac{1}{n}\sum_{i=1}^nM_a(\varepsilon_i)\bigg\}\Big\Vert\wh{\Omega}-\Omega\Big\Vert>\varepsilon CC_n/\sqrt{n}\right],
	\end{equation}
	where $C=1\big/\Big(2C_{K,q}\sqrt{K}\Big)>0$. Recall that $ M_a(\varepsilon_i)$ is the independently and identically distributed random variable with finite moments $E\big\{M_a(\varepsilon_i)\big\}=C_{M_a}<\infty$ and $\vt\big\{M_a(\varepsilon_i)\big\}=C_{M_a}^{(v)}<\infty$. The verification details are given in \sc{Part 4 }\normalfont of Appendix C.1. Then, by Law of Large Numbers, we have $\sum_{i=1}^nM_a(\varepsilon_i)/n\stackrel{p}{\longrightarrow}C_{M_a}$. Thus, the right hand side of (\ref{aa8}) can be upper bounded by $P\Big\{\sum_{i=1}^nM_a(\varepsilon_i)/n>2C_{M_a}\Big\}+P\Big\{\big\Vert\wh{\Omega}-\Omega\big\Vert>\varepsilon CC_n/\big(2\sqrt{n}C_{M_a}\big)\Big\}$. By Chebyshev's Inequality \citep{1867Chebyshev}, we have $P\Big\{\Big|\sum_{i=1}^nM_a(\varepsilon_i)/n-C_{M_a}\Big|>C_{M_a}\Big\}\leq C_{M_a}^{(v)}/\big(nC_{M_a}^2\big)$. Then, we have $ \mD_{a2}^{(j)} $ to be upper bounded by
	\begin{equation}\label{aa9}
		\mD_{a2}^{(j)} \leq
		KP\bigg\{\Big\Vert\wh{\Omega}-\Omega\Big\Vert>\varepsilon C_2C_n/\sqrt{n}\bigg\}+C_3/n,
	\end{equation}
	where $C_2=C/\big(2C_{M_a}\big)>0$ and $C_3= KC_{M_a}^{(v)}/C_{M_a}^2>0$ are some fixed constants. Combining the results of (\ref{aa6}) and (\ref{aa9}), the conclusion in (\ref{aa2}) has been proved.
	
	\sc{Step 2.}\normalfont$\ $We next study (\ref{aa3}). Recall that $\wh{W_b^{(j)}}=2^{-1}\big(uC_n\big)^{\top}\big\{n^{-1}\wh{\ddot{\mL}^{(j)}}(p_j)\big\}\big(uC_n\big)$. Similarly, we can define $W_b^{(j)}=2^{-1}\big(uC_n\big)^{\top}\big\{n^{-1}\ddot{\mL}^{(j)}(p_j)\big\}\big(uC_n\big)$, $W_I^{(j)}= 2^{-1}\big(uC_n\big)^{\top}\big\{-I(p_j)\big\}\big(uC_n\big)$, $ \Delta_{b1}^{(j)}=W_b^{(j)}-W_I^{(j)}$ and $ \Delta_{b2}^{(j)}=\wh{W_b^{(j)}}-W_b^{(j)}$. Then, we have $\wh{W_b^{(j)}}=W_I^{(j)}+\Delta_{b1}^{(j)}+\Delta_{b2}^{(j)}$. Furthermore, $\max\limits_{j}\sup\limits_{\lVert u\lVert=1}\big(\wh{W_b^{(j)}}+C_n^2\nu_{\min}\big)=\max\limits_{j}\sup\limits_{\lVert u\lVert=1}\big(W_I^{(j)}+C_n^2\nu_{\min}\big)+\max\limits_{j}\sup\limits_{\lVert u\lVert=1}\Delta_{b1}^{(j)}+\max\limits_{j}\sup\limits_{\lVert u\lVert=1}\Delta_{b2}^{(j)}$. 
	By \ref{con1}, we have $\lambda_{\min}\big\{I(p_j)\big\}\geq \tau_{\min}>0$. In other words, $\lambda_{\max}\big\{-I(p_j)\big\}\leq -\tau_{\min}$. Note that $ \nu_{\min}<\tau_{\min}$. Then, we can obtain that $P\Big\{C_n^{-2}\max\limits_{j}\sup\limits_{\lVert u\lVert=1}\big(W_I^{(j)}+C_n^2\nu_{\min}\big)>\varepsilon\Big\}=0$. Thus, $P\Big\{C_n^{-2}\max\limits_{j}\sup\limits_{\lVert u\lVert=1}\big(\wh{W_b^{(j)}}$  $+C_n^2\nu_{\min}\big)>\varepsilon\Big\}\leq\mD_{b1}^{(j)}+\mD_{b2}^{(j)}$, where $ \mD_{b1}^{(j)}=P\Big\{C_n^{-2}\max\limits_{j}\sup\limits_{\lVert u\lVert=1}\Delta_{b1}^{(j)}>\varepsilon/2\Big\}$ and $ \mD_{b2}^{(j)}=P\Big\{C_n^{-2}\max\limits_{j}\sup\limits_{\lVert u\lVert=1}\Delta_{b2}^{(j)}>\varepsilon/2\Big\}$. To prove (\ref{aa3}), it suffices to upper bound $\mD_{b1}^{(j)}$ and $\mD_{b2}^{(j)}$ separately. The details are given in the following two sub-steps.
	
	\sc{Step 2.1.}\normalfont$\ $We start with $\mD_{b1}^{(j)}$. Recall that 
	$\ddot{\mL}^{(j)}(p_j)=\big(\ddot{\ell}_{k_1k_2}^{(j)}\big) \in\mR^{K\times K}$, $\ddot{\ell}_{k_1k_2}^{(j)}=-\sum_{i=1}^n\alpha_{ik_1}^{(j)}$ $s(Z_{ij},p_{k_1j})\alpha_{ik_2}^{(j)}s(Z_{ij},p_{k_2j})$, $I(p_j)=\big(\maE_{k_1k_2}^{(j)}\big)\in\mR^{K\times K}$, $\maE_{k_1k_2}^{(j)}=n^{-1}\sum_{i=1}^nE\big(c_{ik_1}$  $c_{ik_2}/M_{ij}\big)$, and $M_{ij}=\sum_{k=1}^Kc_{ik}p_{kj}\sum_{k=1}^Kc_{ik}\big(1-p_{kj}\big)$. Define $\Delta_{bE}^{(j)}=\big(\delta_{k_1k_2}^{(j,b_1)}\big)\in\mR^{K\times K}$, where  $\delta_{k_1k_2}^{(j,b_1)}=n^{-1}\ddot{\ell}_{k_1k_2}^{(j)}-\big(-\maE_{k_1k_2}^{(j)}\big)=-n^{-1}\sum_{i=1}^n\Big\{\alpha_{ik_1}^{(j)}s(Z_{ij},p_{k_1j})\alpha_{ik_2}^{(j)}$  $s(Z_{ij},p_{k_2j})-E\big(c_{ik_1}$  
	$c_{ik_2}/M_{ij}\big)\Big\}$. Equivalently, $\Delta_{bE}^{(j)}=n^{-1}\ddot{\mL}^{(j)}(p_j)-\big\{-I(p_j)\big\}$. Then, we have $\sup\limits_{\lVert u\lVert=1}\Delta_{b1}^{(j)}\leq\sup\limits_{\lVert u\lVert=1}\big|\Delta_{b1}^{(j)}\big|\leq2^{-1}C_n^2\big\Vert\Delta_{bE}^{(j)}\big\Vert$. Thus, we obtain the following inequality as 
	$\mD_{b1}^{(j)}\leq \sum_{j=1}^p$ $P\Big\{\big\Vert\Delta_{bE}^{(j)}\big\Vert>\varepsilon\Big\}$.  Note that $\Delta_{bE}^{(j)}$ is a symmetric matrix. Thus, by the proof of Lemma 1 in \cite{2009Wang}, we have $\lambda_{\max}\big(\Delta_{bE}^{(j)}\big)\leq K\max_{k_1,k_2}\big|\delta_{k_1k_2}^{(j,b_1)}\big|$. Then, $\mD_{b1}^{(j)}$ can be further upper bounded by 
	\begin{equation}\label{aa10}
		\mD_{b1}^{(j)}\leq \sum_{j=1}^p
		P\bigg\{K\max_{k_1,k_2}\Big|\delta_{k_1k_2}^{(j,b_1)}\Big|>\varepsilon\bigg\}\leq\sum_{j=1}^p\sum_{k_1,k_2}
		P\bigg\{\Big|\delta_{k_1k_2}^{(j,b_1)}\Big|>\varepsilon/K\bigg\}.
	\end{equation}
	Recall that $\delta_{k_1k_2}^{(j,b_1)}=-n^{-1}\sum_{i=1}^n\delta_{k_1k_2}^{(j,b_1,i)}$, where $\delta_{k_1k_2}^{(j,b_1,i)}=\alpha_{ik_1}^{(j)}s(Z_{ij},p_{k_1j})\alpha_{ik_2}^{(j)}s(Z_{ij},p_{k_2j})-E\big(c_{ik_1}c_{ik_2}/M_{ij}\big)$. Then, the right hand side of (\ref{aa10}) can be bounded by $K^2\sum_{j=1}^p$  $ P\Big\{\Big|n^{-1}\sum_{i=1}^n\delta_{k_1k_2}^{(j,b_1,i)}\Big|>\varepsilon/K\Big\}$. It can verified that $|\delta_{k_1k_2}^{(j,b_1,i)}|\leq p_m^2$, $E\big(\delta_{k_1k_2}^{(j,b_1,i)}\big)=0$ and $ \vt\big(\delta_{k_1k_2}^{(j,b_1,i)}\big)$  $\leq p_v$, where $ p_v=p_{\min}^{-3}+(1-p_{\min})^{-3}+p_{\max}^{-3}+(1-p_{\max})^{-3}$ is a fixed and positive constant. The verification details are given in \sc{Part 5 }\normalfont of Appendix C.1. Therefore, the Bernstein's Inequality can be applied. Similar to \sc{Step 1.1,}\normalfont$\ $we can obtain the following upper bound of $\mD_{b1}^{(j)}$ as
	\begin{equation}\label{aa11}
		\mD_{b1}^{(j)}\leq K^2\sum_{j=1}^pP\Bigg\{\bigg|\frac{1}{n}\sum_{i=1}^n\delta_{k_1k_2}^{(j,b_1,i)}\bigg|>\frac{\varepsilon}{K}\Bigg\}\leq 2K^2\exp\bigg\{-n\Big(U_1\varepsilon^2-\log(p)/n\Big)\bigg\},
	\end{equation}
	where $ U_1=1\big/\Big(p_vK^2+p_m^2\varepsilon/K\Big)>0$ is a fixed constant.
	
	\sc{Step 2.2.}\normalfont$\ $Next, we study $\mD_{b2}^{(j)}$. Recall that $ \mD_{b2}^{(j)}=P\Big\{C_n^{-2}\max\limits_{j}\sup\limits_{\lVert u\lVert=1}\Delta_{b2}^{(j)}>\varepsilon/2\Big\}$. Define $\ddot{\mL}^{(j)}_{\Delta}=n^{-1}\wh{\ddot{\mL}^{(j)}}(p_j)-n^{-1}\ddot{\mL}^{(j)}(p_j)=\big(\delta_{k_1k_2}^{(j,b_2)}\big)\in\mR^{K\times K}$, where $\delta_{k_1k_2}^{(j,b_2)}=n^{-1}\sum_{i=1}^n\delta_{k_1k_2}^{(j,b_2,i)}$ and $\delta_{k_1k_2}^{(j,b_2,i)}=\wh{\ddot{\ell}_{k_1k_2}^{(j,i)}}-\ddot{\ell}_{k_1k_2}^{(j,i)}$ with $ \wh{\ddot{\ell}_{k_1k_2}^{(j,i)}}=-\wh{\alpha_{ik_1}^{(j)}}\wh{\alpha_{ik_2}^{(j)}}s(Z_{ij},p_{k_1j})s(Z_{ij},p_{k_2j})$ and  $\ddot{\ell}_{k_1k_2}^{(j,i)}=-\alpha_{ik_1}^{(j)}\alpha_{ik_2}^{(j)}s(Z_{ij},p_{k_1j})s(Z_{ij},p_{k_2j})$. Note that $\sup\limits_{\lVert u\lVert=1}\Delta_{b2}^{(j)}\leq\sup\limits_{\lVert u\lVert=1}\big|\Delta_{b2}^{(j)}\big|=\sup\limits_{\lVert u\lVert=1}\big|\wh{W_b^{(j)}}-W_b^{(j)}\big|\leq 2^{-1}C_n^2\big\Vert\ddot{\mL}^{(j)}_{\Delta}\big\Vert$. Then, we have $\mD_{b2}^{(j)}\leq P\Big\{\max\limits_{j}\big\Vert\ddot{\mL}^{(j)}_{\Delta}\big\Vert>\varepsilon\Big\}$. Note that $ \ddot{\mL}^{(j)}_{\Delta}$ is also a symmetric matrix. Similar to \sc{Step 2.1,}\normalfont$\ $we have $\lambda_{\max}\big(\ddot{\mL}^{(j)}_{\Delta}\big)\leq K\max_{k_1,k_2}\big|\delta_{k_1k_2}^{(j,b_2)}\big|$. Thus, $P\Big\{\max\limits_{j}\big\Vert\ddot{\mL}^{(j)}_{\Delta}\big\Vert>\varepsilon\Big\}\leq\sum_{k_1,k_2}P\Big\{\max\limits_{j}\left|\delta_{k_1k_2}^{(j,b_2)}\right|>\varepsilon/K\Big\}$. Similar to \sc{Step 1.2,}\normalfont$\ $we conduct the Taylor's expansion about $\delta_{k_1k_2}^{(j,b_2)}$ for $\wh{\Omega}$ at $\Omega$ as
	\begin{equation}\label{aa12}
		\frac{1}{n}\sum_{i=1}^n\wh{\ddot{\ell}_{k_1k_2}^{(j,i)}}-\frac{1}{n}\sum_{i=1}^n\ddot{\ell}_{k_1k_2}^{(j,i)}=\bigg\{\frac{1}{n}\sum_{i=1}^n\dddot{\ell}_{k_1k_2}^{(j,i)}\big(\wt{\Omega}\big)\bigg\}^{\top}\Big(\wh{\Omega}-\Omega\Big),
	\end{equation}
	where $\dddot{\ell}_{k_1k_2}^{(j,i)}\big(\Omega\big)=\partial\ddot{\ell}_{k_1k_2}^{(j,i)}/\partial\Omega\in\mR^{2K+q+1}$ and $\wt{\Omega}=\alpha\Omega+\big(1-\alpha\big)\wh{\Omega}$ for some $ \alpha\in(0,1)$. One can also verify that $\big\Vert \dddot{\ell}_{k_1k_2}^{(j,i)}\big(\wt{\Omega}\big)\Vert\leq C_{K,q}M_b(\varepsilon_i)$ with 
	$M_b(\varepsilon_i)=M_1^b\varepsilon_i^2+M_2^b|\varepsilon_i|+M_3^b$. Here $M_1^b>0$, $M_2^b>0$ and $M_3^b>0$ are some fixed constants. The verification details are given in \sc{Part 6 }\normalfont of Appendix C.1. Then, we have $\Big\Vert\Big\{\sum_{i=1}^n\dddot{\ell}_{k_1k_2}^{(j,i)}\big(\wt{\Omega}\big)/n\Big\}^{\top}\big(\wh{\Omega}-\Omega\big)\Big\Vert\leq \Big\{\sum_{i=1}^nC_{K,q}M_b(\varepsilon_i)/n\Big\}\big\Vert\wh{\Omega}-\Omega\big\Vert$, which is also independent of $j$. Consequently, we can obtain that 
	\begin{equation}\label{aa13}
		P\left\{\max\limits_{j}\left|\delta_{k_1k_2}^{(j,b_2)} \right|>\varepsilon/K\right\}\leq P\left[\bigg\{\frac{1}{n}\sum_{i=1}^nM_b(\varepsilon_i)\bigg\}\Big\Vert\wh{\Omega}-\Omega\Big\Vert>\varepsilon/K\right],
	\end{equation}
	Similar to \sc{Step 1.2,}\normalfont$\ $we have $E\big\{M_b(\varepsilon_i)\big\}=C_{M_b}<\infty$ and $\vt\big\{M_b(\varepsilon_i)\big\}=C_{M_b}^{(v)}<\infty$. The verification details are given in \sc{Part 7 }\normalfont of Appendix C.1. Then, by Law of Large Numbers, we have $\sum_{i=1}^nM_b(\varepsilon_i)/n\stackrel{p}{\longrightarrow}C_{M_b}$. Thus, the right hand side of (\ref{aa13}) can be upper bounded by $P\Big\{\sum_{i=1}^nM_b(\varepsilon_i)/n>2C_{M_b}\Big\}+P\Big\{\big\Vert\wh{\Omega}-\Omega\big\Vert>\varepsilon/\big(KC_{M_b}\big)\Big\}$. By Chebyshev's Inequality, we have $P\Big\{\Big|\sum_{i=1}^nM_b(\varepsilon_i)/n-C_{M_b}\Big|>C_{M_b}\Big\}\leq C_{M_b}^{(v)}/\big(nC_{M_b}^2\big)$. Then, we have $ \mD_{b2}^{(j)} $ to be upper bounded by
	\begin{equation}\label{aa14}
		\mD_{b2}^{(j)} \leq
		K^2P\bigg\{\Big\Vert\wh{\Omega}-\Omega\Big\Vert>U_2\varepsilon\bigg\}+U_3/n,
	\end{equation}
	where $U_2=1/\big(KC_{M_b}\big)>0$ and $U_3= K^2C_{M_b}^{(v)}/C_{M_b}^2>0$ are some fixed constants. Combining the results of (\ref{aa11}) and (\ref{aa14}), the conclusion in (\ref{aa3}) has been proved.
	
	\sc{Step 3.}\normalfont$\ $Finally, we shall prove (\ref{aa4}). Recall that $\wh{W_c^{(j)}}=2^{-1}\big(uC_n\big)^{\top}n^{-1}\big\{\wh{\ddot{\mL}^{(j)}}(\wt{p}_j)-\wh{\ddot{\mL}^{(j)}}(p_j)\big\}\big(uC_n\big)$. Similar to \sc{Step 1.}\normalfont$\ $and \sc{Step 2,}\normalfont$\ $define $W_c^{(j)}=2^{-1}\big(uC_n\big)^{\top}n^{-1}\big\{\ddot{\mL}^{(j)}(\wt{p}_j)-\ddot{\mL}^{(j)}(p_j)\big\}\big(uC_n\big)$, $\Delta_{c1}^{(j)}=2^{-1}\big(uC_n\big)^{\top}n^{-1}\big\{\wh{\ddot{\mL}^{(j)}}(\wt{p}_j)-\ddot{\mL}^{(j)}(\wt{p}_j)\big\}\big(uC_n\big)$ and $\Delta_{c2}^{(j)}=2^{-1}\big(uC_n\big)^{\top}$  $n^{-1}\big\{\ddot{\mL}^{(j)}(p_j)-\wh{\ddot{\mL}^{(j)}}(p_j)\big\}\big(uC_n\big)$. Then, we can derive that $ \wh{W_c^{(j)}}=W_c^{(j)}+\Delta_{c1}^{(j)}+\Delta_{c2}^{(j)}$. Thus, we have 
	$ P\Big\{C_n^{-2}\max\limits_{j}\sup\limits_{\lVert u\lVert=1}\big|\wh{W_c^{(j)}}\big|>\varepsilon\Big\}\leq \mD_{c1}^{(j)}+\mD_{c2}^{(j)}+\mD_{c3}^{(j)}$, where $ \mD_{c1}^{(j)}=P\Big\{C_n^{-2}\max\limits_{j}\sup\limits_{\lVert u\lVert=1}\big|\Delta_{c1}^{(j)}\big|>\varepsilon/3\Big\}$, $ \mD_{c2}^{(j)}=P\Big\{C_n^{-2}\max\limits_{j}\sup\limits_{\lVert u\lVert=1}\big|\Delta_{c2}^{(j)}\big|>\varepsilon/3\Big\}$ and $ \mD_{c3}^{(j)}=P\Big\{C_n^{-2}\max\limits_{j}\sup\limits_{\lVert u\lVert=1}\big|W_c^{(j)}\big|>\varepsilon/3\Big\}$. We have argued the upper bound of $\mD_{c2}^{(j)}$ in \sc{Step 2.2.}\normalfont$\ $Note that $\mD_{c1}^{(j)}$ is the same as $\mD_{c2}^{(j)}$ but for the parameter $\wt{p}_j$. Thus, we can obtain the following common upper bound for $\mD_{c1}^{(j)}$ and $\mD_{c2}^{(j)}$ as
	\begin{equation}\label{aa15}
		K^2P\bigg\{\Big\Vert\wh{\Omega}-\Omega\Big\Vert>2U_2\varepsilon/3\bigg\}+U_3/n.
	\end{equation}
	Next, we focus on $ \mD_{c3}^{(j)}$.
	
	To upper bound $ \mD_{c3}^{(j)}$, we define $\wt{\ddot{\mL}^{(j)}_{\Delta}}=n^{-1}\ddot{\mL}^{(j)}(\wt{p}_j)-n^{-1}\ddot{\mL}^{(j)}(p_j)=\big(\delta_{k_1k_2}^{(j,c)}\big)\in\mR^{K\times K}$, where $\delta_{k_1k_2}^{(j,c)}=n^{-1}\sum_{i=1}^n\delta_{k_1k_2}^{(j,c,i)}$, $\delta_{k_1k_2}^{(j,c,i)}=\wt{\ddot{\ell}_{k_1k_2}^{(j,i)}}-\ddot{\ell}_{k_1k_2}^{(j,i)}$, $ \wt{\ddot{\ell}_{k_1k_2}^{(j,i)}}=-\wt{\alpha_{ik_1}^{(j)}}\wt{\alpha_{ik_2}^{(j)}}s(Z_{ij},\wt{p}_{k_1j})$ 
	$s(Z_{ij},\wt{p}_{k_2j})$, $\wt{\alpha_{ik}^{(j)}}=c_{ik}\wt{p}_{kj}^{Z_{ij}}\big(1-\wt{p}_{kj}\big)^{1-Z_{ij}}\big/\sum_{k=1}^Kc_{ik}\wt{p}_{kj}^{Z_{ij}}\big(1-\wt{p}_{kj}\big)^{1-Z_{ij}}$, and $ s(Z_{ij},\wt{p}_{kj})=Z_{ij}/\wt{p}_{kj}-\big(1-Z_{ij}\big)/\big(1-\wt{p}_{kj}\big)$. Then, we have $\mD_{c3}^{(j)}\leq P\Big\{\max\limits_{j}\sup\limits_{\lVert u\lVert=1}\big\Vert\wt{\ddot{\mL}^{(j)}_{\Delta}}\big\Vert>2\varepsilon/3\Big\}$. Similar to \sc{Step 2.1,}\normalfont$\ $we have $\lambda_{\max}\big(\wt{\ddot{\mL}^{(j)}_{\Delta}}\big)\leq K\max_{k_1,k_2}\big|\delta_{k_1k_2}^{(j,c)}\big|$. Thus, $P\Big\{\max\limits_{j}\sup\limits_{\lVert u\lVert=1}\big\Vert\wt{\ddot{\mL}^{(j)}_{\Delta}}\big\Vert>2\varepsilon/3\Big\}\leq\sum_{k_1,k_2}P\Big\{\max\limits_{j}\sup\limits_{\lVert u\lVert=1}\left|\delta_{k_1k_2}^{(j,c)}\right|>2\varepsilon/\big(3K\big)\Big\}$. Similar to \sc{Step 1.2,}\normalfont$\ $we conduct the Taylor's expansion about $\delta_{k_1k_2}^{(j,c)}$ for $\wt{p}_j$ at $p_j$ as
	\begin{equation}\label{aa16}
		\frac{1}{n}\sum_{i=1}^n\wt{\ddot{\ell}_{k_1k_2}^{(j,i)}}-\frac{1}{n}\sum_{i=1}^n\ddot{\ell}_{k_1k_2}^{(j,i)}=\bigg\{\frac{1}{n}\sum_{i=1}^n\dddot{\ell}_{k_1k_2}^{(j,i)}\big(\Tilde{\Tilde{p}}_j\big)\bigg\}^{\top}\Big(\wt{p}_j-p_j\Big),
	\end{equation}
	where $\dddot{\ell}_{k_1k_2}^{(j,i)}\big(p_j\big)=\partial\ddot{\ell}_{k_1k_2}^{(j,i)}/\partial p_j\in\mR^K$ and $\Tilde{\Tilde{p}}_j=\beta p_j+\big(1-\beta\big)\wt{p}_j$ for some $\beta\in(0,1)$. One can verify that $\big\Vert\dddot{\ell}_{k_1k_2}^{(j,i)}\big(\Tilde{\Tilde{p}}_j\big)\Vert\leq p_m^3\sqrt{K}$. The verification details are given in \sc{Part 8 }\normalfont of Appendix C.1.
	Then, we have $\Big\Vert\Big\{\sum_{i=1}^n\dddot{\ell}_{k_1k_2}^{(j,i)}\big(\Tilde{\Tilde{p}}_j\big)/n\Big\}^{\top}\big(\wt{p}_j-p_j\big)\Big\Vert\leq p_m^3\sqrt{K}\big\Vert\wt{p}_j-p_j\big\Vert$. Recall that $\wt{p}_j=\alpha_jp_j+\big(1-\alpha_j\big)\big(p_j+uC_n/\sqrt{n}\big)$ for some $\alpha_j\in(0,1)$. Immediately, we have $\sup\limits_{\lVert u\lVert=1}\big\Vert\wt{p}_j-p_j\big\Vert\leq C_n\sqrt{K/n}$. Thus, we can obtain that $ P\Big\{\max\limits_{j}\sup\limits_{\lVert u\lVert=1}\big|\delta_{k_1k_2}^{(j,c)} \big|>2\varepsilon/\big(3K\big)\Big\}\leq P\Big\{Kp_m^3C_n/\sqrt{n}>2\varepsilon/\big(3K\big)\Big\}$. We have $P\Big\{Kp_m^3C_n/\sqrt{n}>2\varepsilon/\big(3K\big)\Big\}=0$ as long as $ n>\Big\{3K^2p_m^3C_n/\big(2\varepsilon\big)\Big\}^2$. Thus, we can obtain that $\mD_{c3}^{(j)}=0$. Combining the results of (\ref{aa15}), the conclusion in (\ref{aa4}) has been proved. As a result, we completes the theorem proof.
	
	\subsection{Proof of Theorem 2}
	
	Write $\varepsilon_{\nu}^p=\exp\big(-\nu p\big)$. We then have $P\Big\{\max_{i,k}\big|\wh{\pi}_{ik}-I(\mK_i=k)\big|>\varepsilon_{\nu}^p\Big\}\leq\sum_{k=1}^KP\Big\{\max_i\big|\wh{\pi}_{ik}-I(\mK_i=k)\big|>\varepsilon_{\nu}^p\Big\}=\sum_{k=1}^K\Big\{\pi_{k}\wh{Q_{\nu}^a}+\big(1-\pi_{k}\big)\wh{Q_{(\nu,k)}^b}\Big\}$, where $ \wh{Q_{(\nu,k)}^a}=P\Big\{\max_i\big|\wh{\pi}_{ik}-1\big|>\varepsilon_{\nu}^p\Big|\mK_i=k\Big\}$  and $\wh{Q_{(\nu,k)}^b}= P\Big\{\max_i\big|\wh{\pi}_{ik}\big|>\varepsilon_{\nu}^p\Big|\mK_i\neq k\Big\}$. Note that $K$ is a fixed integer. Therefore, it suffices to upper bound $ \wh{Q_{(\nu,k)}^a} $ and $ \wh{Q_{(\nu,k)}^b} $ separately. 
	
	\sc{Step 1.}\normalfont$\ $We first focus on $\wh{Q_{(\nu,k)}^a}$. Define $\wh{\mathcal{R}_i^{(m,k)}}=\prod_{j=1}^p\wh{r_{ij}^{(m,k)}}$, where $\wh{r_{ij}^{(m,k)}}=\big(\wh{p}_{mj}/\wh{p}_{kj}\big)^{Z_{ij}}\big\{(1-\wh{p}_{mj})/(1-\wh{p}_{kj})\big\}^{1-Z_{ij}}$. To study $\wh{Q_{(\nu,k)}^a}$, it is important to understand the asymptotic behavior of $ |\wh{\pi}_{ik}-1|$, which is given by $ |\wh{\pi}_{ik}-1|=\Big(\sum_{m\neq k}\wh{c_i^{(m,k)}}\wh{\mathcal{R}_i^{(m,k)}}\Big)\big/$  $\Big(1+\sum_{m\neq k}\wh{c_i^{(m,k)}}\wh{\mathcal{R}_i^{(m,k)}}\Big)\leq\sum_{m\neq k}\wh{c_i^{(m,k)}}\wh{\mathcal{R}_i^{(m,k)}}$. Here $\wh{c_i^{(m,k)}}= \Big[\wh{\pi}_m\exp\Big\{-\big(Y_i-\wh{\gamma}_m-X_i^{\top}\wh{\theta}\big)^2/(2\wh{\sigma}^2)\Big\}\Big]\big/\Big[\wh{\pi}_k\exp\Big\{-\big(Y_i-\wh{\gamma}_k-X_i^{\top}\wh{\theta}\big)^2/(2\wh{\sigma}^2)\Big\}\Big]$. Thus, we can obtain
	\begin{equation}\label{bb1}
		\wh{Q_{(\nu,k)}^a}=P\bigg\{\max_i|\wh{\pi}_{ik}-1|>\varepsilon_{\nu}^p\Big|\mK_i=k\bigg\}\leq \sum_{m\neq k}P\left\{\max_i\wh{c_i^{(m,k)}}\wh{\mathcal{R}_i^{(m,k)}}>\varepsilon_{\nu}^p/K\right\}.
	\end{equation}
	To upper bound $\wh{Q_{(\nu,k)}^a}$, it suffices to upper bound $P^{(m,k)}=P\Big\{\max_i\wh{c_i^{(m,k)}}\wh{\mathcal{R}_i^{(m,k)}}>\varepsilon_{\nu}^p/K\Big\}$. Recall that $\varepsilon_{\nu}^p=\exp\big(-\nu p\big)$. Direct computation leads to $P^{(m,k)}=P\Big\{\max_ip^{-1}$  $\log\Big(\wh{c_i^{(m,k)}}\wh{\mathcal{R}_i^{(m,k)}}\Big)>-\nu-\big(\log K\big)/p\Big\}$. It can be verified that $P^{(m,k)}\leq P\Big\{\max_ip^{-1}\log$  $\Big(\wh{c_i^{(m,k)}}\wh{\mathcal{R}_i^{(m,k)}}\Big)>-2\nu\Big\}$ as long as $ p>\big(\log K\big)/\nu $. Similarly, we can define $\mathcal{R}_i^{(m,k)}=\prod_{j=1}^pr_{ij}^{(m,k)}$, where $r_{ij}^{(m,k)}=\big(p_{mj}/p_{kj}\big)^{Z_{ij}}\big\{(1-p_{mj})/(1-p_{kj})\big\}^{1-Z_{ij}}$. Let $\mathcal{R}_{i\Delta}^{(m,k)}=p^{-1}\log\wh{\mathcal{R}_i^{(m,k)}}-p^{-1}\log\mathcal{R}_i^{(m,k)}$. Note that $\max_ip^{-1}\log\Big(\wh{c_i^{(m,k)}}\wh{\mathcal{R}_i^{(m,k)}}\Big)\leq \max_ip^{-1}\log\wh{c_i^{(m,k)}}$  $+\max_i\mathcal{R}_{i\Delta}^{(m,k)}+\max_ip^{-1}\log\mathcal{R}_i^{(m,k)}$. Thus, we have $\wh{Q_{(\nu,k)}^a}=o(1)$ if we can show that
	\begin{gather}
		\label{bb2} \max\limits_{i}\Big|p^{-1}\log\wh{c_i^{(m,k)}}\Big|=o_p(1), \\ 
		\label{bb3} \max\limits_{i}\Big|\mathcal{R}_{i\Delta}^{(m,k)}\Big|=o_p(1),\\
		\label{bb4}P\bigg\{\max_ip^{-1}\log\mathcal{R}_i^{(m,k)}+2\nu>\varepsilon\bigg\}
		\leq\exp\Big(-C_1p/C_2\Big),
	\end{gather}
	for any given $\varepsilon>0$. Here $C_1=(\varepsilon-2\nu+\Delta_{\min})^2/2>0$ and $ C_2=C_{\sigma}+b^2>0$ are some fixed and positive constants. Those three conclusions are to be proved separately in the following three substeps.
	
	\sc{Step 1.1.}\normalfont$\ $We start with (\ref{bb2}). Recall that $\wh{c_i^{(m,k)}}= \Big[\wh{\pi}_m\exp\Big\{-\big(Y_i-\wh{\gamma}_m-X_i^{\top}\wh{\theta}\big)^2/(2$ $\wh{\sigma}^2)\Big\}\Big]\big/\Big[\wh{\pi}_k\exp\Big\{-\big(Y_i-\wh{\gamma}_k-X_i^{\top}\wh{\theta}\big)^2/(2\wh{\sigma}^2)\Big\}\Big]\leq\Big(\wh{\pi}_m\big/\wh{\pi}_k\Big)\exp\Big\{\big(Y_i-\wh{\gamma}_k-X_i^{\top}\wh{\theta}\big)^2/(2\wh{\sigma}^2)\Big\}$. With $\mK_i=k$, we have $Y_i=\gamma_k+X_i^{\top}\theta+\varepsilon_i$, where $\varepsilon_i$ follows a normal distribution with mean 0 and variance $ \sigma^2$. By Cauchy-Schwarz inequality \citep{1952hardy}, we can obtain that $ \big(Y_i-\wh{\gamma}_k-X_i^{\top}\wh{\theta}\big)^2=\Big\{\gamma_k-\wh{\gamma}_k+X_i^{\top}\big(\theta-\wh{\theta}\big)+\varepsilon_i\Big\}^2\leq 3\big(\gamma_k-\wh{\gamma}_k\big)^2+3\Big\{X_i^{\top}\big(\theta-\wh{\theta}\big)\Big\}^2+3\varepsilon_i^2$. Direct computation leads to $p^{-1}\big|\log\wh{c_i^{(m,k)}}\big|\leq p^{-1}\big|\log\Big(\wh{\pi}_m\big/\wh{\pi}_k\Big)\big|+p^{-1}\big|\big(Y_i-\wh{\gamma}_k-X_i^{\top}\wh{\theta}\big)^2/(2\wh{\sigma}^2)\big|\leq\mC_{1}^{(m,k)}+\mC_{2}^{(m,k)}+\mC_{3}^{(m,k)}+\mC_{4}^{(m,k)}$, where $\mC_{1}^{(m,k)}=p^{-1}\max_i\big|\log\big(\wh{\pi}_m\big/\wh{\pi}_k\big)\big|$, $\mC_{2}^{(m,k)}=3p^{-1}\max_i\big(\wh{\gamma}_k-\gamma_k\big)^2/(2\wh{\sigma}^2)$, $\mC_{3}^{(m,k)}=3p^{-1}\max_i\Big\{X_i^T\big(\wh{\theta}-\theta\big)\Big\}^2\big/\big(2\wh{\sigma}^2\big)$, and $\mC_{4}^{(m,k)}=3p^{-1}\max_i\varepsilon_i^2/(2\wh{\sigma}^2)$. Note that these constants $ C_t^{(m,k)}$ with $1\leq t\leq4 $ are independent of the subscript $i$. Next, it can be verified that $\mC_{1}^{(m,k)}=O_p(1/p)$, $\mC_{2}^{(m,k)}=O_p\big(1/(pn)\big)$, $\mC_{3}^{(m,k)}=O_p\big(1/(pn)\big)$, and $\mC_{4}^{(m,k)}=O_p\big(\log n/p\big)$. The verification details are given in \sc{Steps 1.1.1\textemdash1.1.4}\normalfont$\ $ of Appendix C.2. Combining the above results, (\ref{bb2}) is proved.
	
	\sc{Step 1.2.}\normalfont$\ $Next, we prove (\ref{bb3}). Recall that $\mathcal{R}_{i\Delta}^{(m,k)}=p^{-1}\log\wh{\mathcal{R}_i^{(m,k)}}-p^{-1}\log\mathcal{R}_i^{(m,k)}$, $\wh{\mathcal{R}_i^{(m,k)}}=\prod_{j=1}^p\wh{r_{ij}^{(m,k)}}$, and $\wh{r_{ij}^{(m,k)}}=\big(\wh{p}_{mj}/\wh{p}_{kj}\big)^{Z_{ij}}\big\{(1-\wh{p}_{mj})/(1-\wh{p}_{kj})\big\}^{1-Z_{ij}}$. $\mathcal{R}_i^{(m,k)}$ and $ r_{ij}^{(m,k)} $ are in the same form. Direct computation leads to $\log\wh{\mathcal{R}_i^{(m,k)}}=\sum_{j=1}^p\wh{V_{ij}^{(m,k)}}$, where $\wh{V_{ij}^{(m,k)}}=\log\wh{r_{ij}^{(m,k)}}=Z_{ij}\log\Big(\wh{p}_{mj}\big/\wh{p}_{kj}\Big)+\Big(1-Z_{ij}\Big)\log\Big\{\big(1-\wh{p}_{mj}\big)\big/\big(1-\wh{p}_{kj}\big)\Big\}$. Similarly, $\log\mathcal{R}_i^{(m,k)}=\sum_{j=1}^pV_{ij}^{(m,k)}$, where $V_{ij}^{(m,k)}=\log r_{ij}^{(m,k)}=Z_{ij}\log\Big(p_{mj}\big/p_{kj}\Big)+\Big(1-Z_{ij}\Big)\log\Big\{\big(1-p_{mj}\big)\big/\big(1-p_{kj}\big)\Big\}$. Thus,  $\mathcal{R}_{i\Delta}^{(m,k)}=\sum_{j=1}^p\Big(\wh{V_{ij}^{(m,k)}}-V_{ij}^{(m,k)}\Big)\big/p$. One can verified that $ \Big|\wh{V_{ij}^{(m,k)}}-V_{ij}^{(m,k)}\Big|\leq 2p_m\big|\wh{p}_{mj}-p_{mj}\big|+2p_m\big|\wh{p}_{kj}-p_{kj}\big|$, where $ p_m=p_{\min}^{-1}+(1-p_{\max})^{-1}$. The verification details are given in \sc{Step 1.2.1}\normalfont$\ $of Appendix C.2. Thus, we can obtain that $ \Big|\mathcal{R}_{i\Delta}^{(m,k)}\Big|\leq\max_j\Big|\wh{V_{ij}^{(m,k)}}-V_{ij}^{(m,k)}\Big|\leq 2p_m\max_j\big|\wh{p}_{mj}-p_{mj}\big|+2p_m\max_j\big|\wh{p}_{kj}-p_{kj}\big|$. Note that this is an upper bound uniformly over $1\leq i\leq n$. In the meanwhile by Theorem 1, we have $\max_j\big|\wh{p}_{kj}-p_{kj}\big|=O_p(C_n/\sqrt{n})$. As a result, $ \max_i\Big|\mathcal{R}_{i\Delta}^{(m,k)}\Big|\leq O_p(C_n/\sqrt{n})$. This completes the proof of (\ref{bb3}). 
	
	\sc{Step 1.3.}\normalfont$\ $To prove (\ref{bb4}), we shall focus on $ \log\mathcal{R}_i^{(m,k)}$. Following the definition in \sc{Step 1.2,}\normalfont$\ $we have $\log\mathcal{R}_i^{(m,k)}=\sum_{j=1}^pV_{ij}^{(m,k)}$, where $V_{ij}^{(m,k)}=\log r_{ij}^{(m,k)}=Z_{ij}\log\Big(p_{mj}\big/p_{kj}\Big)+\Big(1-Z_{ij}\Big)\log\Big\{\big(1-p_{mj}\big)\big/\big(1-p_{kj}\big)\Big\}$. Note that $ V_{ij}^{(m,k)}$ is a uniformly bounded random variable with $|V_{ij}^{(m,k)}|\leq b$ with the upper bound $b=\max\Big\{\log\big(p_{\max}$  $/p_{\min}\big), \log\big\{(1-p_{\min})/(1-p_{\max})\big\}\Big\}$. Thus, the Bernstein's Inequality \citep{bernstein1926} can be readily applied as
	\begin{equation}\label{bb5}
		P\bigg[\frac{1}{p}\sum_{j=1}^p\bigg\{V_{ij}^{(m,k)}-E\Big(V_{ij}^{(m,k)}\Big)\bigg\}>\varepsilon_{(\nu,E)}^p\bigg]\leq\exp\Big(-\mathcal{D}_a\big/\mathcal{D}_b\Big),
	\end{equation}
	where $\mathcal{D}_a=p\big(\varepsilon_{(\nu,E)}^p\big)^2/2$, $\mathcal{D}_b=\sum_{j=1}^p\vt\big(V_{ij}^{(m,k)}\big)/p+b\varepsilon_{(\nu,E}^p/3 $, and $\varepsilon_{(\nu,E)}^p=\varepsilon-2\nu-\sum_{j=1}^pE\big(V_j^{(m,k)}\big)\big/p>\varepsilon-2\nu+\Delta_{\min}>0 $. To further bound the right-hand side of (\ref{bb5}), we need to lower bound $\mathcal{D}_a$ and upper bound $ \mathcal{D}_b $ separately. One can verified that $\mathcal{D}_a\geq C_1p$ and $\mathcal{D}_b\leq C_2$, where $C_1=(\varepsilon-2\nu+\Delta_{\min})^2/2>0$ and $ C_2=C_{\sigma}+b^2>0$ are some fixed constants. The verification details are given in \sc{Steps 1.3.1\textemdash 1.3.2}\normalfont$\ $of Appendix C.2. Thus, we can obtain that $ P\Big\{\max_ip^{-1}\log\mathcal{R}_i^{(m,k)}+2\nu>\varepsilon\Big\}\leq\exp\Big(-C_1p/C_2\Big)$. Consequently, we have (\ref{bb4}) rigorously proved. 
	
	Combining the results from (\ref{bb2}), (\ref{bb3}) and (\ref{bb4}), we can obtain that $P^{(m,k)}\leq P\Big\{\max_ip^{-1}\log\Big(\wh{c_i^{(m,k)}}$  $\wh{\mathcal{R}_i^{(m,k)}}\Big)>-2\nu\Big\}\leq P\Big\{o_p(1)+\max_ip^{-1}\log\mathcal{R}_i^{(m,k)}>-2\nu\Big\}
	= o(1)$. Since $K$ is a fixed integer, we then have $\wh{Q_{(\nu,k)}^a}=\sum_{m\neq k} P^{(m,k)}=o(1)$. This completes the proof of \sc{Step 1.}\normalfont$\ $
	
	\sc{Step 2.}\normalfont$\ $Next, we study $\wh{Q_{(\nu,k)}^b}= P\Big\{\max_i\big|\wh{\pi}_{ik}\big|>\varepsilon_{\nu}^p\Big|\mK_i\neq k\Big\}$. Note that $\sum_{k=1}^KI(\mK_i$  $=k)=1$ with $I(\mK_i=k)\in\{0,1\}$. Thus, there should exist only a $k'\in\{1,2,\ldots,K\}$, such that $I(\mK_i=k')=1$ and $ I(\mK_i=k)=0$ for every $k\neq k'$. Then we have $\wh{Q_{(\nu,k)}^b}=\sum_{k'\neq k}\wh{P_{\nu}^{(k',k)}}$, where   $\wh{P_{\nu}^{(k',k)}}=P\Big(\max_{i}\big|\wh{\pi}_{ik}\big|>\varepsilon_{\nu}^p\big|\mK_i=k'\Big)P\big(\mK_i=k'\big)\big/P\big(\mK_i\neq k\big)$.
	Recall that $\sum_{s=1}^K\wh{\pi}_{is}=1$ with $ 0<\wh{\pi}_{is}<1$. Direct computation leads to $ |\wh{\pi}_{ik'}-1|=\sum_{s\neq k'}\wh{\pi}_{is}\geq\wh{\pi}_{ik} $. Then, we have $\wh{P_{\nu}^{(k',k)}}\leq P\Big(\max_{i}\big|\wh{\pi}_{ik'}-1\big|>\varepsilon_{\nu}^p\big|\mK_i=k'\Big)$. By the results of $\wh{Q_{(\nu,k)}^a}$, we can obtain $\wh{P_{\nu}^{(k',k)}}=o(1)$ as long as $ p>\big(\log K\big)/\nu$. Note that $K$ is a fixed integer. Thus, we have $ \wh{Q_{(\nu,k)}^b}=\sum_{k'\neq k}P_{\nu}^{(k',k)}=o(1)$ for any $ p>\big(\log K\big)/\nu$. This completes the discussion of $\wh{Q_{(\nu,k)}^b}$. 
	
	Combining the results of $\wh{Q_{(\nu,k)}^a}$ and $\wh{Q_{(\nu,k)}^b}$ from the \sc{Step 1 \& 2}\normalfont, we can obtain that $P\Big\{\max_{i,k}\big|\wh{\pi}_{ik}-I(\mK_i=k)\big|>\varepsilon_{\nu}^p\Big\}=\sum_{k=1}^K\Big\{\pi_{k}\wh{Q_{(\nu,k)}^a}+\big(1-\pi_{k}\big)\wh{Q_{(\nu,k)}^b}\Big\}=o(1)$. This proves the theorem conclusion and completes the whole theorem proof.

	\subsection{Proof of Theorem 3}
	
	Recall that $\Omega=(\pi^{\top},\gamma^{\top},\theta^{\top},\sigma^2)^{\top}\in\mR^{2K+q+1}$ and $\Phi=(\gamma^{\top},\theta^{\top})^{\top}\in\mR^{K+q}$, we then have $ \big\Vert\wh{\Omega}_{\sr}-\wh{\Omega}_{\so}\big\Vert\leq \big\Vert\wh{\pi}_{\sr}-\wh{\pi}_{\so}\big\Vert+\big\Vert\wh{\Phi}_{\sr}-\wh{\Phi}_{\so}\big\Vert+ \big|\wh{\sigma}^2_{\sr}-\wh{\sigma}^2_{\so}\big|$. Therefore, the theorem conclusion follows, if we can show that $\big\Vert\wh{\pi}_{\sr}-\wh{\pi}_{\so}\big\Vert=o_p(1/\sqrt{n})$, $\big\Vert\wh{\Phi}_{\sr}-\wh{\Phi}_{\so}\big\Vert=o_p(1/\sqrt{n})$ and $\big|\wh{\sigma}^2_{\sr}-\wh{\sigma}^2_{\so}\big|=o_p(1/\sqrt{n})$. Those three conclusions are to be proved separately in the following three steps.
	
	\sc{Step 1.}\normalfont$\ $We start with $\big\Vert\wh{\pi}_{\sr}-\wh{\pi}_{\so}\big\Vert=o_p(1/\sqrt{n})$. Recall that $\wh{\pi}=\big(\wh{\pi}_{1},\ldots,\wh{\pi}_{K}\big)$, $\wh{\pi}_{\sr,k}= \Big(\sum_{i=1}^n\wh{\pi}_{ik}\Big)\big/n $ and $\wh{\pi}_{\so,k} = \Big(\sum_{i=1}^n a_{ik}\Big)\big/n$. Then, we have $\big\Vert\wh{\pi}_{\sr}-\wh{\pi}_{\so}\big\Vert\leq \sum_{i=1}^n\sum_{k=1}^K\big|\wh{\pi}_{ik}-a_{ik}\big|/n \leq K\max_{i,k}\big|\wh{\pi}_{ik}-a_{ik}\big|$. Therefore, $\big\Vert\wh{\pi}_{\sr}-\wh{\pi}_{\so}\big\Vert=o_p(1/\sqrt{n})$ is proved if we can show that $ \max_{i,k}\big|\wh{\pi}_{ik}-a_{ik}\big|=o_p(1/\sqrt{n})$. We then consider for an arbitrary but fixed constant $ \varepsilon>0$ and the evaluation $P_{\varepsilon}^*=P\Big\{\sqrt{n}\max_{i,k}$  $\Big|\wh{\pi}_{ik}-a_{ik}\Big|>\varepsilon\Big\}$. By Theorem 2, we know that $P\Big\{\max_{i,k}\big|\wh{\pi}_{ik}-a_{ik}\big|>\exp\big(-\nu p\big)\Big\}=o(1)$. Thus, we shall compute $ P_{\varepsilon}^* $ as
	\begin{equation}\label{t3.1}
		P_{\varepsilon}^*=P\Big\{\max_{i,k}\Big|\wh{\pi}_{ik}-a_{ik}\Big|>\varepsilon/\sqrt{n}\Big\}\leq P\Big\{\max_{i,k}\big|\wh{\pi}_{ik}-a_{ik}\big|>\exp\big(-\nu p\big)\Big\}=o(1), \tag{B.1}
	\end{equation}
	as long as $n$ is sufficiently large. The inequality in (\ref{t3.1}) is mainly because $\sqrt{n}\exp\big(-\nu p\big)=\exp\Big\{2^{-1}\log(n)-\nu p\Big\}\to 0$ as $ n\to\infty$ by the technical condition \ref{con4}. This completes the proof of $\big\Vert\wh{\pi}_{\sr}-\wh{\pi}_{\so}\big\Vert=o_p(1/\sqrt{n})$.
	
	\sc{Step 2.}\normalfont$\ $Next, we prove $\big\Vert\wh{\Phi}_{\sr}-\wh{\Phi}_{\so}\big\Vert=o_p(1/\sqrt{n})$. Recall that $\wh{\Phi}_{\sr}=\Big(\wh{\Sigma}^{\fxx}_{\sr}\Big)^{-1}$  $\wh{\Sigma}^{\fxy}_{\sr}$ and $ \wh{\Phi}_{\so}=\Big(\wh{\Sigma}^{\fxx}_{\so}\Big)^{-1}\wh{\Sigma}^{\fxy}_{\so}$, where $ \wh{\Sigma}^{\fxx}_{\sr}=n^{-1}\sum_{i=1}^nX_i^{\pi}X_i^{\pi\top}$, $ \wh{\Sigma}^{\fxy}_{\sr}=n^{-1}\sum_{i=1}^n$  $X_i^{\pi}Y_i $, $ \wh{\Sigma}^{\fxx}_{\so}=n^{-1}\sum_{i=1}^nX_i^aX_i^{a\top}$ and $ \wh{\Sigma}^{\fxy}_{\so}=n^{-1}\sum_{i=1}^nX_i^aY_i $.
	We then have $\big\Vert\wh{\Phi}_{\sr}-\wh{\Phi}_{\so}\big\Vert\leq\Big\Vert\Big(\wh{\Sigma}^{\fxx}_{\sr}\Big)^{-1}-\Big(\wh{\Sigma}^{\fxx}_{\so}\Big)^{-1}\Big\Vert\cdot\Big\Vert\wh{\Sigma}^{\fxy}_{\sr}\Big\Vert+\Big\Vert\Big(\wh{\Sigma}^{\fxx}_{\so}\Big)^{-1}\Big\Vert\cdot\Big\Vert\wh{\Sigma}^{\fxy}_{\sr}-\wh{\Sigma}^{\fxy}_{\so}\Big\Vert$. Note that $ \wh{\Sigma}^{\fxx}_{\so}=n^{-1}\sum_{i=1}^nX_i^aX_i^{a\top}\stackrel{p}{\longrightarrow} E\Big(X_i^aX_i^{a\top}\Big)$, which is a positive definite matrix. Therefore, the theorem conclusion follows, if we can show that $ \Big\Vert\wh{\Sigma}^{\fxx}_{\sr}-\wh{\Sigma}^{\fxx}_{\so}\Big\Vert=o_p(1/\sqrt{n})$ and $ \Big\Vert\wh{\Sigma}^{\fxy}_{\sr}-\wh{\Sigma}^{\fxy}_{\so}\Big\Vert=o_p(1/\sqrt{n})$. Since the proof is nearly identical, we shall present the proof details for $\Big\Vert\wh{\Sigma}^{\fxx}_{\sr}-\wh{\Sigma}^{\fxx}_{\so}\Big\Vert$ only. 
	
	By definition, we have $\Big\Vert\wh{\Sigma}^{\fxx}_{\sr}-\wh{\Sigma}^{\fxx}_{\so}\Big\Vert=\Big\Vert n^{-1}\sum_{i=1}^n\big(X_i^{\pi}X_i^{\pi\top}-X_i^aX_i^{a\top}\big)\Big\Vert\leq n^{-1}\sum_{i=1}^n$  $\big\Vert\Delta_{\fxx}^i\big\Vert $, where $\Delta_{\fxx}^i=X_i^{\pi}X_i^{\pi\top}-X_i^aX_i^{a\top} $. Write $\Delta_{\fxx}^i=(\delta_{j_1j_2}^i)\in\mR^{(K+q)\times(K+q)}$.
	Note that $\Delta_{\fxx}^i$ is a symmetric matrix. Thus, by the proof of Lemma 1 in \cite{2009Wang}, we have $\big\Vert \Delta_{\fxx}^i\big\Vert=\lambda_{\max}\big(\Delta_{\fxx}^i\big)\leq(K+q)\max_{j_1,j_2}\big|\delta_{j_1j_2}^i\big|$. It can be verified that $\max_{j_1,j_2}\big|\delta_{j_1j_2}^i\big|\leq\max_k\big|\wh{\pi}_{ik}-a_{ik}\big|\big(2+\big\Vert X_i\big\Vert\big)$. The verification details are given in Appendix C.3. Thus, we can obtain that $n^{-1}\sum_{i=1}^n\big\Vert\Delta_{\fxx}^i\big\Vert\leq n^{-1}\sum_{i=1}^n(K+q)\max_k\big|\wh{\pi}_{ik}-a_{ik}\big|\big(2+\big\Vert X_i\big\Vert\big)\leq(K+q) \Big\{\max_{i,k}\big|\wh{\pi}_{ik}-a_{ik}\big|\Big\}\sum_{i=1}^n\big(2+\big\Vert X_i\big\Vert\big)/n$. By the previous step, we have $ \max_{i,k}\big|\wh{\pi}_{ik}-a_{ik}\big|=o_p(1/\sqrt{n})$. Therefore, $\Big\Vert\wh{\Sigma}^{\fxx}_{\sr}-\wh{\Sigma}^{\fxx}_{\so}\Big\Vert=o_p(1/\sqrt{n})$ is proved. Similarly, we can prove that $ \Big\Vert\wh{\Sigma}^{\fxy}_{\sr}-\wh{\Sigma}^{\fxy}_{\so}\Big\Vert=o_p(1/\sqrt{n})$. Thus, this completes the proof of \sc{Step 2.}\normalfont$\ $
	
	\sc{Step 3.}\normalfont$\ $Finally, we prove $\big|\wh{\sigma}^2_{\sr}-\wh{\sigma}^2_{\so}\big|=o_p(1/\sqrt{n})$. 
	Recall that $\wh{\sigma}^2_{\sr} =\sum_{i=1}^n\Big(Y_i-X_i^{\pi\top}\wh{\Phi}_{\sr}\Big)^2\big/ n$ and $\wh{\sigma}^2_{\so}=\sum_{i=1}^n\Big(Y_i-X_i^{a\top}\wh{\Phi}_{\so}\Big)^2\big/n$. 
	Then, we have $\big|\wh{\sigma}_{\sr}^2-\wh{\sigma}_{\so}^2\big|\leq \Big|\sum_{i=1}^n\Big(Y_i-X_i^{\pi\top}\wh{\Phi}_{\sr}\Big)^2\big/ n-\sum_{i=1}^n\Big(Y_i-X_i^{a\top}\wh{\Phi}_{\so}\Big)^2\big/ n\Big|\leq \mQ_1+\mQ_2$, where $\mQ_1 =\Big|\sum_{i=1}^n\Big\{\Big(Y_i-X_i^{\pi\top}\wh{\Phi}_{\sr}\Big)^2-\Big(Y_i-X_i^{a\top}\wh{\Phi}_{\sr}\Big)^2\Big\}\big/n\Big|$ and $\mQ_2= \Big|\sum_{i=1}^n\Big\{\Big(Y_i-X_i^{a\top}\wh{\Phi}_{\sr}\Big)^2-\Big(Y_i-X_i^{a\top}\wh{\Phi}_{\so}\Big)^2\Big\}\big/n\Big|$. Therefore, it suffices to study $\mQ_1$ and $\mQ_2$ separately.
	
	\sc{Step 3.1.}\normalfont$\ $We start with $\mQ_1$. Direct computation leads to $\mQ_1 =\Big|\sum_{i=1}^n\Big\{\Big(X_i^{\pi\top}\wh{\Phi}_{\sr}\Big)^2$  $-\Big(X_i^{a\top}\wh{\Phi}_{\sr}\Big)^2-2Y_i\Big(X_i^{\pi}-X_i^{a}\Big)^{\top}\wh{\Phi}_{\sr}\Big\}\big/n\Big|\leq \mQ_1^a+\mQ_1^b$, where  $\mQ_1^a=\sum_{i=1}^n\Big|\Big(X_i^{\pi\top}\wh{\Phi}_{\sr}\Big)^2-\Big(X_i^{a\top}\wh{\Phi}_{\sr}\Big)^2\Big|\big/n$ and $\mQ_1^b = 2K\big\Vert\wh{\Phi}_{\sr}\big\Vert\Big\{\max_{i,k}\big|\wh{\pi}_{i,k}-a_{ik}\big|\Big\}\sum_{i=1}^n\big|Y_i\big|/n$. Further, we have $\mQ_1^a =\sum_{i=1}^n\Big|\Big(X_i^{\pi}-X_i^{a}\Big)^{\top}\wh{\Phi}_{\sr}\Big|\cdot\Big|\Big(X_i^{\pi}+X_i^{a}\Big)^{\top}\wh{\Phi}_{\sr}\Big|\big/n\leq 2K\big\Vert\wh{\Phi}_{\sr}\big\Vert^2 \Big\{\max_{i,k}\big|\wh{\pi}_{ik}-a_{ik}\big|\Big\}\sum_{i=1}^n\big(\big\Vert X_i\big\Vert+K\big)/n$. Note that both $\mQ_1^a$ and $\mQ_1^b$ are determined by $\max_{i,k}\big|\wh{\pi}_{ik}-a_{ik}\big|$. By \sc{Step 1,}\normalfont$\ $we have $\max_{i,k}\big|\wh{\pi}_{ik}-a_{ik}\big|=o_p(1/\sqrt{n})$. Thus,  $\mQ_1^a=o_p(1/\sqrt{n})$ and $\mQ_1^b=o_p(1/\sqrt{n})$. Therefore, we have $\mQ_1=o_p(1/\sqrt{n})$.
	
	\sc{Step 3.2.}\normalfont$\ $Next, we study $\mQ_2$. Direct computation leads to $\mQ_2 =\sum_{i=1}^n\Big|\Big\{\Big(X_i^{a\top}$  $\wh{\Phi}_{\sr}\Big)^2$  $-\Big(X_i^{a\top}\wh{\Phi}_{\so}\Big)^2-2Y_iX_i^{a\top}\Big(\wh{\Phi}_{\sr}-\wh{\Phi}_{\so}\Big)\Big\}\Big|\big/n\leq\mQ_2^a+\mQ_2^b$, where  $\mQ_2^a =\sum_{i=1}^n$  $\Big|\Big(X_i^{a\top}\wh{\Phi}_{\sr}\Big)^2-\Big(X_i^{a\top}\wh{\Phi}_{\so}\Big)^2\Big|\big/n$ and 
	$\mQ_2^b = 2\big\Vert\wh{\Phi}_{\sr}-\wh{\Phi}_{\so}\big\Vert \sum_{i=1}^n\big(\big\Vert X_i\big\Vert+1\big)\big|Y_i\big|\big/n$. Similar to the previous substep, we have $\mQ_2^a =\sum_{i=1}^n\Big|X_i^{a\top}\Big(\wh{\Phi}_{\sr}-\wh{\Phi}_{\so}\Big)\Big|\cdot\Big|X_i^{a\top}\Big(\wh{\Phi}_{\sr}+\wh{\Phi}_{\so}\Big)\Big|\big/n \leq \big\Vert\wh{\Phi}_{\sr}-\wh{\Phi}_{\so}\big\Vert\cdot\big\Vert\wh{\Phi}_{\sr}+\wh{\Phi}_{\so}\big\Vert\cdot\sum_{i=1}^n\big(\big\Vert X_i\big\Vert+1\big)^2\big /n$. Note that both $\mQ_2^a$ and $\mQ_2^b$ are determined by $\big\Vert\wh{\Phi}_{\sr}-\wh{\Phi}_{\so}\big\Vert$. By \sc{Step 2,}\normalfont$\ $we have $\big\Vert\wh{\Phi}_{\sr}-\wh{\Phi}_{\so}\big\Vert=o_p(1/\sqrt{n})$. Thus,  $\mQ_2^a=o_p(1/\sqrt{n})$ and $\mQ_2^b=o_p(1/\sqrt{n})$. Therefore, we have $\mQ_2=o_p(1/\sqrt{n})$. Combining the results of $\mQ_1=o_p(1/\sqrt{n})$ and $\mQ_2=o_p(1/\sqrt{n})$, $\big|\wh{\sigma}^2_{\sr}-\wh{\sigma}^2_{\so}\big|=o_p(1/\sqrt{n})$ is proved. Thus, we have $ \big\Vert\wh{\Omega}_{\sr}-\wh{\Omega}_{\so}\big\Vert=o_p(1/\sqrt{n}) $. This completes the proof of Theorem 3. 
	
	\section{Verification details}
	
	In this appendix, we show in detail the calculation process of some results given in Appendix B.
	\subsection{Proof of results in Appendix B.1}
	
	\sc{Part 1.}\normalfont$\ $Recall that 
	\begin{equation*}
		\wh{\mL^{(j)}}(p_j)=\sum_{i=1}^{n}\log\bigg\{\sum_{k=1}^K\wh{c_{ik}}p_{kj}^{Z_{ij}}\Big(1-p_{kj}\Big)^{1-Z_{ij}}\bigg\},
	\end{equation*}	
	where $\wh{c_{ik}}=\wh{\pi}_k\big(\sqrt{2\pi}\wh{\sigma}\big)^{-1}\exp\Big\{-\Big(Y_i-\wh{\gamma}_k-X_i^{\top}\wh{\theta}\Big)^2\big/\big(2\wh{\sigma}^2\big)\Big\}$. Note that $\wh{c_{ik}}$ is independent of $ p_j=(p_{1j},\ldots,p_{2j})\in\mR^K$. To verify that $-\wh{\mL^{(j)}}(p_j)$ is a strictly convex function for $ p_j $, it suffices to show that $-\wh{\ddot{\mL^{(j)}}}(p_j)=-\partial^2\wh{\mL^{(j)}}(p_j)/\partial p_j\partial p_j^{\top}\in\mR^{K\times K}$ is semi-positive definite. Recall that $\wh{\ddot{\mL}^{(j)}}(p_j)=\big(\wh{\ddot{\ell}_{k_1k_2}^{(j)}}\big)\in\mR^{K\times K}$, $\wh{\ddot{\ell}_{k_1k_2}^{(j)}}=-\sum_{i=1}^n\wh{\alpha_{ik_1}^{(j)}}s(Z_{ij},p_{k_1j})\wh{\alpha_{ik_2}^{(j)}}$ 
	$s(Z_{ij},p_{k_2j})$,  $\wh{\alpha_{ik}^{(j)}}=\wh{c_{ik}}p_{kj}^{Z_{ij}}\big(1-p_{kj}\big)^{1-Z_{ij}}\big/\sum_{k=1}^K\wh{c_{ik}}p_{kj}^{Z_{ij}}\big(1-p_{kj}\big)^{1-Z_{ij}}$, $s(Z_{ij},p_{kj})=Z_{ij}/p_{kj}-\big(1-Z_{ij}\big)/\big(1-p_{kj}\big)$, and $\wh{c_{ik}}=\wh{\pi}_k\big(\sqrt{2\pi}\wh{\sigma}\big)^{-1}$  $\exp\Big\{-\Big(Y_i-\wh{\gamma}_k-X_i^{\top}\wh{\theta}\Big)^2\big/\big(2\wh{\sigma}^2\big)\Big\}$. Define $ \wh{\alpha_i^{(j)}}=\Big(\wh{\alpha_{i1}^{(j)}}s(Z_{ij},p_{1j}),\ldots, \wh{\alpha_{iK}^{(j)}}s(Z_{ij},p_{Kj})\Big)\in\mR^K$. Then, $ -\wh{\ddot{\mL}^{(j)}}(p_j) $ can be rewritten as $-\wh{\ddot{\mL}^{(j)}}(p_j)=\sum_{i=1}\wh{\alpha_i^{(j)}}\big(\wh{\alpha_i^{(j)}}\big)^{\top}$. For an arbitrary but non-zero vector $X$, we have $ X^{\top}\big\{-\wh{\ddot{\mL}^{(j)}}(p_j)\big\}X=\sum_{i=1}^n\big\{X^{\top}\wh{\alpha_i^{(j)}}\big\}\big\{X^{\top}\wh{\alpha_i^{(j)}}\big\}^{\top}\geq 0$. Note that $\wh{\alpha_i^{(j)}}\big(\wh{\alpha_i^{(j)}}\big)^{\top}$ is a matrix of rank one. Then, we can obtain that $ -\wh{\ddot{\mL}^{(j)}}(p_j) $ is of full rank. By definition,  $-\wh{\mL^{(j)}}(p_j)$ is a strictly convex function for $ p_j $
	
	\sc{Part 2.}\normalfont$\ $Recall that $\dot{\ell_k}^{(j,i)}=\alpha_{ik}^{(j)}s(Z_{ij},p_{kj})$, $\alpha_{ik}^{(j)}=c_{ik}p_{kj}^{Z_{ij}}\big(1-p_{kj}\big)^{1-Z_{ij}}\big/\sum_{k=1}^Kc_{ik}p_{kj}^{Z_{ij}}$  $\big(1-p_{kj}\big)^{1-Z_{ij}}$, $s(Z_{ij},p_{kj})=Z_{ij}/p_{kj}-\big(1-Z_{ij}\big)/\big(1-p_{kj}\big)$, and $c_{ik}=\pi_k\big(\sqrt{2\pi}\sigma\big)^{-1}$  $\exp\Big\{-\Big(Y_i-\gamma_k-X_i^{\top}\theta\Big)^2\big/\big(2\sigma^2\big)\Big\}$. Note that $ \big|\alpha_{ik}^{(j)}\big|\leq 1$ and $ Z_{ij}\in\{0,1\}$. By \ref{con2}, we then have $ \big|\dot{\ell_k}^{(j,i)}\big|\leq p_m$, where $ p_m=p_{\min}^{-1}+(1-p_{\max})^{-1}$. Next, we calculate $ E\big(\dot{\ell_k}^{(j,i)}\big)=E\Big\{E\big(\dot{\ell_k}^{(j,i)}\big|Y_i,X_i\big)\Big\}$. Direct computation leads to $ P\big(Z_{ij}=1\big|Y_i,X_i\big)=P\big(Z_{ij}=1,Y_i,X_i\big)/P\big(Y_i,X_i\big)=\sum_{k=1}^Kc_{ik}p_{kj}/\sum_{k=1}^Kc_{ik}$. Then, we have $ E\big(\dot{\ell_k}^{(j,i)}\big|Y_i,X_i\big)=\big\{c_{ik}/\sum_{k=1}^Kc_{ik}p_{kj}\big\}\big\{\sum_{k=1}^Kc_{ik}p_{kj}/\sum_{k=1}^Kc_{ik}\big\}-\big\{c_{ik}/\sum_{k=1}^Kc_{ik}(1-p_{kj})\big\}$  $\big\{1-\sum_{k=1}^Kc_{ik}$ $p_{kj}/\sum_{k=1}^Kc_{ik}\big\}=0$. Thus, $ E\big(\dot{\ell_k}^{(j,i)}\big)=0$. Next, we calculate $\vt\big(\dot{\ell_k}^{(j,i)}\big)$. Direct computation leads to $\vt\big(\dot{\ell_k}^{(j,i)}\big)=E\big(\dot{\ell_k}^{(j,i)}\big)^2$. Recall that $\big|\dot{\ell_k}^{(j,i)}\big|\leq p_m$. Thus, we can immediately obtain that $\vt\big(\dot{\ell_k}^{(j,i)}\big)\leq p_m^2$. This completes \sc{Part 2.}\normalfont
	
	\sc{Part 3.}\normalfont$\ $We need to study $\big\Vert \dot{\alpha}_{ik}^{(j)}\big(\wt{\Omega}\big)s\big(Z_{ij},p_{kj}\big)\big\Vert$. We can rewrite $\alpha_{ik}^{(j)}s\big(Z_{ij},p_{kj}\big)=\beta_{ik}^{(j)}s\big(Z_{ij},p_{kj}\big)\big/\sum_{k=1}\beta_{ik}^{(j)}$, where $ \beta_{ik}^{(j)}=c_{ik}p_{kj}^{Z_{ij}}\big(1-p_{kj}\big)^{1-Z_{ij}}$. Then, we need to calculate $\dot{\alpha}_{ik}^{(j)}(\Omega)$, which is the first-order partial derivative about $\alpha_{ik}^{(j)}$ with respect to $\Omega=(\pi^{\top},\gamma^{\top},\theta^{\top},\sigma)^{\top}\in\mR^{2K+q+1}$. We start with $ \pi=(\pi_1,\ldots,\pi_K)^{\top}\in\mR^K$. Direct computation leads to 
	\begin{equation*}
		\left|\dfrac{\partial\alpha_{ik}^{(j)}}{\partial\pi_k}\right|=\left|\dfrac{\beta_{ik}^{(j)}}{\pi_k\sum_{k=1}\beta_{ik}^{(j)}}-\dfrac{\big(\beta_{ik}^{(j)}\big)^2}{\pi_k\big(\sum_{k=1}\beta_{ik}^{(j)}\big)^2}\right|\leq\dfrac{2}{\pi_{\min}}.
	\end{equation*}
	Similarly, we have $\left|\partial\alpha_{ik}^{(j)}/\partial\pi_m\right|\leq \pi_{\min}^{-1}$ for every $m\neq k$. Next, we study $\gamma=(\gamma_1,\ldots,\gamma_K)^{\top}\in\mR^K$. By the assumption assumptions, we have
	\begin{equation*}
		\left|\dfrac{\partial\alpha_{ik}^{(j)}}{\partial\gamma_k}\right|=\left|\left\{\dfrac{\beta_{ik}^{(j)}}{\sum_{k=1}\beta_{ik}^{(j)}}-\dfrac{\big(\beta_{ik}^{(j)}\big)^2}{\big(\sum_{k=1}\beta_{ik}^{(j)}\big)^2}\right\}\dfrac{\big(Y_i-\gamma_{k}-X_i^{\top}\theta\big)}{\sigma^2}\right|\leq \dfrac{2}{\sigma^2}\Big(2\left|\gamma_{\max}\right|+\left|\varepsilon_i\right|\Big).
	\end{equation*}
	Similarly, we have $\left|\partial\alpha_{ik}^{(j)}/\partial\gamma_m\right|\leq \big(2\left|\gamma_{\max}\right|+\left|\varepsilon_i\right|\big)/\sigma^2$  for every $ m\neq k$. We next focus on $\theta\in\mR^q$.  Direct computation leads to 
	\begin{align*}
		\left\Vert\dfrac{\partial\alpha_{ik}^{(j)}}{\partial\theta}\right\Vert=\Bigg\Vert&\left\{\dfrac{\beta_{ik}^{(j)}}{\sum_{k=1}\beta_{ik}^{(j)}}\Big(Y_i-\gamma_k-X_i^{\top}\theta\Big)X_i/\sigma^2\right\}\\
		&-\left\{\dfrac{\beta_{ik}^{(j)}}{\big(\sum_{k=1}\beta_{ik}^{(j)}\big)^2}\sum_{k=1}^K\beta_{ik}^{(j)}\Big(Y_i-\gamma_{k}-X_i^{\top}\theta\Big)X_i/\sigma^2\right\}\Bigg\Vert.
	\end{align*}
	Then, we have $\left\Vert\partial\alpha_{ik}^{(j)}/\partial\theta\right\Vert\leq (K+1)\big(2\left|\gamma_{\max}\right|+\left|\varepsilon_i\right|\big)\Vert X_i\Vert/\sigma^2$. Finally, we take partial derivative about $\alpha_{ik}^{(j)}$ with respect to $\sigma$ as
	\begin{align*}
		\left|\dfrac{\partial\alpha_{ik}^{(j)}}{\partial\sigma}\right|=&\Bigg|\left\{\dfrac{\beta_{ik}^{(j)}}{\sum_{k=1}\beta_{ik}^{(j)}}\Big(\dfrac{1}{\sigma^2}\big(Y_i-\gamma_k-X_i^{\top}\theta\big)^2-1\Big)\big/\sigma\right\}\\
		&-\left\{\dfrac{\beta_{ik}^{(j)}}{\big(\sum_{k=1}\beta_{ik}^{(j)}\big)^2}\sum_{k=1}^K\beta_{ik}^{(j)}\Big(\dfrac{1}{\sigma^2}\big(Y_i-\gamma_k-X_i^{\top}\theta\big)^2-1\Big)\big/\sigma\right\}\Bigg|.
	\end{align*}
	Then, $\left|\partial\alpha_{ik}^{(j)}/\partial\sigma\right|\leq (K+1)\big(2\left|\gamma_{\max}\right|+\left|\varepsilon_i\right|\big)^2/\sigma_{\min}^3+(K+1)/\sigma$. Combining the above results, we can obtain that $\Vert\dot{\alpha}_{ik}^{(j)}(\Omega)s(Z_{ij},p_{kj})\Vert\leq C_{K,q}M_a(\varepsilon_i)$,
	where $C_{K,q}=\sqrt{2K+q+1}>0$ and $M_a(\varepsilon_i)=p_{m}(K+1)\big(2\left|\gamma_{\max}\right|+\left|\varepsilon_i\right|\big)^2/\sigma^3+p_{m}(K+1)\big(2\left|\gamma_{\max}\right|+\left|\varepsilon_i\right|\big)\Vert X_i\Vert/\sigma^2+2p_{m}\big(2\left|\gamma_{\max}\right|+\left|\varepsilon_i\right|\big)/\sigma^2+p_{m}(K+1)/\sigma+2p_{m}/\pi_{\min}$. $M_a(\varepsilon_i)$ can be formulated as $M_a(\varepsilon_i)=M_1^a\varepsilon_i^2+M_2^a|\varepsilon_i|+M_3^a$, where $M_1^a>0$, $M_2^a>0$ and $M_3^a>0$ are some fixed constants.
	
	\sc{Part 4.}\normalfont$\ $Note that $\varepsilon_i$ is the independently and identically distributed residual following the normal distribution with mean 0 and variance $\sigma^2$. One can verify that $E\big(|\varepsilon_i|\big)=\sigma\sqrt{2/\pi}$ and $E\big(\varepsilon_i^2\big)=\sigma^2$. Then, we have
	$E\big\{M_a(\varepsilon_i)\big\}\leq C_{M_a}<\infty$, where $C_{M_a}=M_1^a\sigma^2+M_2^a\sigma\sqrt{2/\pi}+M_3^a$. 
	Next, we calculate $\vt\big\{M_a(\varepsilon_i)\big\}$. Direct computation leads to $M_a^2(\varepsilon_i)=\big(M_1^a\big)^2\varepsilon_i^4+\big(M_2^a\big)^2\varepsilon_i^2+\big(M_3^a\big)^2+2M_1^aM_2^a\varepsilon_i^2|\varepsilon_i|+2M_1^aM_3^a\varepsilon_i^2+2M_2^aM_3^a|\varepsilon_i|$.
	One can compute that $E\big(\varepsilon_i^2|\varepsilon_i|\big)=2\sigma^3\sqrt{2/\pi}$ and $E\big(\varepsilon_i^4\big)=3\sigma^4$. Then, we have $\vt\big\{M_a(\varepsilon_i)\big\}=E\big\{M_a^2(\varepsilon_i)\big\}-\big\{EM_a(\varepsilon_i)\big\}^2\leq C_{M_a}^{(v)}$, where $C_{M_a}^{(v)}=3\big(M_1^a\big)^2\sigma_{\max}^4+\big(M_2^a\big)^2\sigma_{\max}^2+\big(M_3^a\big)^2+4M_1^aM_2^a\sigma^3\sqrt{2/\pi}+2M_1^aM_3^a\sigma^2+2M_2^aM_3^a\sigma\sqrt{2/\pi}$ is a positive and fixed constant.
	
	\sc{Part 5.}\normalfont$\ $Recall that 
	$\delta_{k_1k_2}^{(j,b,i)}=\alpha_{ik_1}^{(j)}s(Z_{ij},p_{k_1j})\alpha_{ik_2}^{(j)}s(Z_{ij},p_{k_2j})-E\big(c_{ik_1}c_{ik_2}/M_{ij}\big)$, where $\alpha_{ik}^{(j)}=c_{ik}p_{kj}^{Z_{ij}}\big(1-p_{kj}\big)^{1-Z_{ij}}\big/\sum_{k=1}^Kc_{ik}p_{kj}^{Z_{ij}}\big(1-p_{kj}\big)^{1-Z_{ij}}$, $c_{ik}=\pi_k\big(\sqrt{2\pi}\sigma\big)^{-1}\exp\Big\{-\Big(Y_i-\gamma_k-X_i^{\top}\theta\Big)^2\big/$  $\big(2\sigma^2\big)\Big\}$, $ s(Z_{ij},p_{kj})=Z_{ij}/p_{kj}-\big(1-Z_{ij}\big)/\big(1-p_{kj}\big)$ and $M_{ij}=\sum_{k=1}^Kc_{ik}p_{kj}\sum_{k=1}^Kc_{ik}\big(1-p_{kj}\big)$. Note that $ \alpha_{ik}^{(j)}\leq 1$ for every $k$. And $\big|s(Z_{ij},p_{k_1j})s(Z_{ij},$ $p_{k_2j})\big|\leq p_{\min}^{-2}+(1-p_{\max})^{-2}$. For $E\big(c_{ik_1}c_{ik_2}/M_{ij}\big)$, we can obtain that $ \big|E\big(c_{ik_1}c_{ik_2}/M_{ij}\big)\big|\leq p_{\min}^{-1}(1-p_{\max})^{-1}$. Thus, we have $|\delta_{k_1k_2}^{j,b,i}|\leq p_m^2$, where $p_m^2=\big\{p_{\min}^{-1}+\big(1-p_{\max}\big)^{-1}\big\}^2$. Obviously, $E\big(\delta_{k_1k_2}^{(j,b_1,i)}\big)=0$. Next, we calculate $ \vt\big(\delta_{k_1k_2}^{(j,b_1,i)}\big)=E\big(\delta_{k_1k_2}^{(j,b,i)}\big)^2=E\Big[E\big\{\big(\delta_{k_1k_2}^{(j,b,i)}\big)^2\big|Y_i,$ $X_i\big\}\Big]$. Direct computation leads to 
	\begin{align*}
		E\left\{\Big(\delta_{k_1k_2}^{(j,b,i)}\Big)^2\big|Y_i,X_i\right\}=E\bigg[\Big\{&\alpha_{ik_1}^{(j)}s(Z_{ij},p_{k_1j})\alpha_{ik_2}^{(j)}s(Z_{ij},p_{k_2j})\Big\}^2\big|Y_i,X_i\bigg]\\
		&-\left\{E\Big(\dfrac{c_{ik_1}c_{ik_2}}{M_{ij}}\big|Y_i,X_i\Big)\right\}^2.
	\end{align*}
	Then, we can compute $\vt\big(\delta_{k_1k_2}^{(j,b,i)}\big)$ as
	\begin{equation*}	
		\vt\big(\delta_{k_1k_2}^{(j,b,i)}\big)=E\left[\dfrac{c_{ik_1}^2c_{ik_2}^2}{\big(\sum_{k=1}^Kc_{ik}\big)^4}\left\{p^{-3}_{Z_{ij}|Y_i,X_i}+\big(1-p_{Z_{ij}|Y_i,X_i}\big)^{-3}\right\}\right]-\left\{E\Big(\dfrac{c_{ik_1}c_{ik_2}}{M_{ij}}\Big)\right\}^2\leq p_v,
	\end{equation*}
	where $ p_{Z_{ij}|Y_i,X_i}=P(Z_{ij}=1|Y_i,X_i)=\big(\sum_{k=1}^Kc_{ik}p_{kj}\big)/\big(\sum_{k=1}^Kc_{ik}\big)$ and $p_v=p_{\min}^{-3}+(1-p_{\min})^{-3}+p_{\max}^{-3}+(1-p_{\max})^{-3}$ is a fixed and positive constant. 
	
	\sc{Part 6.}\normalfont$\ $Here we study $\big\Vert \dddot{\ell}_{k_1k_2}^{(j,i)}\big(\wt{\Omega}\big)\Vert$. Recall that $ \ddot{\ell}_{k_1k_2}^{(j,i)}=s(Z_{ij},p_{k_1j})s(Z_{ij},p_{k_2j})\beta_{ik_1}^{(j)}$  $\beta_{ik_2}^{(j)}\big/\big(\sum_{k=1}\beta_{ik}^{(j)}\big)^2$, where $ \beta_{ik}^{(j)}=\pi_k\big(\sqrt{2\pi}\sigma\big)^{-1}\exp\Big\{-\big(Y_i-\gamma_k-X_i^{\top}\theta\big)^2/\big(2\sigma^2\big)\Big\}p_{kj}^{Z_{ij}}\Big(1-p_{kj}\Big)^{1-Z_{ij}}$ and 
	$s(Z_{ij},p_{kj})=\Big(Z_{ij}/p_{kj}-(1-Z_{ij})/(1-p_{kj})\Big)$. Then, we need to calculate $\dddot{\ell}_{k_1k_2}^{(j,i)}(\Omega)$, which is the first-order partial derivative about $\ddot{\ell}_{k_1k_2}^{(j,i)}$ with respect to $\Omega=(\pi^{\top},\gamma^{\top},\theta^{\top},\sigma)^{\top}\in\mR^{2K+q+1}$. Define $S_{k_1k_2}^{Z_ij}=s(Z_{ij},p_{k_1j})s(Z_{ij},p_{k_2j})$, and then we have $\left|S_{k_1k_2}^{Z_ij}\right|\leq p_m$ with $ p_m=p_{\min}^{-2}+\big(1-p_{\min}\big)^{-2}$. We start with $ \pi=(\pi_1,\ldots,\pi_K)^{\top}\in\mR^K$. Direct computation leads to 
	\begin{equation*}
		\left|\dfrac{\partial\ddot{\ell}_{k_1k_2}^{(j,i)}}{\partial\pi_{k_1}}\right|=\left|\dfrac{\beta_{ik_1}^{(j)}\beta_{ik_2}^{(j)}S_{k_1k_2}^{Z_ij}}{\pi_{k_1}\big(\sum_{k=1}\beta_{ik}^{(j)}\big)^2}-\dfrac{2\big(\beta_{ik_1}^{(j)}\big)^2\beta_{ik_2}^{(j)}S_{k_1k_2}^{Z_ij}}{\pi_{k_1}\big(\sum_{k=1}\beta_{ik}^{(j)}\big)^3}\right|\leq \dfrac{3p_m}{\pi_{\min}}.
	\end{equation*}
	Similarly, we have $\left|\partial\ddot{\ell}_{k_1k_2}^{(j,i)}/\partial\pi_{k_2}\right|\leq 3p_m\pi_{\min}^{-1}$ and $\left|\partial\ddot{\ell}_{k_1k_2}^{(j,i)}/\partial\pi_{k}\right|\leq 2p_m\pi_{\min}^{-1}$ for every $k\neq k_1,k_2$. Next, we study $\gamma=(\gamma_1,\ldots,\gamma_K)^{\top}\in\mR^K$. Here we write $\mK_i$ be the real class membership of the $i$th observation (i.e., $Y_i=\gamma_{\mK_i}+X_i^{\top}\theta+\varepsilon_i$). By the theorem assumptions, we have
	\begin{equation*}
		\left|\dfrac{\partial\ddot{\ell}_{k_1k_2}^{(j,i)}}{\partial\gamma_{k_1}}\right|=\left|\left\{\dfrac{\beta_{ik_1}^{(j)}\beta_{ik_2}^{(j)}S_{k_1k_2}^{Z_ij}}{\big(\sum_{k=1}\beta_{ik}^{(j)}\big)^2}-\dfrac{2\big(\beta_{ik_1}^{(j)}\big)\beta_{ik_2}^{(j)}S_{k_1k_2}^{Z_ij}}{\big(\sum_{k=1}\beta_{ik}^{(j)}\big)^3}\right\}\dfrac{\big(Y_i-\gamma_{k_1}-X_i^{\top}\theta\big)}{\sigma^2}\right|\leq \dfrac{3p_m}{\sigma^2}\Big(2\left|\gamma_{\max}\right|+\left|\varepsilon_i\right|\Big).
	\end{equation*}
	Similarly, we have $\left|\partial\ddot{\ell}_{k_1k_2}^{(j,i)}/\partial\gamma_{k_2}\right|\leq 3p_m\big(2\left|\gamma_{\max}\right|+\left|\varepsilon_i\right|\big)/\sigma^2$ and $\left|\partial\ddot{\ell}_{k_1k_2}^{(j,i)}/\partial\gamma_{k}\right|\leq 2p_m\big(2\left|\gamma_{\max}\right|+\left|\varepsilon_i\right|\big)/\sigma^2$ for every $ k\neq k_1,k_2$. We next focus on $\theta\in\mR^q$.  Direct computation leads to 
	\begin{align*}
		\left\Vert\dfrac{\partial\ddot{\ell}_{k_1k_2}^{(j,i)}}{\partial\theta}\right\Vert=\Bigg\Vert&\left\{\dfrac{\beta_{ik_1}^{(j)}\beta_{ik_2}^{(j)}S_{k_1k_2}^{Z_ij}}{\big(\sum_{k=1}\beta_{ik}^{(j)}\big)^2}\Big(2Y_i-\gamma_{k_1}-\gamma_{k_2}-2X_i^{\top}\theta\Big)X_i/\sigma^2\right\}\\
		&-\left\{\dfrac{2\beta_{ik_1}^{(j)}\beta_{ik_2}^{(j)}S_{k_1k_2}^{Z_ij}}{\big(\sum_{k=1}\beta_{ik}^{(j)}\big)^3}\sum_{k=1}^K\beta_{ik}^{(j)}\Big(Y_i-\gamma_{k}-X_i^{\top}\theta\Big)X_i/\sigma^2\right\}\Bigg\Vert.
	\end{align*}
	Then, we have $\left\Vert\partial\ddot{\ell}_{k_1k_2}^{(j,i)}/\partial\theta\right\Vert\leq 2p_m(K+1)\big(2\left|\gamma_{\max}\right|+\left|\varepsilon_i\right|\big)\Vert X_i\Vert/\sigma^2$. Finally, we take partial derivative about $\ddot{\ell}_{k_1k_2}^{(j,i)}$ with respect to $\sigma$ as
	\begin{align*}
		\left|\dfrac{\partial\ddot{\ell}_{k_1k_2}^{(j,i)}}{\partial\sigma}\right|=&\Bigg|\left\{\dfrac{\beta_{ik_1}^{(j)}\beta_{ik_2}^{(j)}S_{k_1k_2}^{Z_ij}}{\big(\sum_{k=1}\beta_{ik}^{(j)}\big)^2}\Big(\dfrac{1}{\sigma^2}\big(Y_i-\gamma_{k_1}-X_i^{\top}\theta\big)^2+\dfrac{1}{\sigma^2}\big(Y_i-\gamma_{k_2}-X_i^{\top}\theta\big)^2-2\Big)\big/\sigma\right\}\\
		&-\left\{\dfrac{2\beta_{ik_1}^{(j)}\beta_{ik_2}^{(j)}S_{k_1k_2}^{Z_ij}}{\big(\sum_{k=1}\beta_{ik}^{(j)}\big)^3}\sum_{k=1}^K\beta_{ik}^{(j)}\Big(\dfrac{1}{\sigma^2}\big(Y_i-\gamma_k-X_i^{\top}\theta\big)^2-1\Big)\big/\sigma\right\}\Bigg|.
	\end{align*}
	Then, $\left|\partial\ddot{\ell}_{k_1k_2}^{(j,i)}/\partial\sigma\right|\leq 2p_m(K+1)\big(2\left|\gamma_{\max}\right|+\left|\varepsilon_i\right|\big)^2/\sigma^3+2p_m(K+1)/\sigma$. Combining the above results, we can obtain that $\Vert\dddot{\ell}_{k_1k_2}^{(j,i)}(\Omega)\Vert\leq C_{K,q}M_b(\varepsilon_i)$,
	where $C_{K,q}=\sqrt{2K+q+1}>0$ is a positive constant depending on $(K,q)$ and $M_b(\varepsilon_i)=2p_m(K+1)\big(2\left|\gamma_{\max}\right|+\left|\varepsilon_i\right|\big)^2/\sigma^3+2p_m(K+1)\big(2\left|\gamma_{\max}\right|+\left|\varepsilon_i\right|\big)\Vert X_i\Vert/\sigma^2+3p_m\big(2\left|\gamma_{\max}\right|+\left|\varepsilon_i\right|\big)/\sigma^2+2p_m(K+1)/\sigma+3p_m/\pi_{\min}$. $M_b(\varepsilon_i)$ can be formulated as $M_b(\varepsilon_i)=M_1^b\varepsilon_i^2+M_2^b|\varepsilon_i|+M_3^b$, where $M_1^b>0$, $M_2^b>0$ and $M_3^b>0$ are some fixed constants.
	
	\sc{Part 7.}\normalfont$\ $By the calculation process in \sc{Part 4}\normalfont, we can directly obtain that
	$E\big\{M_b(\varepsilon_i)\big\}$  $=C_{M_1}<\infty$, where $C_{M_b}=M_1^b\sigma^2+M_2^b\sigma\sqrt{2/\pi}+M_3^b$. 
	Similarly, we can calculate $\vt\big\{M_b(\varepsilon_i)\big\} $ as $\vt\big\{M_b(\varepsilon_i)\big\}=E\big\{M_b^2(\varepsilon_i)\big\}-\big\{EM_b(\varepsilon_i)\big\}^2\leq C_{M_b}^{(v)}$, where $C_{M_b}^{(v)}=3\big(M_1^b\big)^2$ $\sigma^4+\big(M_2^b\big)^2\sigma^2+\big(M_3^b\big)^2+4M_1^bM_2^b\sigma^3\sqrt{2/\pi}+2M_1^bM_3^b\sigma^2+2M_2^bM_3^b\sigma\sqrt{2/\pi}$ is a positive and fixed constant.
	
	\sc{Part 8.}\normalfont$\ $Recall that 
	$\ddot{\ell}_{k_1k_2}^{(j,i)}=\alpha_{ik_1}^{(j)}s(Z_{ij},p_{k_1j})\alpha_{ik_2}^{(j)}s(Z_{ij},p_{k_2j})$, where $\alpha_{ik}^{(j)}=c_{ik}p_{kj}^{Z_{ij}}\big(1-p_{kj}\big)^{1-Z_{ij}}\big/\sum_{k=1}^Kc_{ik}p_{kj}^{Z_{ij}}\big(1-p_{kj}\big)^{1-Z_{ij}}$, $c_{ik}=\pi_k\big(\sqrt{2\pi}\sigma\big)^{-1}\exp\Big\{-\Big(Y_i-\gamma_k-X_i^{\top}\theta\Big)^2\big/$  $\big(2\sigma^2\big)\Big\}$, and $ s(Z_{ij},p_{kj})=Z_{ij}/p_{kj}-\big(1-Z_{ij}\big)/\big(1-p_{kj}\big)$. Direct computation leads to $\dddot{\ell}_{k_1k_2}^{(j,i)}\big(p_j\big)=\partial\ddot{\ell}_{k_1k_2}^{(j,i)}/\partial p_j=\Big(\dddot{\ell}_{k_1k_21}^{(j,i)}\big(p_j\big),\ldots,\dddot{\ell}_{k_1k_2K}^{(j,i)}\big(p_j\big)\Big)\in\mR^K$, where  $\dddot{\ell}_{k_1k_2k_3}^{(j,i)}\big(p_j\big)=\alpha_{ik_1}^{(j)}s(Z_{ij},p_{k_1j})\alpha_{ik_2}^{(j)}s(Z_{ij},p_{k_2j})\alpha_{ik_3}^{(j)}s(Z_{ij},p_{k_3j})$. By the calculation process in \sc{Part 5.}\normalfont, we have $\Big|\dddot{\ell}_{k_1k_2k_3}^{(j,i)}\big(p_j\big)\Big|\leq p_m^3$. Thus, we can obtain that $\big\Vert\dddot{\ell}_{k_1k_2}^{(j,i)}\big(\Tilde{\Tilde{p}}_j\big)\Vert\leq p_m^3\sqrt{K}$.
	
	\subsection{Proof of results in Appendix B.2}
	
	\sc{Step 1.1.1}\normalfont$\ $We start with $\mC_{1}^{(m,k)}$. One can be verify that $ \big|\log\big(\wh{\pi}_m\big/\wh{\pi}_k\big)\big|\leq \big|\log\wh{\pi}_m-\log\pi_m\big|+\big|\log\wh{\pi}_k-\log\pi_k\big|+2\big|\log\pi_{\max}\big|$. We can conduct the Taylor's expansion at $\pi_m$ as $\big|\log\wh{\pi}_m-\log\pi_m\big|=\big|\tilde{\pi}_m^{-1}\big|\big|\wh{\pi}_m-\pi_m\big|\leq\pi_{\min}^{-1}\big|\wh{\pi}_m-\pi_m\big|$, where $ \tilde{\pi}_m=\alpha\pi_m+(1-\alpha)\wh{\pi}_m$ for some $ \alpha\in(0,1)$. Similarly, we have $\big|\log\wh{\pi}_k-\log\pi_k\big|\leq\pi_{\min}^{-1}\big|\wh{\pi}_k-\pi_k\big|$. Then, we have $ \mC_{i1}^{(m,k)}=p^{-1}\max_i\big|\log\big(\wh{\pi}_m\big/\wh{\pi}_k\big)\big|\leq\big(p\pi_{\min}\big)^{-1}\big|\wh{\pi}_m-\pi_m\big| +\big(p\pi_{\min}\big)^{-1}\big|\wh{\pi}_k-\pi_k\big|+2p^{-1}\big|\log\pi_{\max}\big|$.  Note that $\wh{\pi}_k-\pi_k=O_p(1/\sqrt{n})$ for every $ 1\leq k\leq K$. Thus, we have $\mC_{1}^{(m,k)}=O_p\big(1/(p\sqrt{n})\big)+O_p(1/p)=O_p(1/p)$.
	
	\sc{Step 1.1.2}\normalfont$\ $Next, we study $\mC_{2}^{(m,k)}$. Recall that $ \mC_{2}^{(m,k)}=3p^{-1}\max_i\big(\wh{\gamma}_k-\gamma_k\big)^2/(2\wh{\sigma}^2)=3p^{-1}\max_i\Big\{\big(\wh{\gamma}_k-\gamma_k\big)^2/(2\sigma^2)\Big\}\big(\sigma/\wh{\sigma}\big)^2$. By the theorem assumption, we have $\big(\wh{\gamma}_k-\gamma_k\big)^2/(2\sigma^2)\leq \big(\wh{\gamma}_k-\gamma_k\big)^2/(2\sigma_{min}^2)$. Note that $\wh{\gamma}_k-\gamma_k=O_p(1/\sqrt{n})$ for every $ 1\leq k\leq K$ and $\wh{\sigma}-\sigma=O_p(1/\sqrt{n})$. Then, we have $\big|\sigma/\wh{\sigma}-1\big|=O_p(1/\sqrt{n})$. Note that these are all independent of the subscript $i$. Thus, we can obtain that $\mC_{2}^{(m,k)}=O_p\big(1/(pn)\big)$.
	
	\sc{Step 1.1.3}\normalfont$\ $We then study $\mC_{3}^{(m,k)}$. Note that $ X_i=\big(X_{i1},\ldots,X_{iq}\big)^{\top}\in\mR^q$ is the independently and identically distributed random vector. By the theorem assumption, $X_{ij}$ follows a sub-Guassian distribution. By definition, we know that $P\Big\{\big|X_{ij}\big|>a\Big\}\leq C_1\exp\big(-C_2a^2\big)$ for any $ a>0$, where $ C_1>0$ and $ C_2>0 $ are some fixed constants. Note that $\big\Vert X_i\big\Vert=\Big(\sum_{j=1}^qX_{ij}^2\Big)^{1/2}\leq\sqrt{q}\max_j\big|X_{ij}\big|$. Then, we can obtain that $P\Big\{\max_i\Vert X_i\Vert>a\Big\}\leq P\Big\{\max_i\max_j\big|X_{ij}\big|>a/\sqrt{q}\Big\}\leq nqC_1\exp\{-C_2a^2/q\}$. Note that $\mC_{3}^{(m,k)}= 3p^{-1}\sqrt{q}\max_i$ $\max_j\big|X_{ij}\big|^2\big\Vert\wh{\theta}-\theta\big\Vert^2\big/\big(2\wh{\sigma}^2\big)$. Then using the same technique in \sc{Step 1.1.2}\normalfont, we can obtain that $\mC_{3}^{(m,k)}=O_p\big(1/(pn)\big)$.
	
	\sc{Step 1.1.4}\normalfont$\ $Finally, we study $\mC_{4}^{(m,k)}$. Recall that $\mC_{4}^{(m,k)}=3p^{-1}\max_i\varepsilon_i^2/(2\wh{\sigma}^2)=3/(2p)\max_i\big(\varepsilon_i/\sigma\big)^2\big(\sigma/\wh{\sigma}\big)^2$. To this end, we shall focus on the asymptotic behavior of $\max_i$  $p^{-1}\big(\varepsilon_i/\sigma\big)^2$. Note that $\varepsilon_i $ follows a normal distribution with mean $0$ and variance $\sigma^2$. By the Chernoff's bound \citep{chernoff1952}, we can obtain that $ P\Big\{p^{-1}\big(\varepsilon_i/\sigma\big)^2>\varepsilon\Big\}\leq 2\exp\big(-\varepsilon p/2\big)$ for any given $\varepsilon>0$. Then, we have $ P\Big\{\max_ip^{-1}\big(\varepsilon_i/\sigma\big)^2>\varepsilon\Big\}\leq\sum_{i=1}^nP\Big\{p^{-1}\big(\varepsilon_i/\sigma\big)^2>\varepsilon\Big\}\leq2\exp\Big(-p\varepsilon/2+\log n\Big)$. Thus, we have $ \max_ip^{-1}\big(\varepsilon_i/\sigma\big)^2=O_p(\log n/p)$. We have argued that $\big|\sigma/\wh{\sigma}-1\big|=O_p(1/\sqrt{n})$ in \sc{Step 1.1.2}\normalfont. Thus, we can obtain that $\max_i\mC_{4}^{(m,k)}=O_p(\log n/p)$. 
	
	\sc{Step 1.2.1}\normalfont$\ $Recall that $\wh{V_{ij}^{(m,k)}}=Z_{ij}\log\Big(\wh{p}_{mj}\big/\wh{p}_{kj}\Big)+\Big(1-Z_{ij}\Big)\log\Big\{\big(1-\wh{p}_{mj}\big)\big/\big(1-\wh{p}_{kj}\big)\Big\}$ and $V_{ij}^{(m,k)}=Z_{ij}\log\Big(p_{mj}\big/p_{kj}\Big)+\Big(1-Z_{ij}\Big)\log\Big\{\big(1-p_{mj}\big)\big/\big(1-p_{kj}\big)\Big\}$. Direct computation leads to $\wh{V_{ij}^{(m,k)}}-V_{ij}^{(m,k)}=Z_{ij}\Big\{\big(\log\wh{p}_{mj}-\log p_{mj}\big)-\big(\log\wh{p}_{kj}-\log p_{kj}\big)\Big\}+\big(1-Z_{ij}\big)\Big[\big\{\log(1-\wh{p}_{mj})-\log \big(1-p_{mj}\big)\big\}-\big\{\log(1-\wh{p}_{kj})-\log \big(1-p_{kj}\big)\big\}\Big]$. By the argument of standard M-estimation, we have $ \wh{p}_{kj}-p_{kj}=O_p(1/\sqrt{n})$ for every $ 1\leq k\leq K$. Then, we can conduct the Taylor's expansion about $\log\wh{p}_{kj}-\log p_{kj} $ at $ p_{kj}$ as $ \log\wh{p}_{kj}-\log p_{kj} =\wt{p}_{kj}^{-1}\big(\wh{p}_{kj}-p_{kj}\big)$, where $ \wt{p}_{kj}\in\big(p_{kj},\wh{p}_{kj}\big) $. By the theorem assumption, we have $\big|\wt{p}_{kj}^{-1}\big|\leq p_{\min}^{-1}$. As a result, $ \big|\log\wh{p}_{kj}-\log p_{kj}\big|\leq p_{\min}^{-1}\big|\wh{p}_{kj}-p_{kj}\big|$ for every $ 1\leq k\leq K$. Similarly, we have $ \big|\log(1-\wh{p}_{kj})-\log \big(1-p_{kj}\big)\big|\leq \big(1-p_{\max}\big)^{-1}\big|\wh{p}_{kj}-p_{kj}\big|$ for every $ 1\leq k\leq K$. Note that $ Z_{ij}\in\{0,1\}$ is a binary random variable. Thus, we can obtain that $\Big|\wh{V_{ij}^{(m,k)}}-V_{ij}^{(m,k)}\Big|\leq 2p_m\big|\wh{p}_{mj}-p_{mj}\big|+2p_m\big|\wh{p}_{kj}-p_{kj}\big|$, where $ p_m= p_{\min}^{-1}+\big(1-p_{\max}\big)^{-1}$.
	
	\sc{Step 1.3.1}\normalfont$\ $Recall that $\mathcal{D}_a=p\big(\varepsilon_{(\nu,E)}^p\big)^2/2$ and $ \varepsilon_{(\nu,E)}^p=\varepsilon-2\nu-\sum_{j=1}^pE\big(V_{ij}^{(m,k)}\big)/p $, which is determined by $\sum_{j=1}^pE\big(V_{ij}^{(m,k)}\big)/p$. Direct computation leads to $ E\big(V_{ij}^{(m,k)}\big)=p_{kj}\log\big(p_{mj}$  $/p_{kj}\big)$ $+\big(1-p_{kj}\big)\log\big\{(1-p_{mj})/(1-p_{kj})\big\}$. By the theorem assumptions, we shall have $ 0<\Delta_{\min}\leq -\sum_{j=1}^pE\big(V_{ij}^{(m,k)}\big)/p\leq\Delta_{\max}\leq b$. Here $\Delta_{\max}=p_{\max}\log\big(p_{\max}/p_{\min}\big)+\big(1-p_{\min}\big)\log\big\{(1-p_{\min})/(1-p_{\max})\big\}$. Thus, we obtain that $\mathcal{D}_a\geq C_1p$, where $ C_1=(\varepsilon-2\nu+\Delta_{\min})^2/2>0$ is a fixed constant.
	
	\sc{Step 1.3.2}\normalfont$\ $By definition, $\mathcal{D}_b$ depends on both $E\big(V_{ij}^{(m,k)}\big)$ and $\vt\big(V_{ij}^{(m,k)}\big)$. We have argued $E\big(V_{ij}^{(m,k)}\big)$ in \sc{Step 1.3.1.}\normalfont$\ $Next, we shall focus on $\vt\big(V_{ij}^{(m,k)}\big)$. By the theorem assumption (\Rmnum{2}), we have $\vt\big(V_{ij}^{(m,k)}\big)\leq C_{\sigma}$, where $C_{\sigma}=\log^2\Big[p_{\max}(1-p_{\min})/\big\{p_{\min}(1-p_{\max})\big\}\Big]\Big/4$. Therefore, we can obtain $\sum_{j=1}^p\vt\big(V_{ij}^{(m,k)}\big)\leq pC_{\sigma}$. Note that $ 0<\Delta_{\min}\leq -\sum_{j=1}^pE\big(V_{ij}^{(m,k)}\big)/p\leq\Delta_{\max}\leq b$. Then, we have $ \mathcal{D}_b=\sum_{j=1}\vt\big(V_{ij}^{(m,k)}\big)/p+b\varepsilon_{(\nu,E)}^p/3\leq C_{\sigma}+b\Big\{\varepsilon-2\nu-\sum_{j=1}^pE\big(V_{ij}^{(m,k)}\big)/p\Big\}\big/3\leq C_{\sigma}+b^2$. Thus, we can directly obtain the upper bound of $\mathcal{D}_b$ as $\mathcal{D}_b\leq C_2$, where $ C_2=C_{\sigma}+b^2>0$ is a fixed constant.
	
	\subsection{Proof of results in Appendix B.3}
	
	Recall that $\Delta_{\fxx}^i=X_i^{\pi}X_i^{\pi\top}-X_i^aX_i^{a\top} =(\delta_{j_1j_2}^i)\in\mR^{(K+q)\times(K+q)}$. Here $ X_i^a=(a_i^{\top},X_i^{\top})^{\top}\in\mR^{K+q}$ with $ a_i=(a_{i1},\ldots,a_{iK})^{\top}\in\mR^K $ and $ X_i^{\pi}=(\wh{\pi}_i^{\top},X_i^{\top})^{\top}\in\mR^{K+q}$ with $ \wh{\pi}_i=(\wh{\pi}_{i1},\ldots,\wh{\pi}_{iK})^{\top}\in\mR^K $ . We need to prove that $\max_{j_1,j_2}\big|\delta_{j_1j_2}^i\big|\leq\max_k\big|\wh{\pi}_{ik}-a_{ik}\big|\big(2+\big\Vert X_i\big\Vert\big)$. Note that $ \Delta_{\fxx}^i $ is a symmetric matrix. Then, $ \Delta_{\fxx}^i $ can be divided into three parts. First for $ 1\leq j_1,j_2\leq K $,  direct computation leads to $\delta_{j_1j_2}^i=a_{ij_1}a_{ij_2}-\wh{\pi}_{ij_1}\wh{\pi}_{ij_2}$. Note that $ \big|a_{ik}\big|\leq 1$ and $ \big|\wh{\pi}_{ik}\big|\leq 1$. Then, we have $ \big|\delta_{j_1j_2}^i\big|\leq \big|a_{ij_1}-\wh{\pi}_{ij_1}\big|+\big|a_{ij_2}-\wh{\pi}_{ij_2}\big|\leq 2\max_k\big|\wh{\pi}_{ik}-a_{ik}\big|$. Second for $ K+1\leq j_1\leq K+q $ and $ 1\leq j_2\leq K $, $\delta_{j_1j_2}^i=\big(a_{ij_2}-\wh{\pi}_{ij_2}\big)X_{i(j_1-K)}$. Thus, we have $ \big|\delta_{j_1j_2}^i\big|\leq \max_k\big|\wh{\pi}_{ik}-a_{ik}\big|\cdot\big\Vert X_i\big\Vert$. Third for $ K+1\leq j_1,j_2\leq K+q $, we can immediately obtain that $ \delta_{j_1j_2}^i=0 $ by definition. Combining the above results of those three parts, we completes this verification details for $\max_{j_1,j_2}\big|\delta_{j_1j_2}^i\big|\leq\max_k\big|\wh{\pi}_{ik}-a_{ik}\big|\big(2+\big\Vert X_i\big\Vert\big)$.
\end{appendix}


\end{document}